\documentclass[linenumbers]{aastex631}
\nolinenumbers
\usepackage{threeparttable}
\usepackage{rotating}
\usepackage{subfigure}

\begin{document}

\title{The Star Formation History of Nearby Galaxies: A Machine Learning Approach}

\correspondingauthor{Yujiao Yang}
\email{yjyang@bao.ac.cn}
\author[0000-0002-3180-2327]{Yujiao Yang}
\affiliation{Key Laboratory of Space Astronomy and Technology, National Astronomical Observatories, Chinese Academy of Sciences, Beijing 100101, People's Republic of China}

\author[0000-0002-1802-6917]{Chao Liu}
\affiliation{Key Laboratory of Space Astronomy and Technology, National Astronomical Observatories, Chinese Academy of Sciences, Beijing 100101, People's Republic of China}
\affiliation{University of Chinese Academy of Sciences, Beijing 100049, People's Republic of China}

\author[0000-0001-8247-4936]{Ming Yang}
\affiliation{Key Laboratory of Space Astronomy and Technology, National Astronomical Observatories, Chinese Academy of Sciences, Beijing 100101, People's Republic of China}

\author[0009-0002-9472-3033]{Yun Zheng}
\affiliation{Zhejiang Lab, Hangzhou, Zhejiang, 311121, People's Republic of China}

\author[0000-0003-3347-7596]{Tian Hao}
\affiliation{Key Laboratory of Space Astronomy and Technology, National Astronomical Observatories, Chinese Academy of Sciences, Beijing 100101, People's Republic of China}
\affiliation{Institute for Frontiers in Astronomy and Astrophysics, Beijing Normal University, Beijing, 102206, People's Republic of China}

\begin{abstract}
Reproducing color-magnitude diagrams (CMDs) of star-resolved galaxies is one of the most precise methods for measuring the star formation history (SFH) of nearby galaxies back to the earliest time. The upcoming big data era poses challenges to the traditional numerical technique in its capacity to deal with vast amounts of data, which motivates us to explore the feasibility of employing machine learning networks in this field. In this study, we refine the synthetic CMD method with a state-of-the-art theoretical stellar evolution model to simulate the properties of stellar populations, incorporate the convolutional neural network (CNN) in the fitting process to enhance the efficiency, and innovate the initial stellar mass estimation to improve the flexibility. The fine-tuned deep learning network, named \texttt{SFHNet}, has been tested with synthetic data and further validated with photometric data collected from the Hubble Space Telescope (\textit{HST}). The derived SFHs are largely in accordance with those reported in the literature. Furthermore, the network provides detailed insights into the distribution of stellar density, initial stellar mass, and star formation rate (SFR) over the age-metallicity map. The application of the deep learning network not only measures the SFH accurately but also enhances the synthetic CMD method's efficiency and flexibility, thereby facilitating a more comprehensive and in-depth understanding of nearby galaxies.

\end{abstract}


\section{Introduction} \label{sec:intro}
The star formation history (SFH), defined as the star formation rate (SFR) as a function of lookback time and metallicity, is a critical parameter for understanding the formation and evolution of galaxies. The SFH characterizes the process of star formation, evolution, and death, providing insights into the underlying physical mechanisms governing galaxies. In the case of star-resolved galaxies, the color-magnitude diagrams (CMDs) record the temperature and luminosity of individual stars and display the fossils of stellar evolution. Therefore, deciphering the information contained in CMDs provides a way to get the most accurate SFH that can be traced back to the earliest times \citep[][and references therein] {2005ARA&A..43..387G,2009ARA&A..47..371T,2022NatAs...6...48A}. The widely used method, synthetic CMD, has been well developed and examined over the past three decades. This has benefited from the high resolution of the Hubble Space Telescope (\textit{HST}) and the improvement of the stellar evolution models in the theoretical field.

The synthetic CMD method is employed for the measurement of nearby galaxies' SFHs through the analysis of stellar density over the CMD space. The basic assumption is that a galaxy is constituted of a mixture of stellar populations with different parameters (e.g., age, metallicity, reddening, binary fraction, etc.). These attributes define the overall CMD features. The tight relationship between CMD features and the parameters of stellar populations enables the estimation of the SFH. For example, the luminosities of the main sequence turn-off (MSTO) and subgiant branch decline with age. The slope of the red giant branch (RGB) is a reliable indicator of metallicity. The properties of the RGB bump are influenced by stellar mass and metallicity. The morphology of the horizontal branch provides insight into the impact of mass loss during the RGB phase. However, some CMD features are insensitive to the parameters mentioned above, including the overlap between young and old populations in the lower-mass main sequence stage, the ambiguity between young massive stars and blue straggler stars (BSSs), and the degeneracy of metallicity, age, and dust extinction of RGB stars. These factors complicate the derivation of the SFH.

\cite{1991AJ....102..951T} first used the synthetic CMD method to sketch recent SFHs in nearby galaxies. After that, many groups improved this general method \citep{1992ApJ...388..400B, 1996ApJ...462..672T, 1997NewA....2..397D, 1997AJ....114..669A, 1999MNRAS.304..705H, 2000MNRAS.317..831H}. The synthetic CMD method has been developed in many perspectives to improve the goodness of fit. Such as exploring different maximum likelihood estimators, comparing a series of CMDs observed in multiple filters, changing the shape and size of the cells in the Hess diagrams (HDs), varying the weights of cells according to their sensitivity to the parameters of stellar populations, using additional metallicity information from spectroscopic measurements \citep[see][for reviews]{2005ARA&A..43..387G,2009ARA&A..47..371T}. These methods can generally be divided into two categories, depending on whether the CMD is parameterized. The main parameterization codes are \texttt{MATCH} \citep{2002MNRAS.332...91D,2012ApJ...751...60D, 2013ApJ...775...76D}, \texttt{IAC}\citep{2004AJ....128.1465A, 2009AJ....138..558A} and \texttt{Cole} \citep{2003ApJ...596..253S,2007ApJ...659L..17C}. These codes use different merit functions and algorithms in the solving process, but all have the same general steps. Their descendants \texttt{TALOS} \citep[with the spectroscopic metallicity distributions;][]{2012A&A...539A.103D, 2012A&A...544A..73D}, \texttt{MORGOTH} \citep[include the mass loss of RGB;][]{2018MNRAS.480.1587S}, and \texttt{SFERA} \citep[combine advantages of the three main codes;][]{2015ApJ...811...76C,2018ApJ...857...63S,2024MNRAS.527.5339B} include additional information. The nonparametric approach is provided by Bayesian inference \citep[e.g.,][]{1999MNRAS.304..705H, 2000MNRAS.317..831H,2013MNRAS.435.2171W}. While these techniques differ in detail, their robustness has been verified and they yield consistent SFHs for many galaxies \citep{2003ApJ...596..253S,2005ARA&A..43..387G,2010ApJ...722.1864M,2010ApJ...720.1225M}.

Although the above techniques show great consistency in measuring SFHs of Local Group galaxies \citep[][e.g.,]{2009ARA&A..47..371T,2014ApJ...789..147W,2017ApJ...846..145W,2017ApJ...837..102S,2023ApJ...956...86S} and beyond that \citep[e.g.,][]{2003AJ....126..187D,2008ApJ...689..160W,2010ApJ...721..297M,2011ApJ...739....5W,2011A&A...530A..59C}, the efficiency need to be improved. The efficiency of numerical methods is primarily limited by the optimization method (Markov Chain Monte Carlo is most commonly used). For example, 50,000 CPU hours are required for \texttt{MATCH} to derive the spatially resolved SFHs of M31 \citep{2015ApJ...805..183L}. In addition to the age and metallicity, many other parameters (such as the varied initial mass function (IMF), the metallicity dispersion, the binary fraction, the interactions between binaries, the helium content, the overshooting process, and the degeneracy between IMF and SFR) also contribute to shaping the CMDs of galaxies. A nearby galaxy is an excellent laboratory to test these effects. Instead of optimizing the numerical and Bayesian methods, we provide a novel approach by applying deep learning networks in this field.

Another motivation is the high efficiency required in the era of big data. Although the high-resolution data of \textit{HST} makes the SFH measurement possible, its small field of view (FoV) hinders the complete understanding of galaxies. For example, The Panchromatic Hubble Andromeda Treasury \citep[PHAT;][]{2012ApJS..200...18D,2014ApJS..215....9W,2023ApJS..268...48W} and its extension to the Triangulum Galaxy, known as PHATTER \citep{2021ApJS..253...53W}, are surveys focused on the resolved stellar populations in M31 and M33, respectively. But these surveys only cover a portion of the galaxies ($\sim$ 0.5 $deg^2$ for M31, and $\sim$ 0.1 $deg^2$ for M33 ). The upcoming space telescopes, such as the China Space Station Telescope  \citep[\textit{CSST};][]{2011SSPMA..41.1441Z, 2018cosp...42E3821Z, Zhan2021}, \textit{Euclid}  \citep{2024arXiv240513491E} and the Roman Space Telescope \citep{2019arXiv190205569A}, will collect vast amounts of data ranging from the ultraviolet to the near-infrared. For example,  \textit{CSST} is a 2-meter space telescope with a large FoV of 1.1 $deg^2$. The main \textit{CSST} survey will scan 17,500 $deg^2$ of the sky and provide seven-band photometric data with a detection limit in the $g$ band of 26.3 mag. The \textit{Euclid} wide survey will cover $\sim$ 15,000 $deg^2$ of the sky, and its early release observations show great capacity for studying nearby galaxies \citep{2024arXiv240513499H}. It can be predicted that the number of star-resolved galaxies and the homogeneity of individual galaxies will improve significantly. Therefore, a high-efficiency method is needed to quantitatively measure the SFHs of large star-resolved galaxies.

In recent years, there has been a notable increase in the application of machine learning networks in the field of astronomy. Convolutional neural networks \citep[CNNs,][]{FUKUSHIMA1982455, 726791} show significant advantages in addressing computer vision tasks, such as image classification, object detection, and semantic segmentation. From \texttt{AlexNet} \citep{10.1145/3065386}, the winner of 2012 ImageNet competition, the accuracy and efficiency of CNNs have been improved from different perspectives, such as introducing different convolutional kernel sizes \citep[Inception Network;][]{2014arXiv1409.4842S}, tuning the network width, depth and resolutions \citep[EfficientNet;][]{2015arXiv151200567S, 2019arXiv190511946T}, connecting the residuals \citep[ResNet;][]{2015arXiv151203385H,2016arXiv160207261S}, changing the way of normalization \citep{2015arXiv150203167I} and so on. CNNs are composed of a series of convolutional layers and can theoretically approximate any continuous functions based on the Universal Approximation Theorem \citep[e.g.,][]{Cybenko1989ApproximationBS, FUNAHASHI1989183}. Therefore, we improve the synthetic CMD method with CNNs to measure SFHs of galaxies. In comparison to traditional numerical methods, deep learning models demonstrate superior efficiency, flexibility, and scalability in processing vast volumes of data. In particular, their ability to automatically extract image features reduces the need to choose statistical estimators and optimization algorithms. Moreover, deep learning models can handle complicated, non-linear relationships between the input and the output variables. The application of the machine learning technique may shed light on the issues that are not yet fully understood, such as the effects of interacting binary stars, the outliers in CMD that can not be explained by present-day stellar evolution models, mass loss in the post-MS stage, and so forth. Additionally, deep learning models can scale to large and high-dimensional data, which is a critical capability for evaluating the impacts of additional parameters on SFH estimation, such as varied IMF, differential reddening, and binary distribution. It is a good opportunity to improve the synthetic CMD method with the state-of-the-art stellar evolution model and a high-efficiency data processing method. 

The deep learning method has been validated on eight dwarf galaxies, namely Ursa Minor, Tucana, NGC 185, NGC 205, NGC 6822, Sextans A, Phoenix, and Pegasus. A summary of the observational properties is provided in Table \ref{table:results}. These galaxies belong to four morphological types and have distances ranging from 0.07 to 1.4 $Mpc$. The morphology of a galaxy is tightly related to its SFH. The distance determines the data quality and the depth to which the CMD features can be observed, which is the primary factor affecting the accuracy of the SFH. Therefore, we simply select the nearest and farthest galaxies for each morphological type within the available data set.

This paper is organized as follows. Section \ref{sec:syncmd} briefly introduces the main steps of the synthetic CMD method. Then, we describe our deep learning approach in Section \ref{sec:mln}, including the network architecture, the generation of mock data, and network training and validation. Section \ref{sec:data} presents the SFHs of eight dwarf galaxies predicted by \texttt{SFHNet}, serving as testbeds for the development of machine learning methodology. We discuss the potential benefits and shortcomings of our approach and provide a quick summary in Section \ref{sec:sum}.

\section{Synthetic CMD method}\label{sec:syncmd}
The synthetic CMD method is a sophisticated technique that combines theoretical models with observational data to reconstruct the SFH of galaxies. The main steps can be summarized as follows: (1) Create synthetic CMDs with stellar evolution models and take observational effects (uncertainties and incompleteness) into account. Age and metallicity are the major parameters that determine the distribution of stars in CMD space. Additionally, the age-metallicity relation, IMF, binary fraction, helium content, differential reddening, and line-of-sight depth may also vary the CMD features. (2) Construct a statistical estimator of the degree of resemblance between a synthetic CMD constructed from an assumed SFH and the observed CMD. The most commonly used techniques are the $\chi^2$ technique, likelihood statistic, and Bayesian inference. (3) Select the synthetic CMD that maximizes/minimizes the estimator. (4) Deriving the SFH from the best-fit synthetic CMD. 

The fitting process is a pivotal step in the synthetic CMD method, which aims to find the synthetic CMD (with known SFH) that most closely resembles to the observed one. In numerical/Bayesian methods, the fitting process is to find the minimum statistical estimators/maximum likelihood. The optimization process is primarily affected by the observational effects (photometric uncertainties and incompleteness). The artificial star test (AST) is the most accurate method to simulate these observational effects, which is a method to measure the accuracy and reliability of photometric stars by injecting simulated stars into astronomical images and evaluating their recovery. The varied sensitivities of CMD features to ages or metallicities may also affect the optimization process. Many methods have been employed in the literature to improve the goodness of fit, including adjusting the fitted CMD region for different age ranges, over-weight, de-weight, excluding some CMD features, or using custom HD cells. In this work, we follow the main steps of the synthetic CMD method and improve the fitting process with a deep learning network.
 
\section{The Machine Learning Approach}\label{sec:mln}
The key step of the synthetic CMD method is to find the best-fit simulated CMD with known SFH. Machine learning involves training algorithms on labeled images to learn patterns and make predictions. During training, the network adjusts its weights and biases to minimize the prediction errors, aiming to achieve a model that can predicate the SFH based on the input CMDs with high accuracy. In this section, we describe the routine of our machine learning approach step by step, including the network architecture, the generation of synthetic CMDs, and the network training, validation, and performance evaluation.

\begin{figure}[!htb]
\centering
\includegraphics[width=1\textwidth]{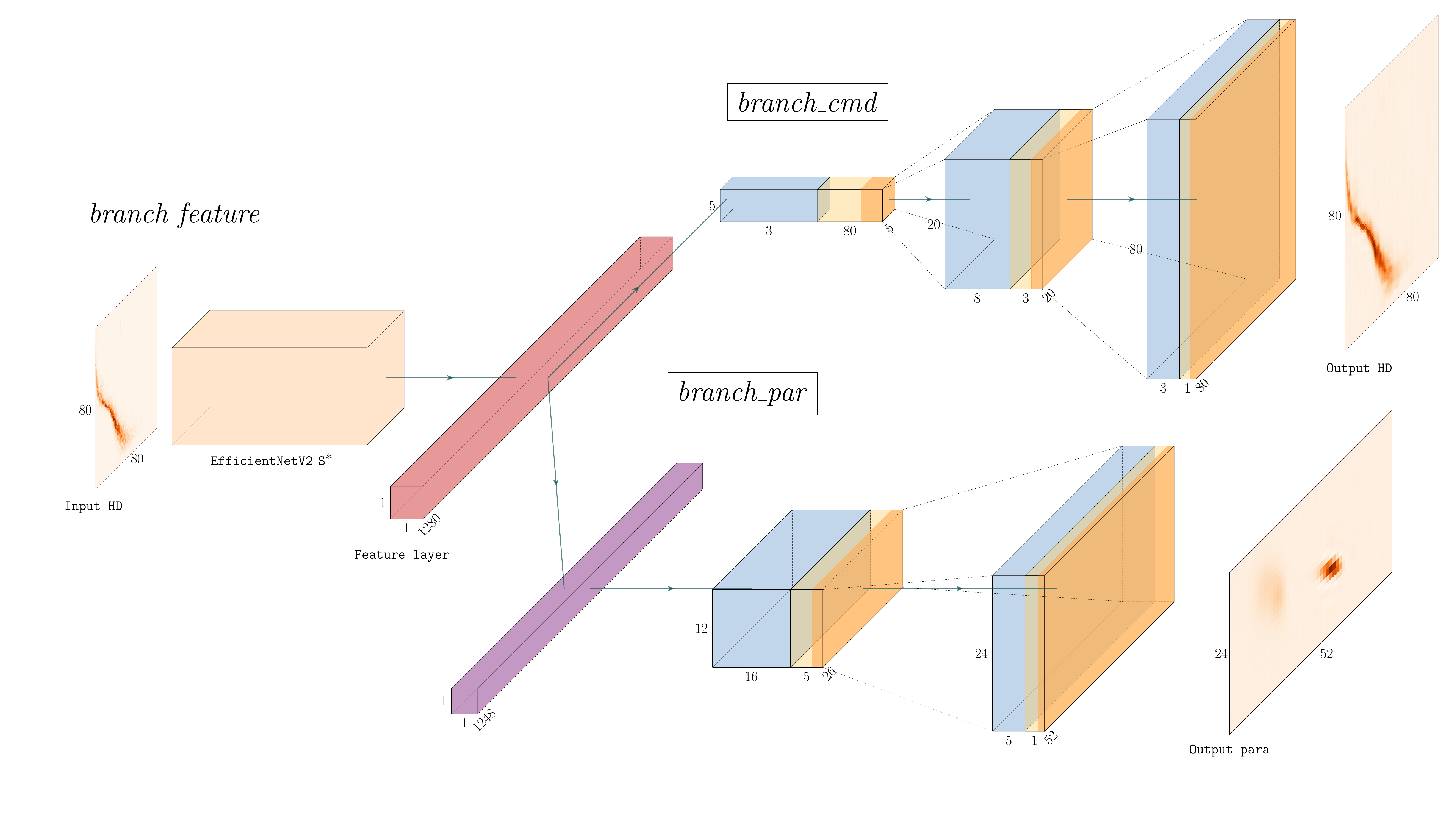}
\caption{The architecture of \texttt{SFHNet}. The \textit{branch$\_$feature} includes the input data (the HDs of synthetic galaxies), the main part of \texttt{EfficientNetV2\_S}, and the extracted feature layer. The \textit{branch$\_$cmd} comprises three sets of (Upsampling layer + Convolution layer) combinations; the \textit{branch$\_$par} includes a global average pooling layer and a fully connected layer,  and two sets of (Upsampling layer + Convolution layer) combinations. The activation function \texttt{SiLU} is applied in the entire network.}
\label{fig:ML_network}
\end{figure}

\subsection{Network Architecture}\label{subsec:model}
Instead of constructing the network from scratch, we use \texttt{EfficientNetV2} as the feature extractor. \texttt{EfficientNetV2} \citep{2021arXiv210400298T} uses the compound scaling method and neural architecture search method to jointly optimize training speed and parameter efficiency. The compound scaling method can uniformly scale the network's depth, width, and resolution. Besides that, the different sizes of convolutional filters  (1$\times$1, 3$\times$3, 5$\times$5) enable the network to capture image features at varying spatial scales. The majority of \texttt{EfficientNetV2\_S} is employed as the feature extractor, replacing the last average pooling layer and classification layer with two branches to predict the HD and the age-metallicity parameters behind the input HD. The modified network is named \texttt{SFHNet}.

Figure \ref{fig:ML_network} illustrates the architecture of \texttt{SFHNet}, which comprises three branches. The \textit{branch$\_$feature} is the main part of \texttt{EfficientNetV2\_S}, responsible for extracting image features from the input HDs. In detail, this branch comprises three Fused-MBConv \citep{2020arXiv200302838G}, three MBConv\citep{2018arXiv180104381S, 2019arXiv190511946T}, and two convolution layers. The training-aware neural architecture search and scaling combination make \texttt{EfficientNetV2\_S} significantly outperform previous models; see the details in \cite{2019arXiv190511946T,2021arXiv210400298T}. The \textit{branch$\_$cmd} is composed of three sets (Upsampling layer + Convolution layer), aiming to reproduce the input HD. The \textit{branch$\_$par} starts with a global average pooling layer and a fully connected layer, followed by two sets of (Upsampling layer + Convolution layer) and ends with a $24\times52$ array. The output of \textit{branch$\_$par} is the distribution of stellar density across the age-metallicity map, from which we can derive the stellar mass of each population and solve the SFH.

The input HDs of mock galaxies (in the form of an 80$\times$80 array) and the true age-metallicity map behind the HDs (in the form of a 24$\times$60 array) are processed through the \textit{branch$\_$feature} and extracted as the feature layer. \texttt{EfficientNetV2\_S} is wide and deep enough to capture the basic features of images and to reveal the more abstract concepts that underlie them. Predictions of HDs and age-metallicity distributions are based on this feature layer. During the training process, three losses are used, including two losses for two output branches and their sum. The loss value is the mean square error between the input and output.

\subsection{Mock Data}\label{subsec:data}
Deep learning networks require a substantial number of galaxies with known SFHs for training. Since the number of star-resolved galaxies is limited, we generate $10^5$ mock galaxies extracted from stellar evolution models and convolved with observational effects of the test galaxies. The generation of training data is achieved through two main steps, i.e., synthetic simple stellar populations (SSPs) and synthetic galaxies.

\subsubsection{Synthetic Simple Stellar Populations}\label{subsubsec:step1}
Generally, the SSPs are extracted from isochrones with IMF, binary fraction, distance, and observation effects. In this work, we use isochrones from PAdova and TRieste Stellar Evolution Code \citep[PARSEC;][]{2012MNRAS.427..127B}\footnote{http://stev.oapd.inaf.it/cgi-bin/cmd}. In order to include the full range of ages and metallicities, theoretical isochrones with 52 age bins (0.1 dex from log($t$) = 6.6 to 8.6, and 0.05 dex from log($t$) = 8.6 to 10.15) and 24 metallicity bins (0.1 dex from [M/H] = -2.0 to 0.3) are adopted. 

In detail, an SSP is generated through three steps. Firstly, a population of $10^5$ stars is generated, with their masses following the Kroupa IMF \citep{2001MNRAS.322..231K}. The minimum and maximum masses are constrained by the observational magnitude limits. The theoretical absolute magnitudes are derived through interpolation based on the mass-luminosity relationship obtained from the isochrone. Secondly, 35$\%$  of the stars are assigned as the primary components in binary systems. The binary fraction $f_{\rm b} = 0.35$ is a typical value used in the literature \citep{2014ApJ...789..147W, 2024ApJ...961...16M}, and the effect of binaries will be further discussed in Section \ref{subsec:errors}. The secondary components are generated following the same IMF but with the masses restricted to MS stars. The magnitude of a binary system is equal to the sum of the luminosity of the two randomly paired components \citep{2009A&A...493..979K}. Distance modulus from \cite{2014ApJ...789..147W} are adopted to get the theoretical apparent magnitude. Thus, we get an SSP containing $10^5$ stars, of which 35$\%$ are binaries. The third step is to convolve the theoretical apparent magnitude with observational effects. Regarding magnitude uncertainties, a polynomial function is employed to fit the relation between $\sigma_{\rm mag}$ and the magnitude for each filter. The observed apparent magnitude is randomly selected from a Gaussian distribution with the mean $\mu$ as the theoretical apparent magnitude, and the standard deviation $\sigma$ as the corresponding uncertainty derived from the fitted function. Regarding incompleteness, the distribution of completeness across the color-magnitude space is first calculated by interpolating the completeness curves of two filters, which are derived from ASTs. Stars allocated to the color-magnitude cells with completeness values less than the adopted completeness cut (0.3) are removed. For the remaining cells, stars are randomly removed to make the completeness consistent with the completeness map. The completeness cut is selected to reconcile the competing demands for deep CMDs and high-quality photometric data. Following the three steps mentioned above, we generate 1,248 synthetic SSPs with 35$\%$ binaries and incorporate the associated uncertainties and incompleteness of the observed galaxy.

Here, we briefly discuss the influences of our simplified binary model on the fitting process. The most important factor is the overlap between young MS stars and blue straggler stars (BSSs). The latter are hotter and brighter than MSTO stars in star clusters and are thought of as the products of interacting binaries. Therefore, BSSs will be treated as young MS stars, resulting in an overestimation of the recent SFR. For example, Figure \ref{fig:obs} shows Ursa Minor contains well-populated blue plume stars above the MSTO region and these stars have been proved to be BSSs \citep{2002AJ....123..813D}. The effects of young MS-BSSs ambiguity will be further discussed in Section \ref{subsec:errors}. We also assume that the secondary components are MS stars, partly following the fact that the companions of post-MS stage stars are either more massive and have evolved into the white dwarf phase or less massive in the low MS phase. If we remove the mass constraint, high mass-ratio binaries generated from our simplified binary model can significantly change the width of RGB. To better constrain recent star formation, a more exquisite binary population model with binary interaction is necessary.

\subsubsection{Synthetic Galaxies}\label{subsubsec:step2}
The goal of this step is to generate a group of stellar populations that mimic realistic galaxies. Following the basic assumption of the synthetic CMD method, the synthetic galaxies are generated by combining SSPs. Some functions (such as delta function, constant, exponentially declining, delayed exponentially declining, and lognormal) are used in works \citep[e.g.,][]{2018MNRAS.480.4379C} that measure the SFHs of distant galaxies through fitting the observed spectral energy distributions (SEDs). These functions are not applicable to the case of nearby galaxies. Environment plays a vital role in complicating SFHs. Violent events like mergers may result in multiple episodes of star formation  \citep{2017ApJ...838..127I,2019ApJ...879..116I}. The observational results \citep{2011ApJ...739....5W} show the SFHs of dwarf galaxies are complex, and that the mean values are inconsistent with a simple functional form. \cite{2019ApJ...873...44C} and \cite{2019ApJ...876....3L} compared the effects of using parametric and non-parametric models in measuring galaxy SED-based SFHs. These works concluded that the parametric method is unable to adequately describe the diverse SFHs observed in galaxies. Therefore, we apply the more flexible non-parametric method to measure galaxies with some assumptions.

Firstly, we assume that the number of star formation episodes varies from 1 to 4. This maximum value is inspired by the experiment conducted in \cite{2019ApJ...876....3L}, which shows that the resulting SFHs are insensitive to episode numbers larger than 4. The corresponding metallicities follow a monotonic increasing function, ensuring that the young populations are more metal-rich than their old counterpart. Then, we assume the parameters of SSPs within a star formation episode follow a bivariate normal distribution, with a negative correlation (randomly chosen from -0.1 to -0.9) and dispersions (randomly chosen from 0.1 dex to 0.5 dex). The sum of the number densities of individual star formation episodes is normalized to 1. For mock galaxies comprising multiple star formation episodes, the weights are randomly assigned, with the total number of stars equal to $10^5$. Thus, we parameterize a mock galaxy with ages, metallicities, and the corresponding number of stars. According to these parameters, we get mock galaxies by combining stars randomly selected from the 1,248 SSPs generated in Section \ref{subsubsec:step1}. Finally, we generate $10^5$ synthetic galaxies to include as many combinations as possible. Their parameters and normalized HDs are saved for training and validating the deep learning network.

\subsection{Training and Validation Methodology}\label{subsec:train}
\begin{figure}[!htb]
\centering
\includegraphics[width=1.0\textwidth]{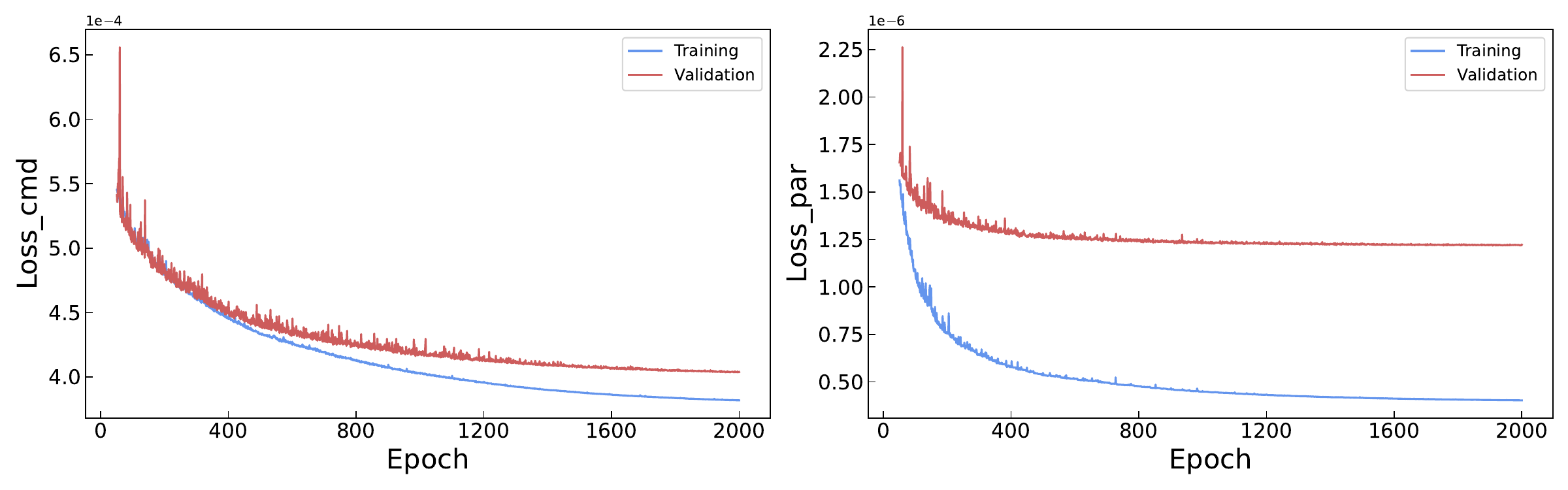}
\caption{The loss curves for \textit{branch$\_$cmd} (left panel) and \textit{branch$\_$par} (right panel). The blue and red lines represent the training set and validation set, respectively.}
\label{fig:loss}
\end{figure}

\texttt{SFHNet} is built with the \texttt{PyTorch} framework and trained on a single NVIDIA A100 GPU. The training process comprises a series of steps to teach the network to make predictions based on the input data. Firstly, the mock galaxies are divided into a training dataset and a validation dataset, with a ratio of 7:3. Then, we feed the training data into \texttt{SFHNet} and tune the model's hyperparameters to minimize the loss function. In detail, the batch size is 100, which defines the number of training examples utilized to calculate the gradient and update the model's weights in one iteration; the \texttt{AdamW} optimizer with an initial learning rate of 0.0005 and a \texttt{ReduceLROnPlateau} schedule with a decay factor of 0.95 and patience of 20 epochs is used, which is used to adjust the model's weights to minimize the loss function; the dropout rate of 0.2 is set for \textit{branch$\_$par}, which is used to prevent overfitting. The training process is executed for 2,000 epochs. At last, we monitor the change of loss values over time and select the best model. In principle, a well-trained model should have loss curves that decrease steadily and converge to low values. Figure \ref{fig:loss} presents the loss values of the training dataset (blue line) and validation dataset (red line) over the number of training epochs. The loss values steadily decline over time and reach a plateau at epoch 2000 for both output branches. Meanwhile, the variance of loss values also gets smaller over time. We save the model weights at epoch 2,000 as the well-trained model and use it to predict the SFHs of dwarf galaxies.

\begin{figure}[!htb]
\centering
\includegraphics[width=0.9\textwidth]{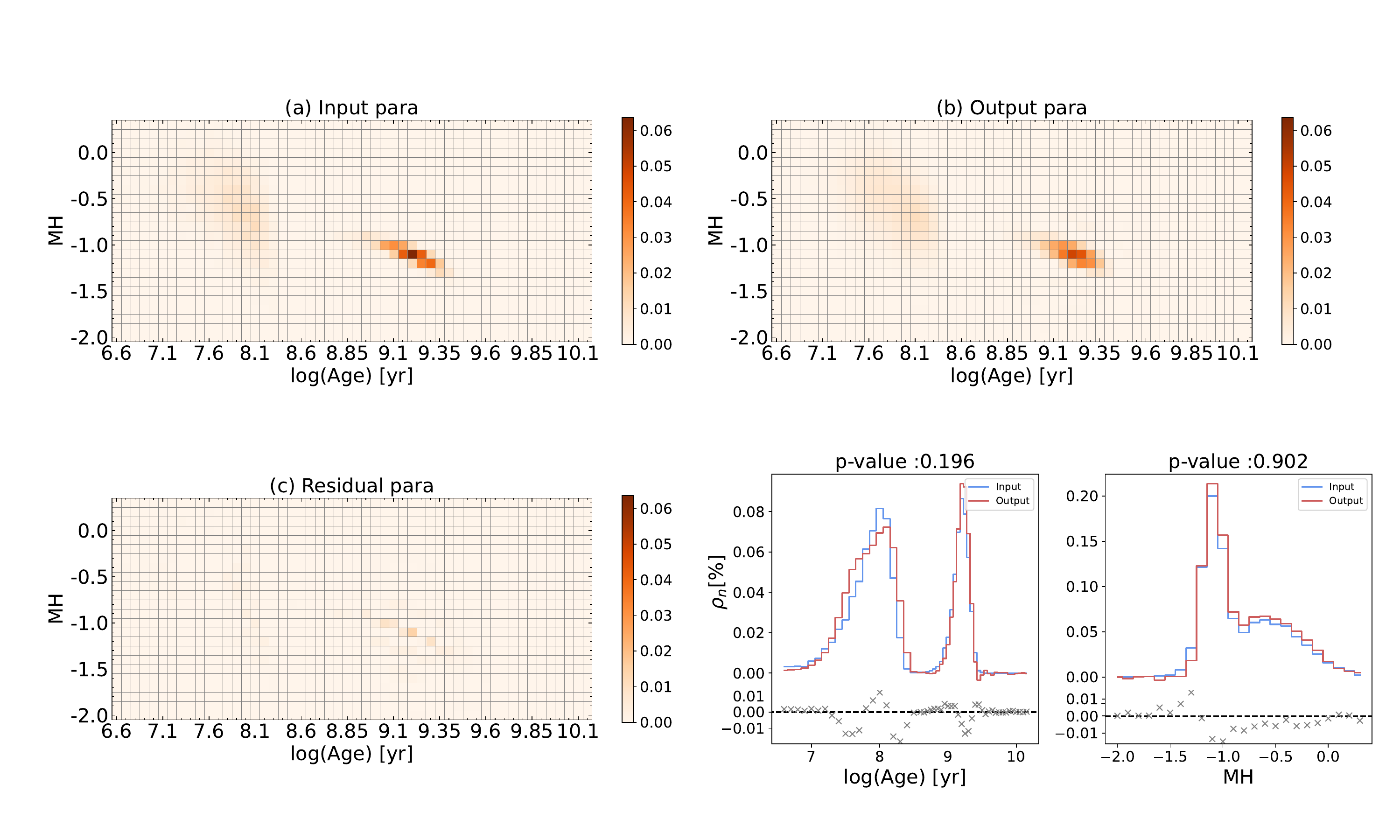}
\caption{Performance evaluation of  \textit{branch\_par}. The upper panels show the distribution of parameters behind the corresponding input HD and the output of \textit{branch\_par}, respectively. Their residual is shown in the lower left panel. The lower right panel shows the projection of parameters over age and metallicity (the blue/red line indicates the input/output value), with the p-values of the two-sided KS test displayed in the title.}
\label{fig:res_par}
\end{figure}

\begin{figure}[!htb]
\centering
\includegraphics[width=0.9\textwidth]{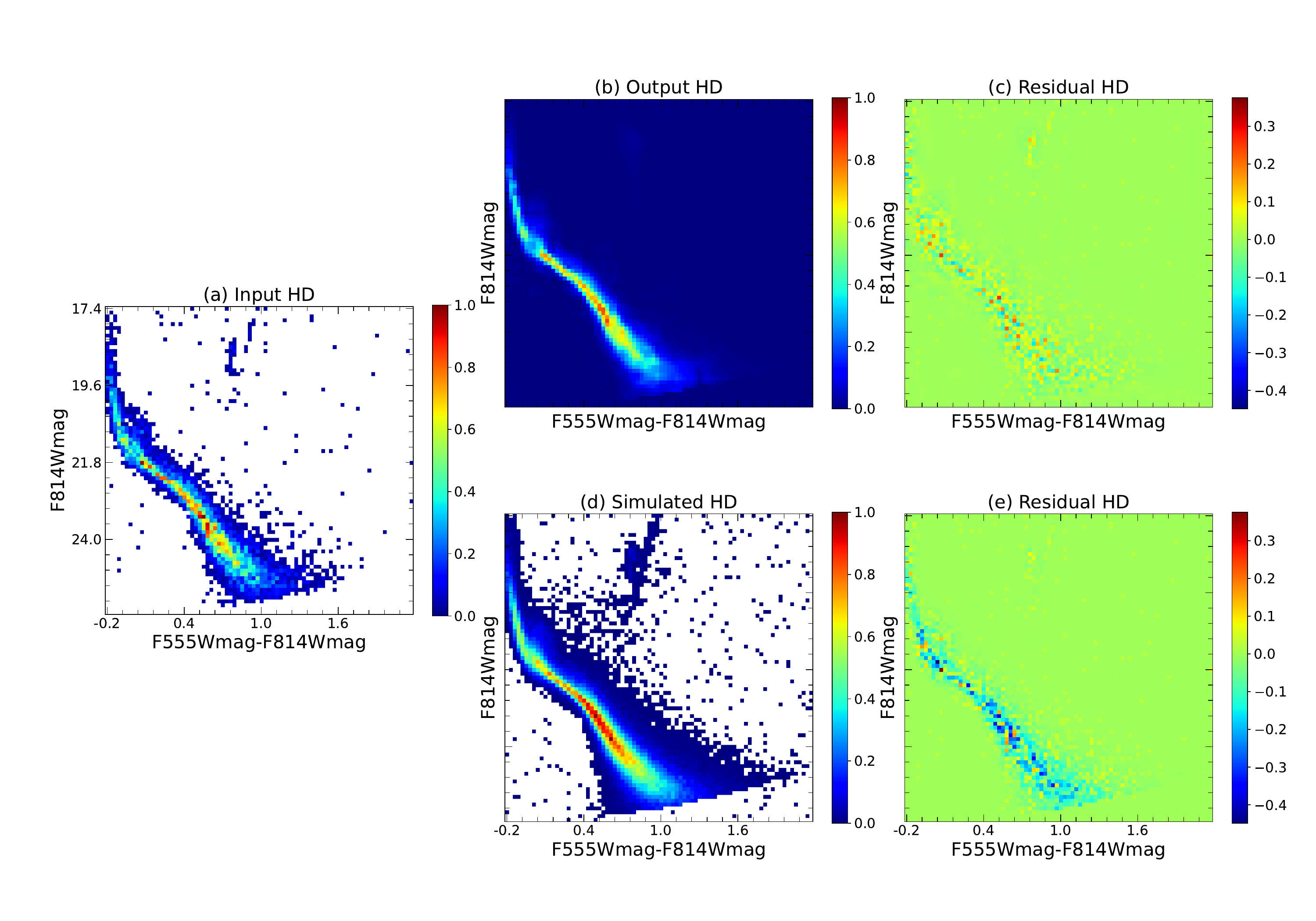}
\caption{Performance evaluation of \textit{branch\_cmd}. Panel (a) shows the input HD; Panel (b) is the output of  \textit{branch\_cmd}; Panel (c) is the residual between the previous two panels. The HD generated from the output parameters is shown in Panel (d), and the residual between the input and simulated HD is shown in Panel (e).}
\label{fig:res_cmd}
\end{figure}

\subsection{Network Performance Evaluation}
To evaluate the performance of the best deep-learning model, we generate an additional 1,000 mock galaxies following the methodology outlined in Section \ref{subsec:data}, which is named the test dataset. For illustrative purposes, we select one mock galaxy from the test dataset and compare its input and output parameters in Figure \ref{fig:res_par} and HDs in Figure \ref{fig:res_cmd}, respectively.

In detail, Figure \ref{fig:res_par} presents the comparison between parameters. Panels (a)-(c) show the input, output, and residuals across the age-metallicity space; the lower right panel shows their projection over individual parameters. The input parameters encoded the number density of each SSP, with the total normalized to 1. Overall, the output parameters share the same distribution as the input parameters, with the former displaying a less concentrated distribution than the latter. The lower right panel depicts the projection of the age-metallicity map, with input/output data indicated as a blue/red line. The p-values of the Kolmogorov-Smirnov (KS) test of the two samples are displayed in the title. At the confidence level of 95\%, we can not reject the null hypothesis that the two samples were drawn from the same distribution since the p-values (0.196 and 0.902) exceed the threshold of 0.05. For the entire test dataset, the proportions of good prediction(defined by KS tests) are 73.3\% and 86.6\% for age and metallicity, respectively. The proportion of good prediction for the KS test simultaneously satisfied for two parameters is 67.2\%.

Figure \ref{fig:res_cmd} illustrates the comparison between HDs. Panel (a)/(d) shows the HD corresponding to the input/output parameters presented in Figure \ref{fig:res_par} (a)/(b), with the highest density cell normalized to 1. Panel (b) is the output of \textit{branch\_cmd}. The corresponding residuals are presented in Panel (c) and Panel (e). The simulated HD is generated from the output parameters. The simulation process is described as follows. First, we get the number of stars of each simulated stellar population by rounding up the product of the total number of stars of the observed galaxy and its probability. The rounding error can be significant when the total number of stars is small. Then we randomly select the stars from the corresponding template SSP (generated in Section \ref{subsubsec:step1}). Finally, we assemble the simulated stellar populations to form the simulated CMD and the simulated HD. We include the simulated HD obtained from the fitted parameters here to qualitatively show the uncertainties introduced by the simulation process. From the visual inspection, the output HD presented in Panel (b) appears to be more similar to the true HD presented in Panel (a), and the stellar sequence in Panel (d) is less concentrated than the input HD. We quantify the similarity between the two HD pairs (Input-Output and  Input-Simulated) with the mean absolute error (MAE) and R-vector (RV) coefficient \citep[which is a matrix correlation for high-dimensional data and can be interpreted in the same way as Pearson's correlations;][]{Escoufier1973LETD,f46022f3-ac12-367a-9986-5b7830d42a7b} for the full test dataset. The results show the average MAEs/RV coefficients for the two pairs are  0.009/0.991 and 0.011/0.980, respectively. The smaller MAE and larger RV coefficient of the input-output pair indicate that the output HD is more similar to the input than the simulated HD.

Among the eight test galaxies, the MSTO of the nearest galaxy (Ursa Minor) is nearly four magnitudes brighter than the detection limit, which imposes strong restrictions on the stellar ages. The shallower photometry of other galaxies decreases the accuracy of \texttt{SFHNet}. The validation of \textit{branch\_par} shows that the proportion of the test dataset passing the KS test has decreased by approximately ten percent (Tucana, NGC 185, NGC 205, NGC 6822). The validation of \textit{branch\_cmd} shows that the average MAEs are around 0.01 and the similarity between HDs has also deteriorated. Take the model of the farthest galaxy (Sextans A) as an example, the average MAEs/RV coefficients for the two pairs are  0.013/0.988 and 0.014/0.965, respectively.

\section{Application to dwarf galaxies}\label{sec:data}
We apply \texttt{SFHNet} to measure the SFH of eight dwarf galaxies for further validation. The observed data were collected from the Local Group Stellar Photometry Archive \footnote{http://astronomy.nmsu.edu/logphot} \citep[LOGPHOT;][]{2006ApJS..166..534H}, which includes 40 nearby galaxies data observed by the \textit{HST}/Wide Field Planetary Camera 2 (WFPC2). The data have been uniformly calibrated and processed by HSTPHOT \citep{2000PASP..112.1383D}, with uncertainties and incompleteness derived from ASTs. The observational details, along with the adopted distance moduli and reddening values,  are presented in Table \ref{table:results}. The rationale behind the choice of these galaxies is that we want to conduct a direct comparison with the numerical results from \cite{2014ApJ...789..147W}. To test the efficiency of our method in dealing with galaxies of different morphological types and varying photometric depth, we simply select the nearest and farthest galaxies for each morphological type within the available dataset.

\begin{table}[!htb]
\centering
\caption{Observational Properties and Measured Parameters of 8 Galaxies}
\begin{tabular}{cccc|cc|cc}
\hline
\hline
Galaxy Name & Field ID & Morphological& Filters & $(M-m)_0^{\dag}$ & $A_{\rm V}$ & $M_{\star}$ & $M_{\star}^{\dag}$ \\
& & Type & & & [mag] & [$10^6$ $M_\odot$] & [$10^6$ $M_\odot$] \\[1ex]
\hline
(1) & (2) & (3) & (4) & (5) & (6) & (7) & (8) \\[1ex]
\hline
Ursa Minor & u2pb01 & dSph   & F555W,F814W & $19.31\pm0.1 $ & $0.090\pm0.001$ & $0.015\pm0.001$  & $0.02^{+0.02}_{-0.01}$\\[1ex]
Tucana     & u2cw02 & dSph   & F555W,F814W & $24.73\pm0.12$ & $0.085\pm0.002$ & $1.044\pm0.060$  & $3.22^{+0.61}_{-3.11}$\\[1ex]
\hline
NGC 185    & u3kl04 & dE     & F555W,F814W & $24.09\pm0.09$ & $0.525\pm0.014$ & $12.838\pm0.524$ & $11.85^{+5.49}_{-10.17}$\\[1ex]
NGC 205    & u3kl10 & dE     & F555W,F814W & $24.54\pm0.07$ & $0.182\pm0.003$ & $14.857\pm1.519$ & $13.97^{+14.79}_{-10.66}$\\[1ex]
\hline
NGC6822    & u37h05 & dIrr   & F555W,F814W & $23.35\pm0.08$ & $0.605\pm0.006$ & $4.923\pm0.710$  & $3.46^{+3.51}_{-3.35}$\\[1ex]
Sextans A  & u5x502 & dIrr   & F555W,F814W & $25.72\pm0.08$ & $0.124\pm0.002$ & $7.309\pm1.003$  & $17.54^{+13.88}_{-16.58}$\\[1ex]
\hline
Phoenix    & u64j01 & dTran  & F555W,F814W & $23.10\pm0.1 $ & $0.044\pm0.001$ & $1.489\pm0.104$  & $1.28^{+0.85}_{-0.98}$\\[1ex]
Pegasus    & u2x504 & dTran  & F555W,F814W & $24.94\pm0.07$ & $0.186\pm0.003$ & $10.319\pm0.833$ & $7.68^{+7.22}_{-7.28}$\\[1ex]
\hline
\end{tabular}
\begin{tablenotes}
\item \textbf{Notes.} The left columns are the observational properties of each galaxy. The middle columns are the adopted distance moduli and reddening values. The right columns are the calculated initial stellar mass (with the total uncertainties) of the observed fields. The symbol $\dag$ denotes the physical parameters from \cite{2014ApJ...789..147W}.
\end{tablenotes}
\label{table:results}
\end{table}

\subsection{Data Pre-processing}
The observed catalog needs to go through three pre-processing steps (i.e., remove the Milky Way foreground stars, correct the reddening, and get the number density map) before feeding to \texttt{SFHNet}. We describe the details of each step as follows.

\begin{figure}[!htb]
\centering
\includegraphics[width=0.9\textwidth]{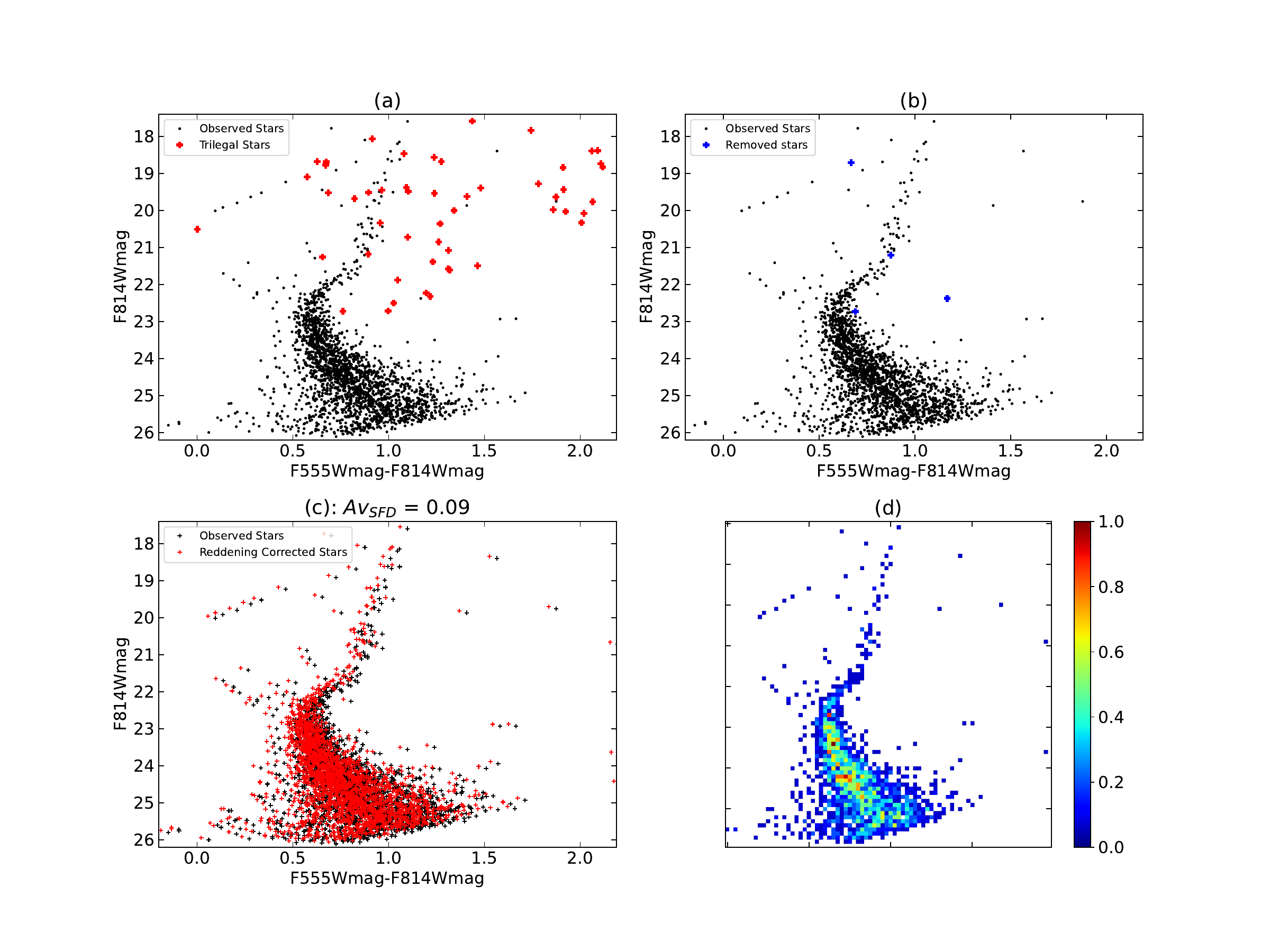}
\caption{Illustrate the pre-processing steps applied to dwarf galaxy Ursa Minor. The upper row shows the CMDs before and after removing the Milky Way foreground stars. The black dots are observed stars, and red and blue pluses highlight the foreground stars from \texttt{TRILEGAL} and the removed stars, respectively. Panel (c) shows the CMD before (black dots) and after (red dots) the reddening correction, with the adopted $A_{\rm v}$ displayed in the subtitle. Panel (d) is the HD, with the cell containing the largest number of stars normalized to 1.}
\label{fig:obs}
\end{figure}

For extra-galactic galaxies, galactic stars may lie within the line of sight and be treated as members of the target galaxy. Therefore, reducing the contamination of the Milky Way foreground stars is essential to get accurate SFHs. The decontamination process needs a galactic star catalog at the same spatial location. We use the result from \texttt{TRILEGAL}, which is a star counts model based on the population synthesis approach \citep{2005A&A...436..895G}. Then, a statistical subtraction method is employed. Specifically, we measure the distance between observed stars and simulated stars generated by \texttt{TRILEGAL} in the color-magnitude space. For each simulated star, if the closest observed star has a distance smaller than $3 \times \sqrt{\sigma_{\rm color}^2 + \sigma_{\rm mag}^2} $, the observed star is classified as a foreground star and be removed from the observational catalog. Figure \ref{fig:obs} illustrates the decontamination process, taking the dwarf galaxy Ursa Minor as an example. Panel (a) shows the observed stars (black dots) and the simulated Milky Way foreground stars from \texttt{TRILEGAL} (red pluses); Panel (b) illustrates the stars that have been removed (blue pluses) through statistical subtraction. The following steps of reddening correction and pixelating CMD are applied to the decontaminated catalog.  

In the aspect of reddening correction, we obtain the total extinction $A_{\rm v}$ from \cite{1998ApJ...500..525S} (SFD) dust reddening map through the Python package \texttt{dustmap}\footnote{https://dustmaps.readthedocs.io/en/latest/index.html}. The extinction curve with $R_{\rm v}=3.1$ and relative extinction coefficients for two filters, i.e., $A_{\rm F555W}/A_{\rm v} = 1.03065$, $A_{\rm F814W}/A_{\rm v} = 0.59696$ are adopted \citep{1989ApJ...345..245C,1994ApJ...422..158O}. Since the low resolution of the SFD extinction map,  the value of $A_{\rm v}$ at the center position is used for the correction. Panel (c) of Figure \ref{fig:obs} compares the CMDs before and after reddening correction. Regarding image pixelation, the CMD is divided into an 80$\times$80 array, with the highest density cell normalized to 1. The bin size for color/magnitude is around 0.05/0.1 mag, with little variation for different galaxies. Panel (d) of Figure~\ref{fig:obs} shows the HD and the color bar indicates the density.

\subsection{Deriving Star Formation Histories}
The major step of deriving SFH is to calculate the initial stellar mass. In previous studies, the synthetic stellar populations were extracted from a fixed total mass budget. Instead of using the top-down method, we estimate the total initial mass from the number of observed stars (corrected with a completeness map) and then extrapolate to the full mass spectrum (0.08-150 $M_\odot$, used in this work). The IMF describes the distribution of initial stellar mass in a newly formed population. We firstly generated $10^6$ stars following the Kroupa IMF across the full mass spectrum to calculate the average stellar mass of a population, which is presented as $M_{\rm pop} = \frac{\sum_{i=0}^{i=n}m_{\rm i}}{n}$, where $n$ is the number of stars and $m_{\rm i}$ is the stellar mass of the individual star. The huge number of stars can reduce the effects of different sampling methods and Poisson noise. After that, we estimate the total initial mass of a stellar population by multiplying $M_{\rm pop}$ and the total number of stars. This approximation leaves the derivation of the total initial mass of the galaxy only about the star numbers.

\begin{figure}[!htb]
\centering
\includegraphics[width=0.8\textwidth]{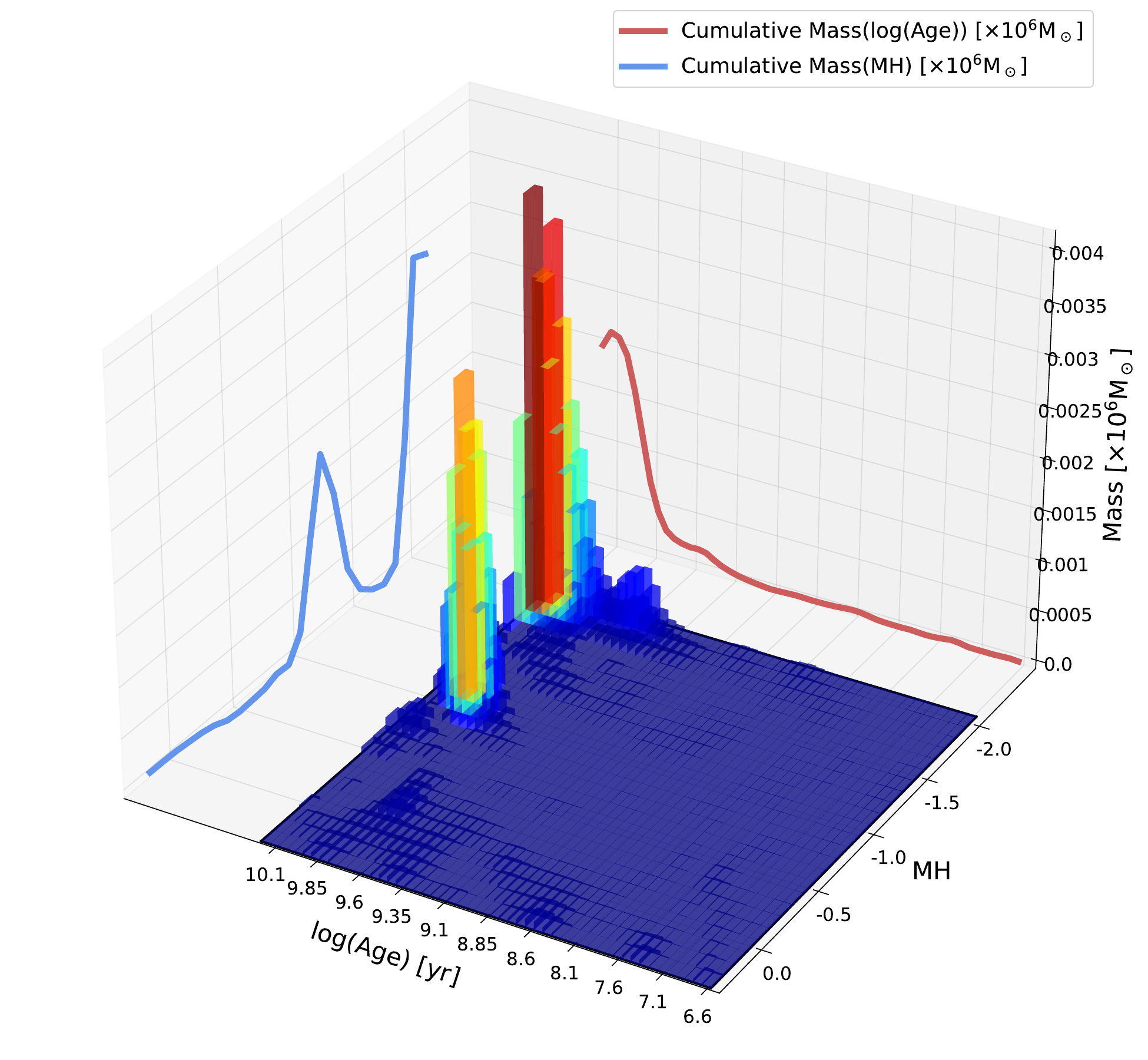}
\caption{Illustrate the initial mass of stellar populations over the age-metallicity map, taking Ursa Minor as an example. The three axes are age, metallicity, and initial stellar mass. The blue/red line indicates the projection of mass over the age/metallicity axis.}
\label{fig:3dmass}
\end{figure}

The basic assumption of the synthetic CMD method is that a galaxy is a mixture of SSPs. The output of  \textit{branch$\_$par}, $para_{\rm j}$ is the proportion of the $j$th population relative to the total galaxy (with the sum of 1). Source blending in a crowded environment may result in an underestimation of the star number, which can be corrected with the completeness coefficient. We get the completeness curves of individual filters through ASTs and interpolate them to the color-magnitude space, with the cell size following the same segmentation scheme of the observed CMD. The resulting completeness map is presented as $C$. The observed CMD records a single snapshot of the evolutionary history of the galaxy. The massive stars have already died while the less massive stars are tracking down their evolutionary path. To obtain the initial number of stars, we multiply the number of observed stars by the ratio of the integration of IMF ($\xi(m)$) over the full mass spectrum and the mass range limited by the CMD frame. Thus, the total initial mass of the observed field, denoted by $M_{\rm galaxy}$, can be calculated as follows : 
\begin{equation}\label{eq:mass}
M_{\rm galaxy} = \sum_{j=0}^{j=N} M_{\rm pop} \times \frac{N_{\rm obs}\times para_{\rm j}}{C} \times \frac {\int_{0.08}^{150} \xi(m){\rm d}m}{\int_{m_{\rm min}}^{m_{\rm max}}\xi(m){\rm d}m},
\end{equation}
where $M_{\rm pop}$ is the average initial mass of a stellar population; $N_{\rm obs}$ is the number of observed stars; $para_{\rm j}$ is the proportion of the $j$th stellar population over the all $N$ populations; $C$ is the completeness map; $m_{\rm min}$ and $m_{\rm max}$ are the masses corresponding to faintest and brightest magnitude limits of the CMD frame; the last term is the ratio between the number of stars over the full mass spectrum and the number of stars in the detectable mass range. Based on this equation, we can estimate the initial mass of each stellar population and construct a mass distribution map over the age-metallicity space. For example, Figure \ref{fig:3dmass} presents the initial mass distribution of Ursa Minor, which exhibits two distinct groups with minor fluctuations in the remaining regions. This suggests that the majority stellar mass of Ursa Minor is allocated during the early star formation episode.

\begin{figure}[!htpb]
\subfigure{\includegraphics[width=0.333\textwidth,trim=0.2cm 0.5cm 0.2cm 0.4cm,clip]{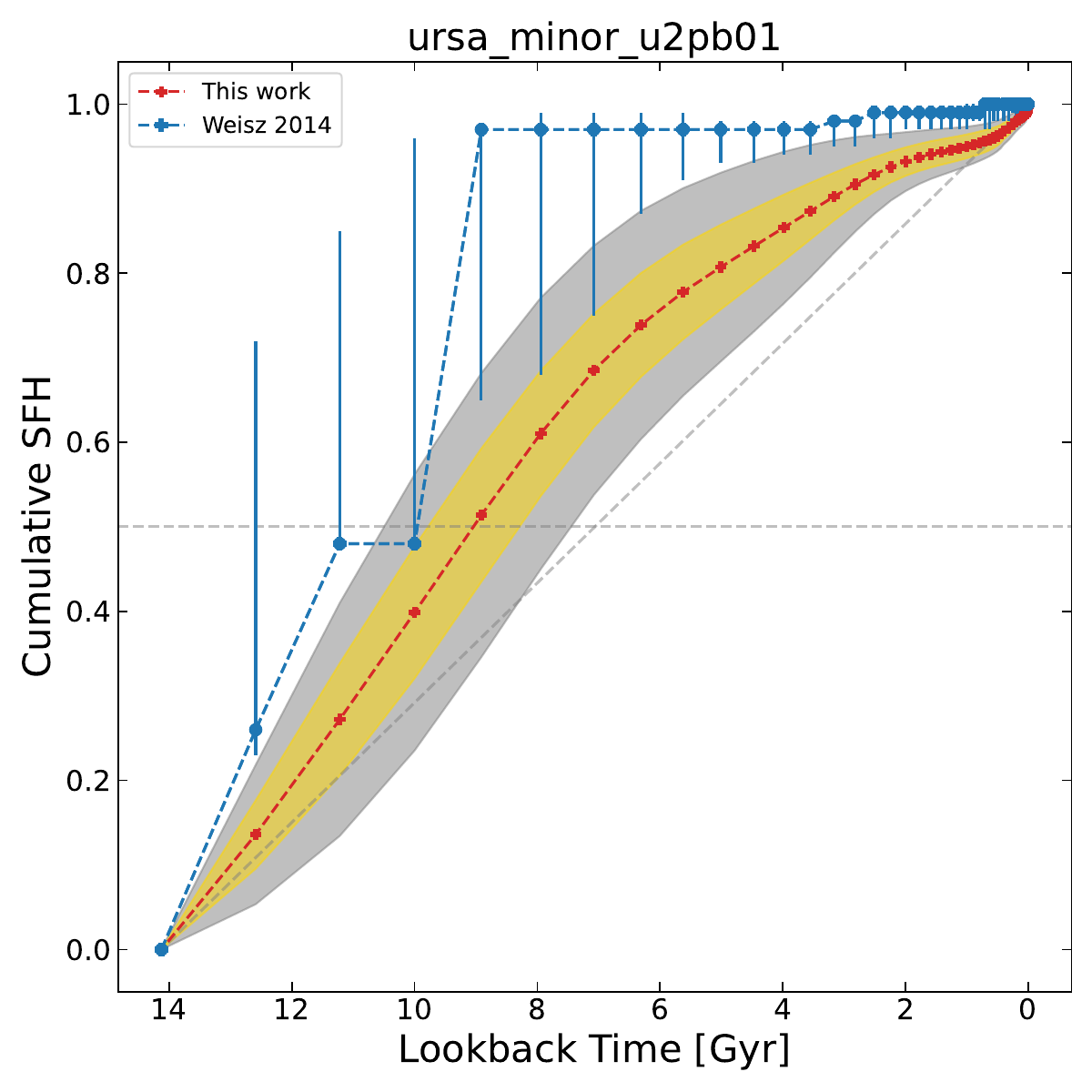}}
\subfigure{\includegraphics[width=0.333\textwidth,trim=0.2cm 0.5cm 0.2cm 0.4cm,clip]{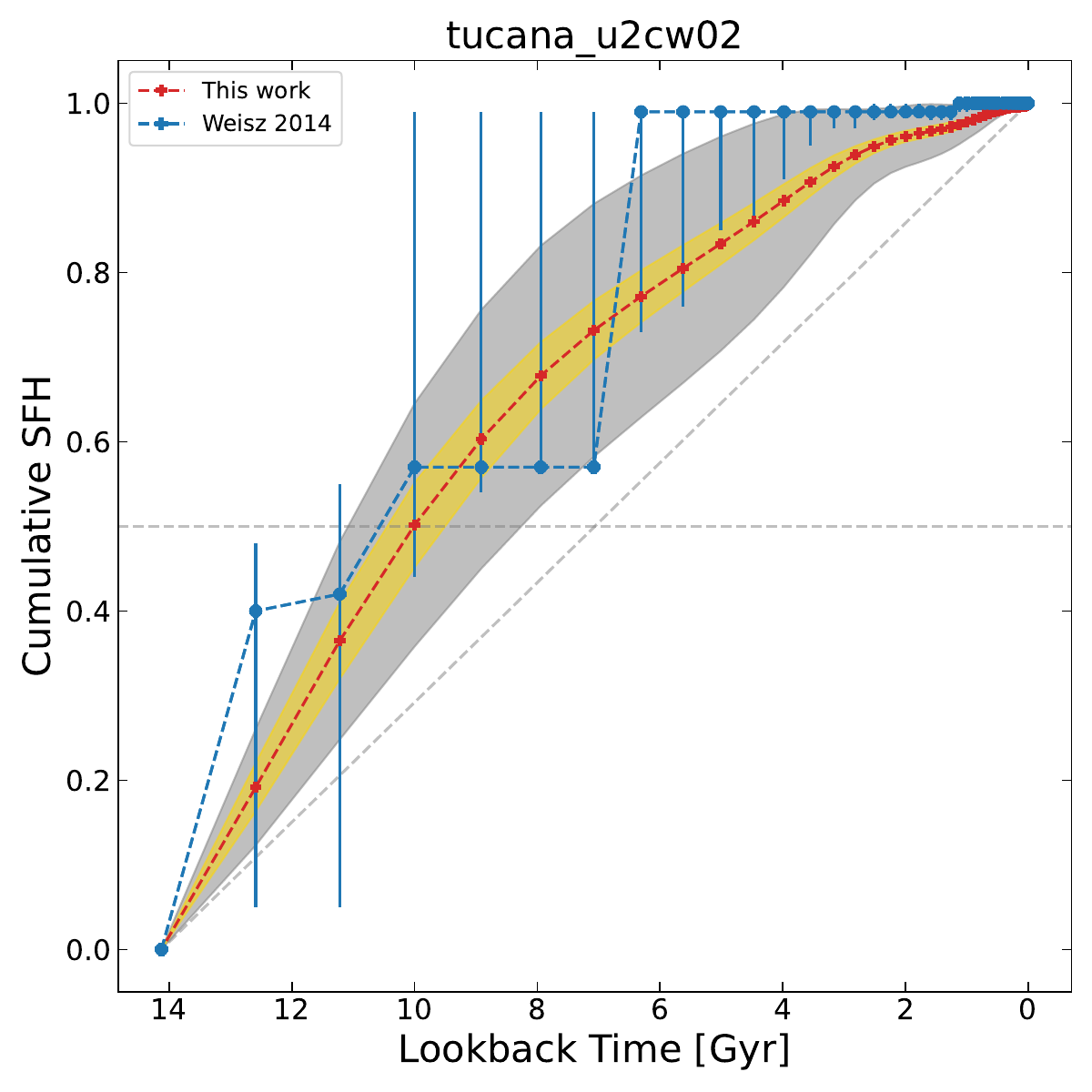}}
\subfigure{\includegraphics[width=0.333\textwidth,trim=0.2cm 0.5cm 0.2cm 0.4cm,clip]{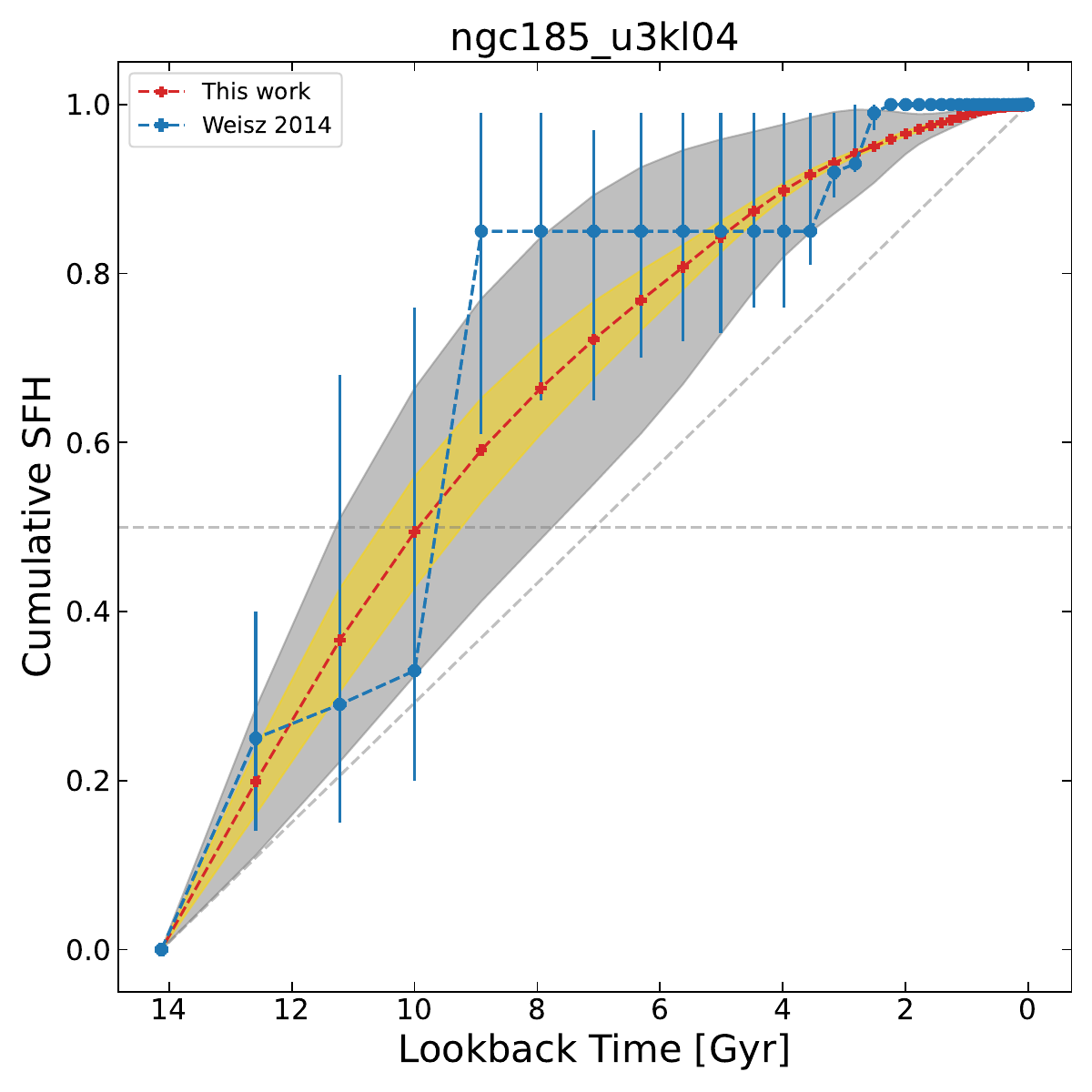}}\\
\subfigure{\includegraphics[width=0.333\textwidth,trim=0.2cm 0.5cm 0.2cm 0.4cm,clip]{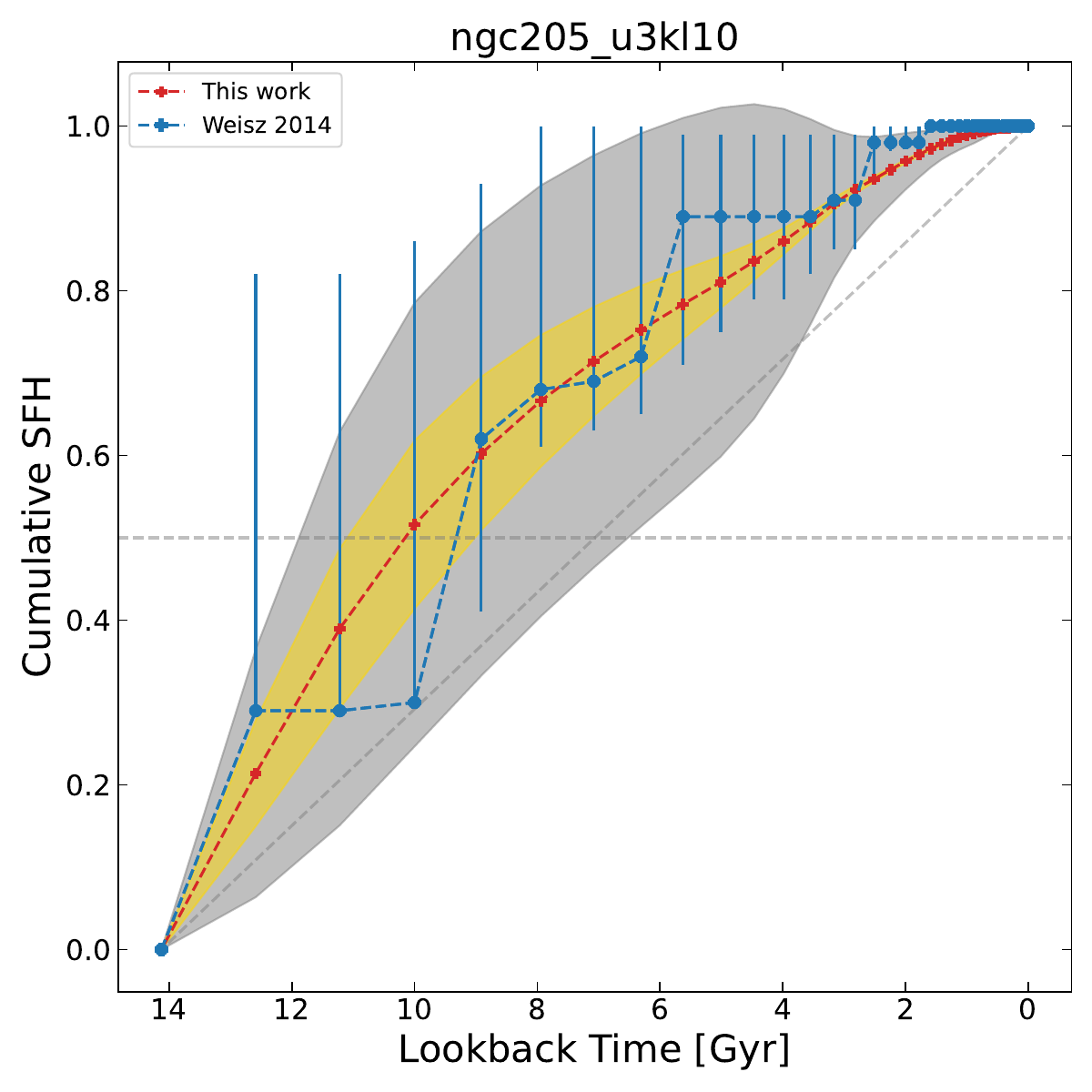}}
\subfigure{\includegraphics[width=0.333\textwidth,trim=0.2cm 0.5cm 0.2cm 0.4cm,clip]{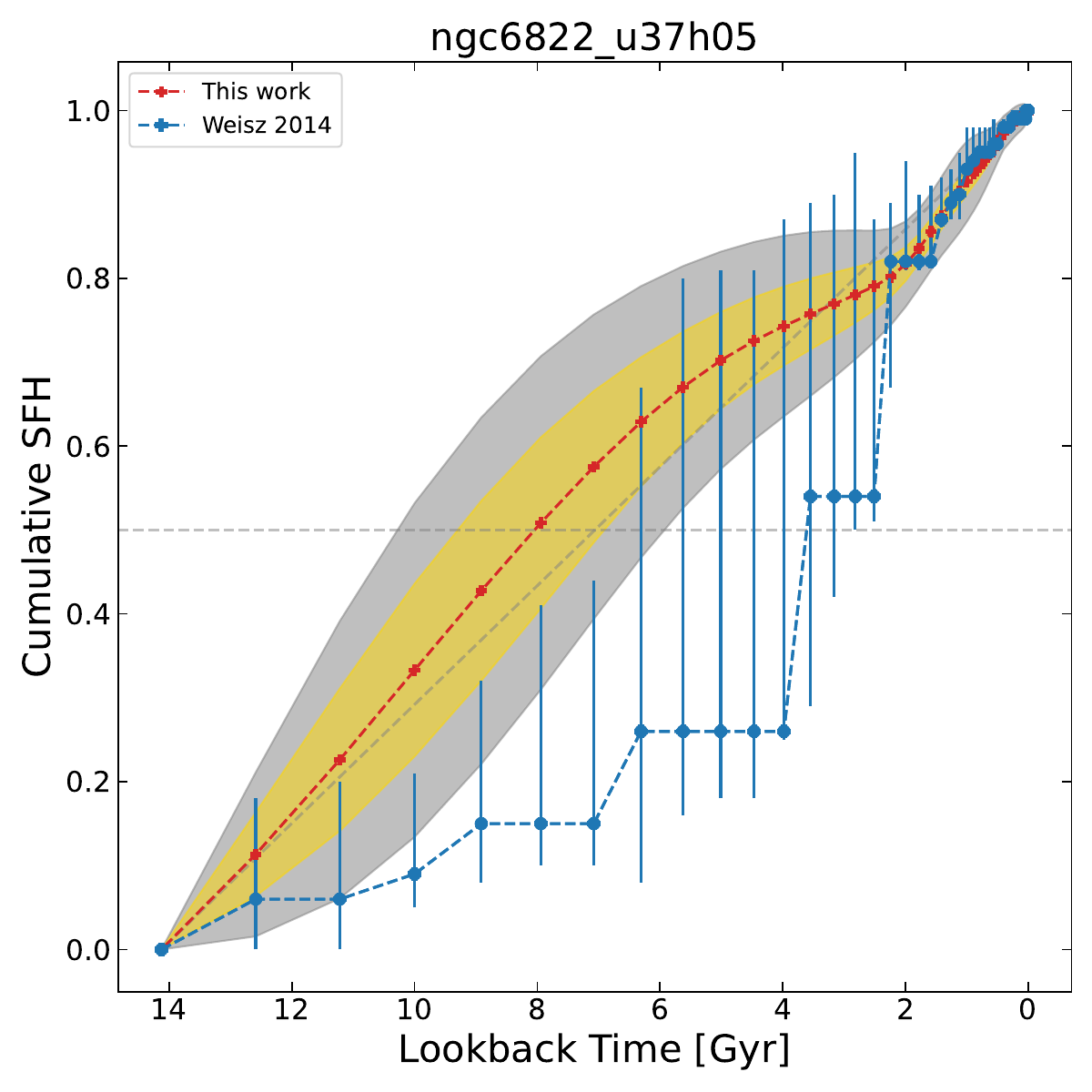}}
\subfigure{\includegraphics[width=0.333\textwidth,trim=0.2cm 0.5cm 0.2cm 0.4cm,clip]{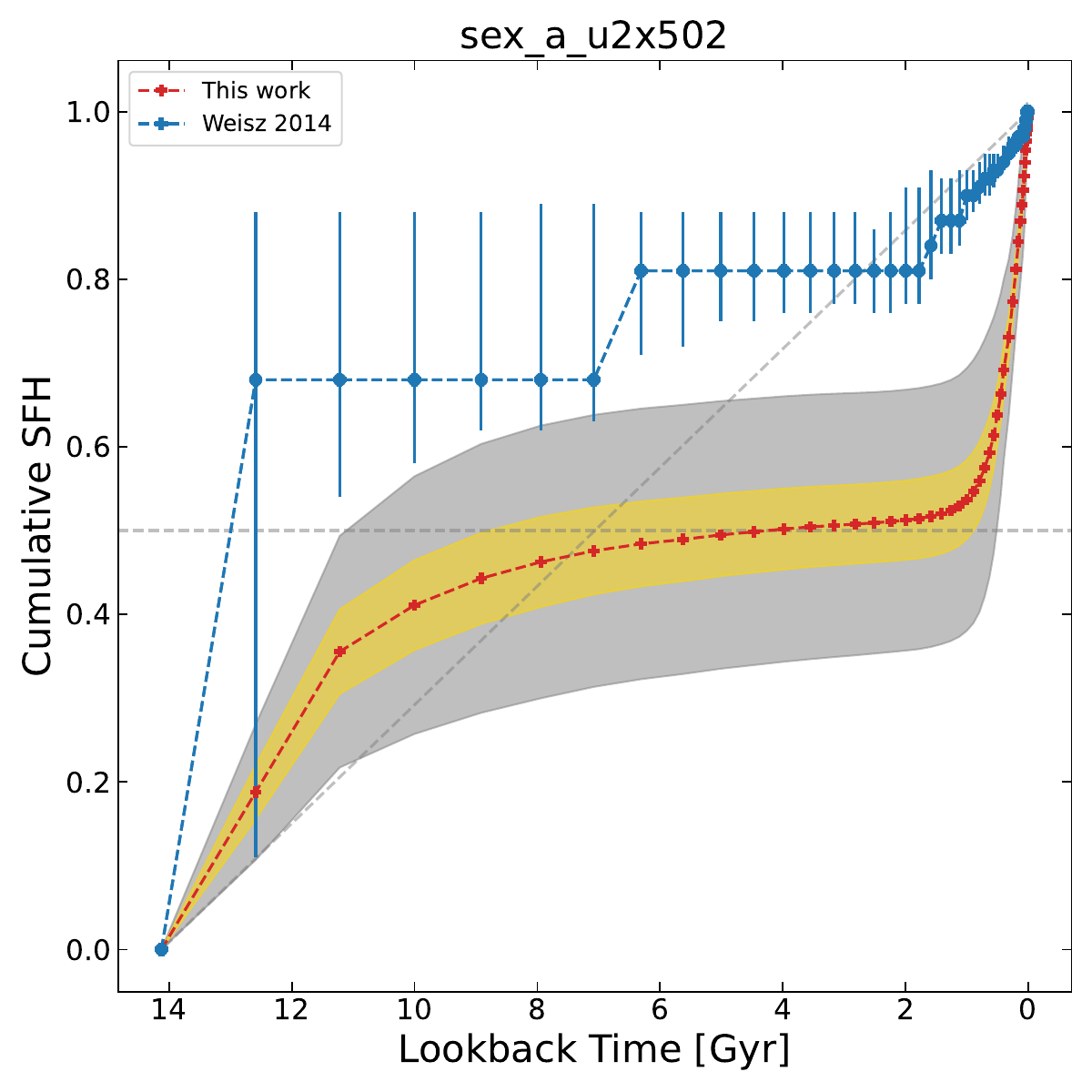}}\\
\subfigure{\includegraphics[width=0.333\textwidth,trim=0.2cm 0.5cm 0.2cm 0.4cm,clip]{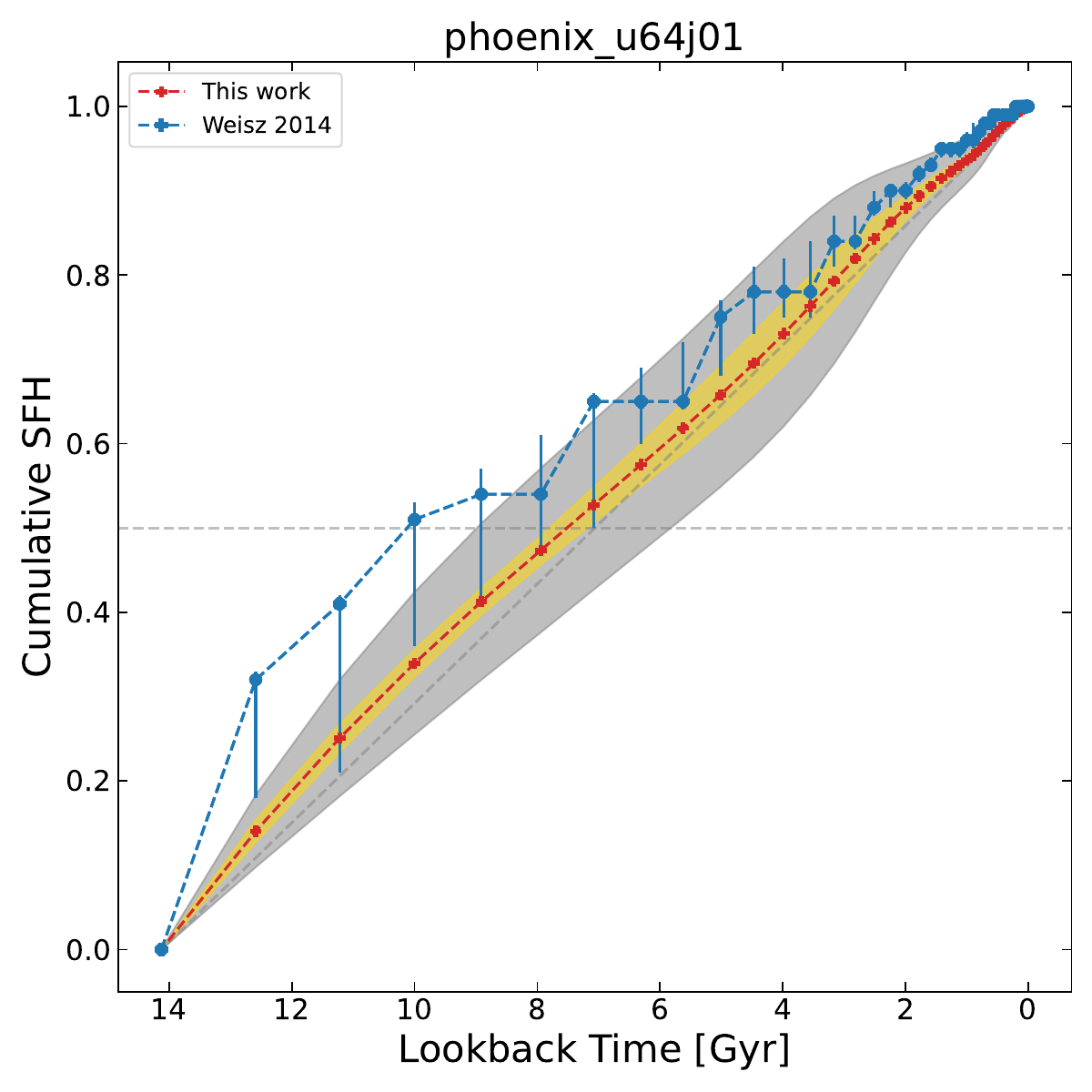}}
\subfigure{\includegraphics[width=0.333\textwidth,trim=0.2cm 0.5cm 0.2cm 0.4cm,clip]{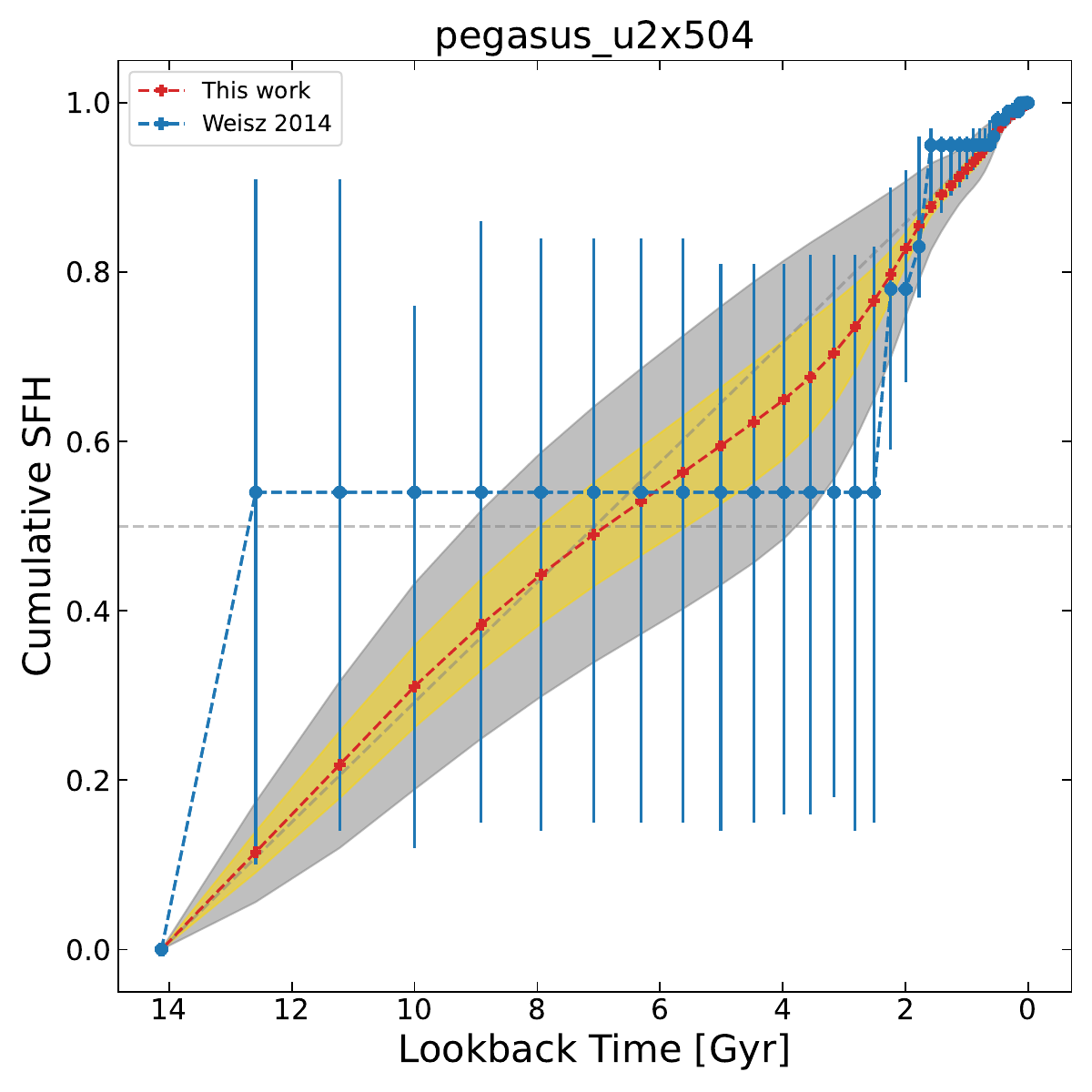}}
\caption{SFHs of eight dwarf galaxies, with galaxy name and the field ID displayed on the title. The red dot line shows the measured initial stellar mass over lookback time; the yellow shadow indicates the random uncertainties and the grey shadow indicates the total uncertainties (sum of random and system uncertainties). The diagonal gray dashed line indicates the mass growth with a constant SFR over time. The blue dot line displays the SFH from \cite{2014ApJ...789..147W}.}
\label{fig:result_all}
\end{figure}

\subsection{Characterizing Uncertainties}
\label{subsec:errors}
In this section, we discuss the uncertainties associated with the derived SFH, including (1) the systematic uncertainties introduced by stellar evolution models and the young MS-BSSs ambiguity; (2) the random uncertainties introduced by Poisson error and fixed parameters (reddening and distance modulus). 

Different stellar evolution models contribute to the main systematic uncertainties. Theoretical isochrones with the same age and metallicity but different initial parameters (such as the adopted solar metal, opacity, convection, or the mass loss efficiency during the red giant branch phase) trace different CMD regions. In this work, we estimate the systematic uncertainties by varying the observed CMDs rather than applying different theoretical isochrone sets in the fitting process. As demonstrated in \cite{2012ApJ...751...60D}, shifting synthetic populations' effective temperature ($\sigma_{\rm log T_{\rm eff}}$) and luminosity ($\sigma_{\rm M_{\rm bol}}$) can mimic the uncertainties introduced by different stellar evolution models. The displacements are found to be correlated with photometric depth and filter combination. Table 4 in \cite{2012ApJ...751...60D} summarizes the variation of $\sigma_{\rm log T_{\rm eff}}$ and $\sigma_{\rm M_{\rm bol}}$ as a function of photometric depth in the filter combination of V and I bands, which are similar to the bands F555W and F814W used in this study. The variations of $\sigma_{\rm log T_{\rm eff}}$ are transformed into color shifts ($\sigma_{\rm color}$) by \texttt{PySynphot} \citep[which is a synthetic photometry package developed for HST;][]{2013ascl.soft03023S}. The displacements in color ($\sigma_{\rm color}$) and magnitude ($\sigma_{\rm mag}$) for the nearest galaxy (Ursa Minor and Phoenix) are approximately 0.05 and 0.19 mag, respectively. For the remaining galaxies, the corresponding values are around 0.09 and 0.31 mag. Therefore, we move the observed CMD along the color-axis/magnitude-axis by values that are normally distributed with the mean equal to zero and the standard deviation equal to $\sigma_{\rm color}$/$\sigma_{\rm mag}$. The process is repeated 10,000 times, and the standard deviation of the derived 10,000 SFHs is the systematic error.

Another source of systematic uncertainties is the ambiguity between BSSs and the young MS stars. BSSs in star clusters are hotter and brighter than their MSTO counterparts, which are believed as the products of binary internal interaction (mass transfer or merger). In the context of a galaxy, BSSs of old populations and the recently formed young MS stars overlap in the CMD space, which is referred to as the young MS-BSS ambiguity \citep[][and references therein]{2007A&A...468..973M,2015ASSL..413..129M}. In this study, the blue plume stars are interpreted as young MS stars due to the lack of BSSs in our synthetic galaxies (see Panels (b) in Figure \ref{fig:res_num_data1}), which results in an overestimation of recent star formation. A more accurate and comprehensive binary population synthetic (BPS) method will improve the accuracy of recent SFH, such as BSE \citep{2002MNRAS.329..897H}, the Yunnan model \citep{2002MNRAS.336..449H,2003MNRAS.341..669H}, BPASS \citep{2008MNRAS.384.1109E}. While we note several pivotal challenges in binary evolution remain unresolved, such as common envelope evolution and mass transfer \citep{2024PrPNP.13404083C}. Furthermore, a detailed BPS approach means to follow the evolution of both components and deal with their interaction for million times, which requires a substantial computational resource. Thus, we do not include the detailed synthetic binaries in this work. Although the young MS-BSS ambiguity may result in an overestimation of the recent SFR, its effect on the overall SFH is minimal. As \cite{2014ApJ...789..147W} claimed, the blue plume stars typically constitute 5\% - 7\% of the total initial stellar mass ($\sim$ 3.6\% for Ursa Minor, as estimated in this study).

The Poisson noise of the number of observed stars and uncertainties in fixed parameters (distance modulus and reddening) constitute the random uncertainties of SFH \citep{2013ApJ...775...76D}, which can be estimated through resampling the observed stars. To directly compare deep learning and traditional numerical methods, we use the distance modulus from  \cite{2014ApJ...789..147W} and the reddening from the SFD map. The flexibility of \texttt{dustmap} allows us to obtain the mean and standard deviation of the reddening simultaneously. Errors in distance vary the CMD vertically up or down, and errors in reddening affect the distribution of stars along the reddening direction. In addition, the variation of reddening across the space, i.e., differential reddening, can also broaden the stellar sequences.

The estimation of random uncertainties on SFHs follows these steps. Firstly, we repeat the de-reddening process with a few modifications. We divide stars into groups according to their spatial distribution and separately correct the reddening for each group. The reddening value for each group is randomly extracted from a Gaussian distribution with the mean and standard deviation obtained from \texttt{dustmap}. After that, we vary the magnitudes with a fluctuation that is consistent with the standard deviation of the distance modulus. The projection effect is negligible since the dwarf galaxies are small. We repeat the previous two steps 100 times to get resampled CMDs. Finally, we measure the uncertainties due to Poisson noise by resampling the number of stars in each magnitude-color cell, with 100 iterations. Thus, we obtain 10,000 modified HDs and measure SFH of them. The standard deviation of these SFHs is the random uncertainty (highlighted as the orange shadow in Figure \ref{fig:result_all}), which is inversely proportional to the number of stars. The more populated galaxies (such as Tucana and Phoenix) show relatively smaller random uncertainties.

\subsection{The goodness of fit}
\label{subsece:gooodness}
We calculate the Saha $W$ statistic \citep{1998AJ....115.1206S} to evaluate the goodness of fit. For two CMDs following the same segmentation scheme, the $W$ statistic is calculated as follows:
\begin{equation}
W = \prod_{i}^{\rm B}\frac{(m_{\rm i} + n_{\rm i})!}{m_{\rm i}!n_{\rm i}!},
\label{eq:saha_w}
\end{equation}
where $m_{\rm i}$ and $n_{\rm i}$ are the number of stars in the $i$th cell of the two CMDs; $\rm B$ is the total number of cells, which is 6,400 in this work. The degree of similarity of the two CMDs is proportional to the value of $W$, which approaches the maximum value when the two CMDs are identical. 

We measure the SFHs of eight galaxies for further validation based on the photometric and artificial data from the LOGPHOT survey. These galaxies belong to four morphological types, i.e., dwarf spheroidal galaxies (dSphs), dwarf elliptical galaxies (dEs), dwarf irregular galaxies (dIrrs), and transition dwarf galaxies (dTrans). The nomenclature is consistent with \cite{2011ApJ...739....5W,2014ApJ...789..147W} and reference therein. There is a tight correlation between galaxy morphology and its SFH. Two galaxies of each morphology type are the nearest and the farthest in the LOGPHOT survey. The observational properties and measured parameters are summarized in Table \ref{table:results}.

Figure \ref{fig:result_all} summarizes SFHs of eight dwarf galaxies. In general, dSphs and dEs formed the majority of their stars early in their evolution; dIrrs show significant recent star formation, and dTrans is approximately consistent with a constant SFR over the galaxy's entire lifetime. In comparison to the SFHs derived from the numerical method (indicated as blue dot lines with error bars), dIrrs and dTrans show less similarity to the results presented in \cite{2014ApJ...789..147W}. Our profiles are smoother, which is a consequence of the high flexibility of \texttt{SFHNet} and the fewer constraints on the mass of individual stellar populations. A more comprehensive discussion of the star-forming properties of each morphological type is provided in Section \ref{subsec:csfh}. 

\begin{figure}[!htb]
\subfigure{\includegraphics[width=0.5\textwidth]{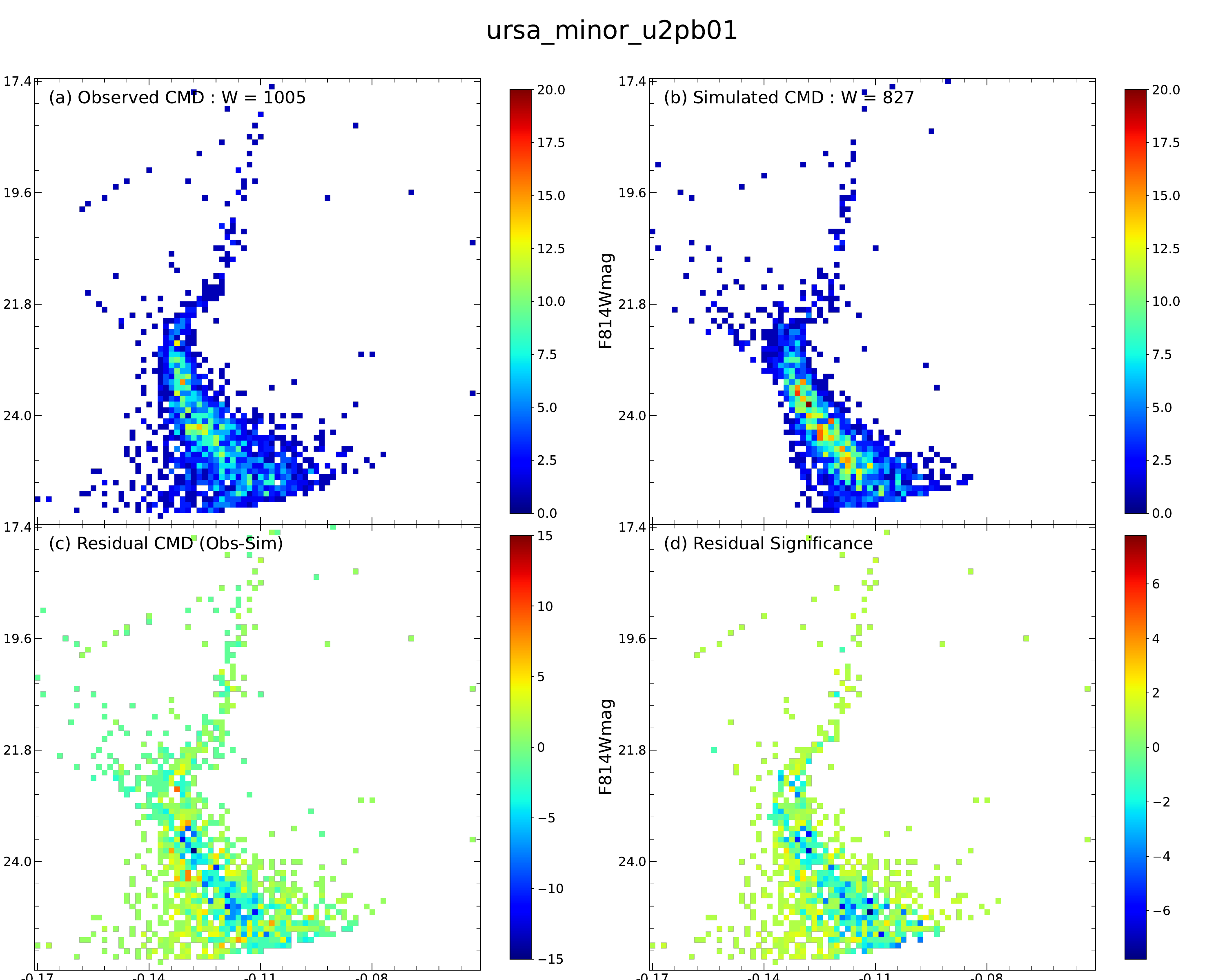}}
\subfigure{\includegraphics[width=0.5\textwidth]{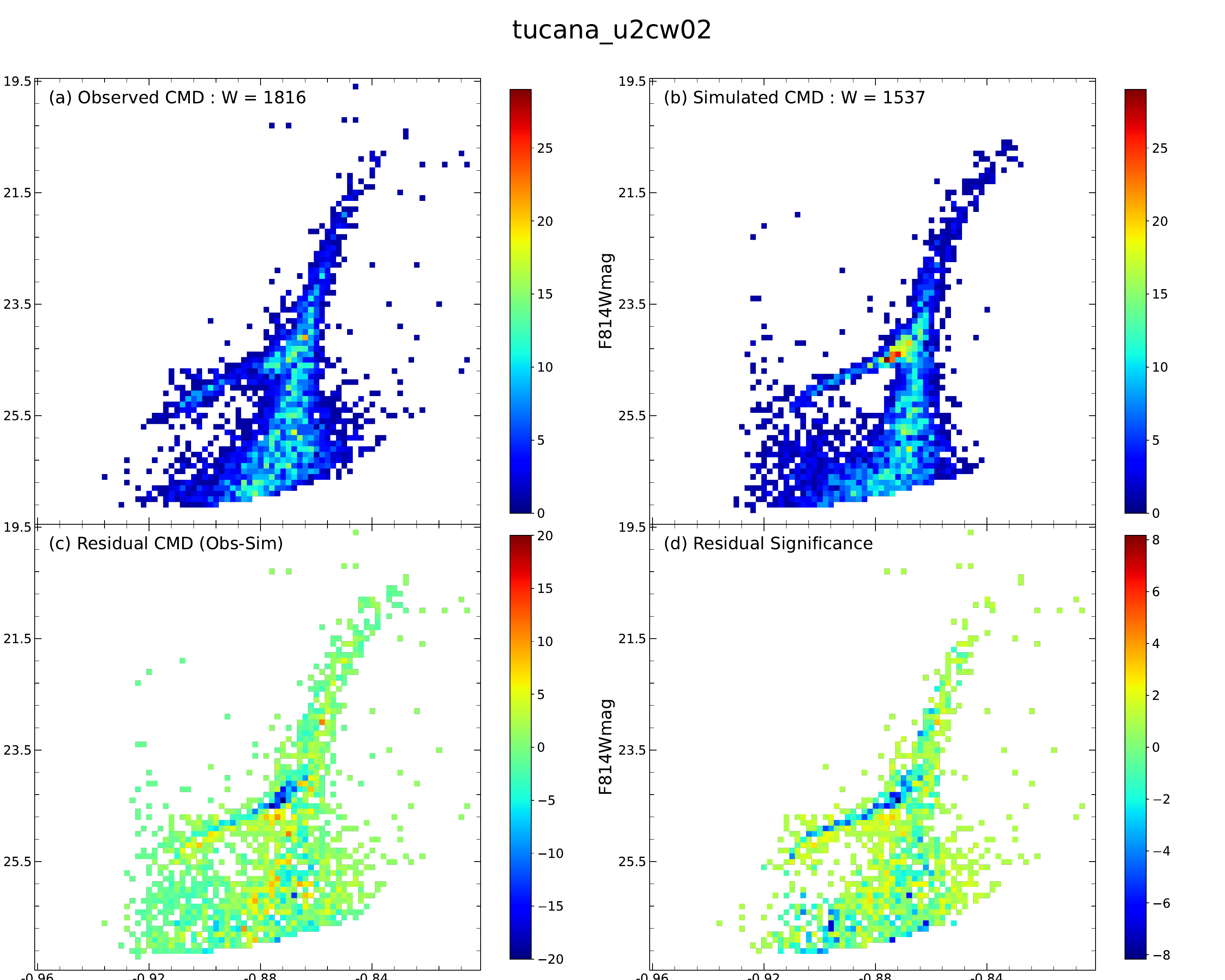}}\\
\subfigure{\includegraphics[width=0.5\textwidth]{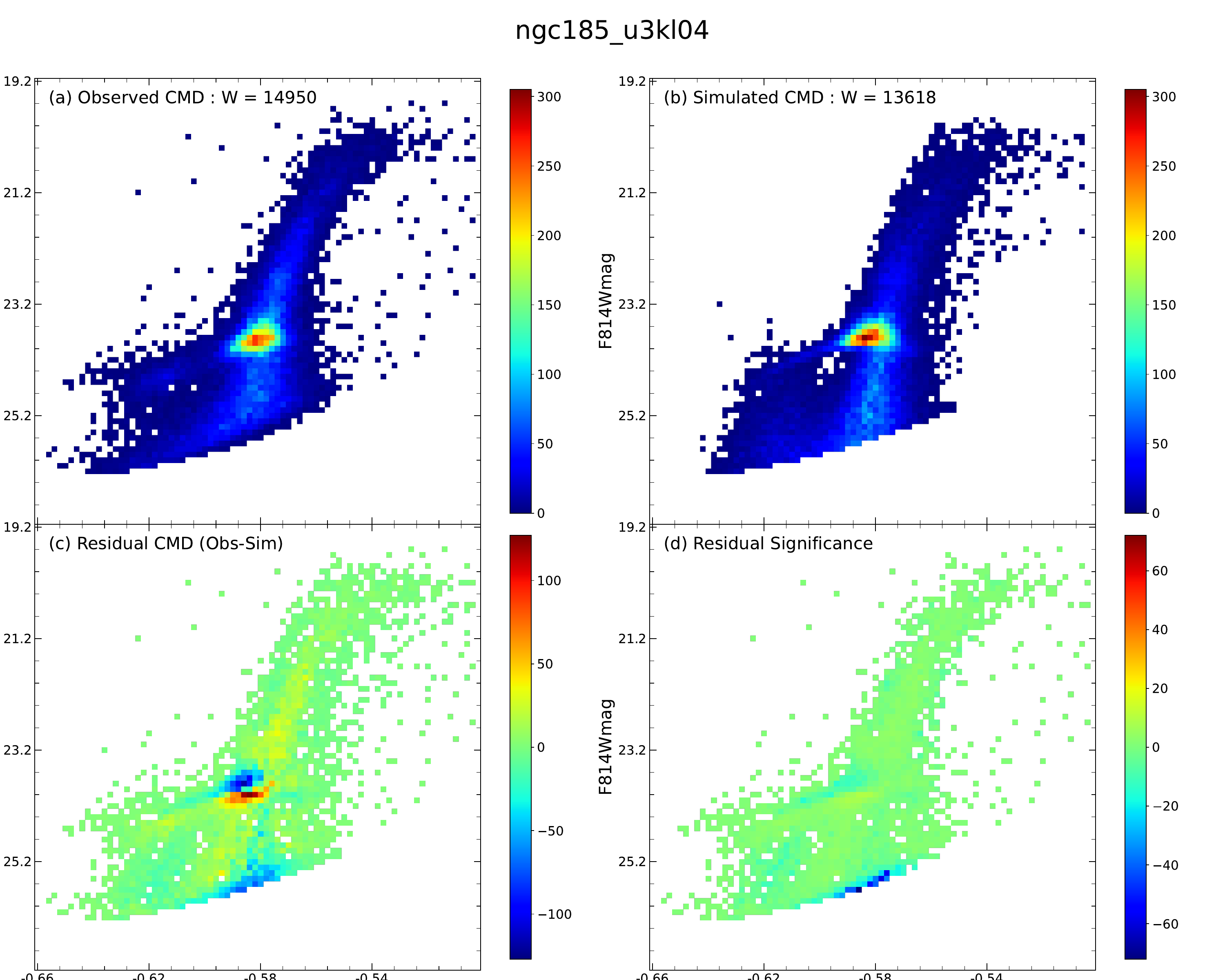}}
\subfigure{\includegraphics[width=0.5\textwidth]{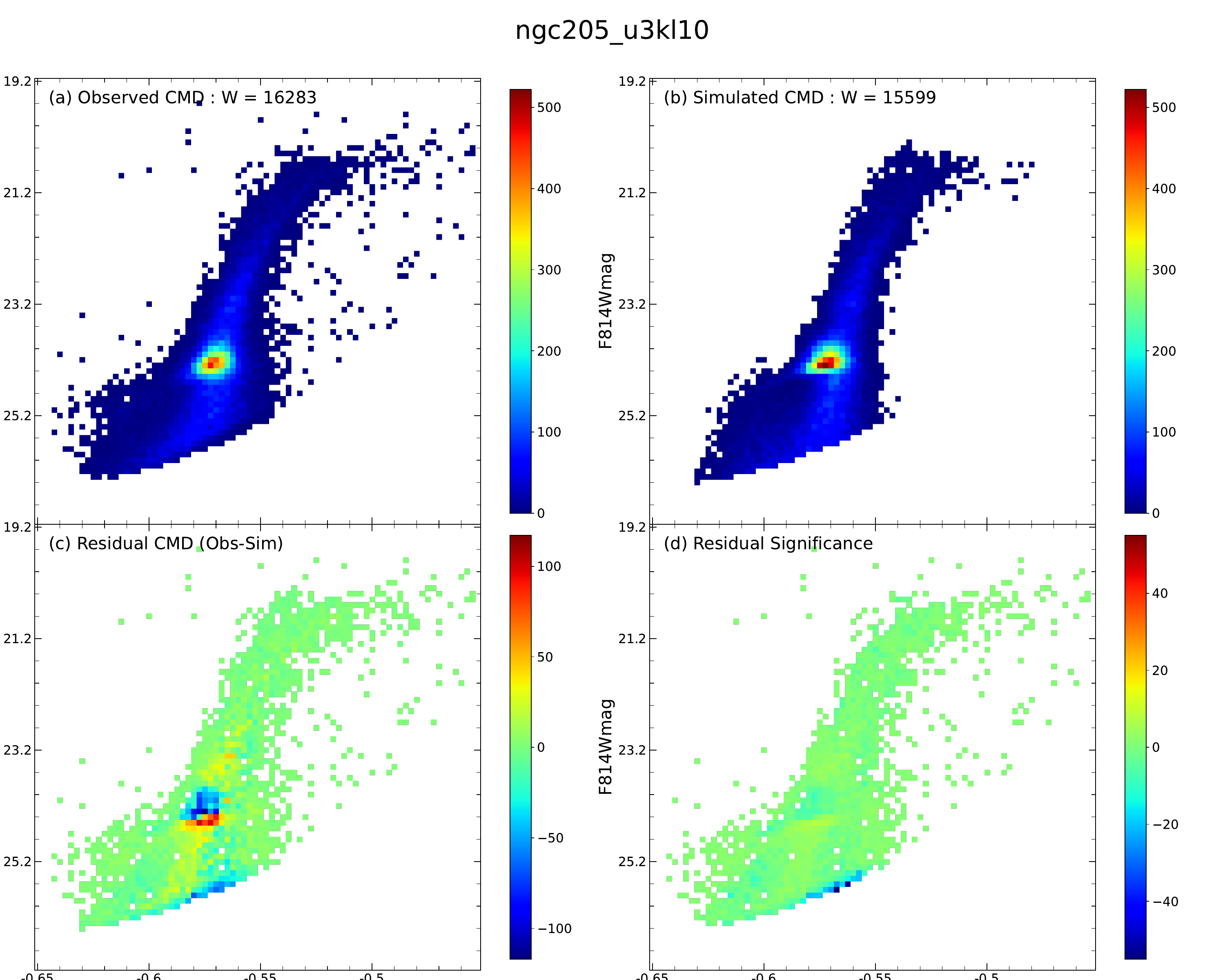}}\\
\caption{Illustration of the goodness of fit. Panels (a) and (b) show the observed CMD and the simulated CMD generated from \texttt{SFHNet} output parameters. The color bar indicates the number of stars in color-magnitude cells. Panel (c) depicts the residual CMD, which equals the observed CMD minus the simulated CMD. Panel (d) shows the residual significance (calculated as the residual of star numbers weighted by the standard deviation of the Poisson distribution).}
\label{fig:res_num_data1}
\end{figure}

\begin{figure}[!htb]
\subfigure{\includegraphics[width=0.5\textwidth]{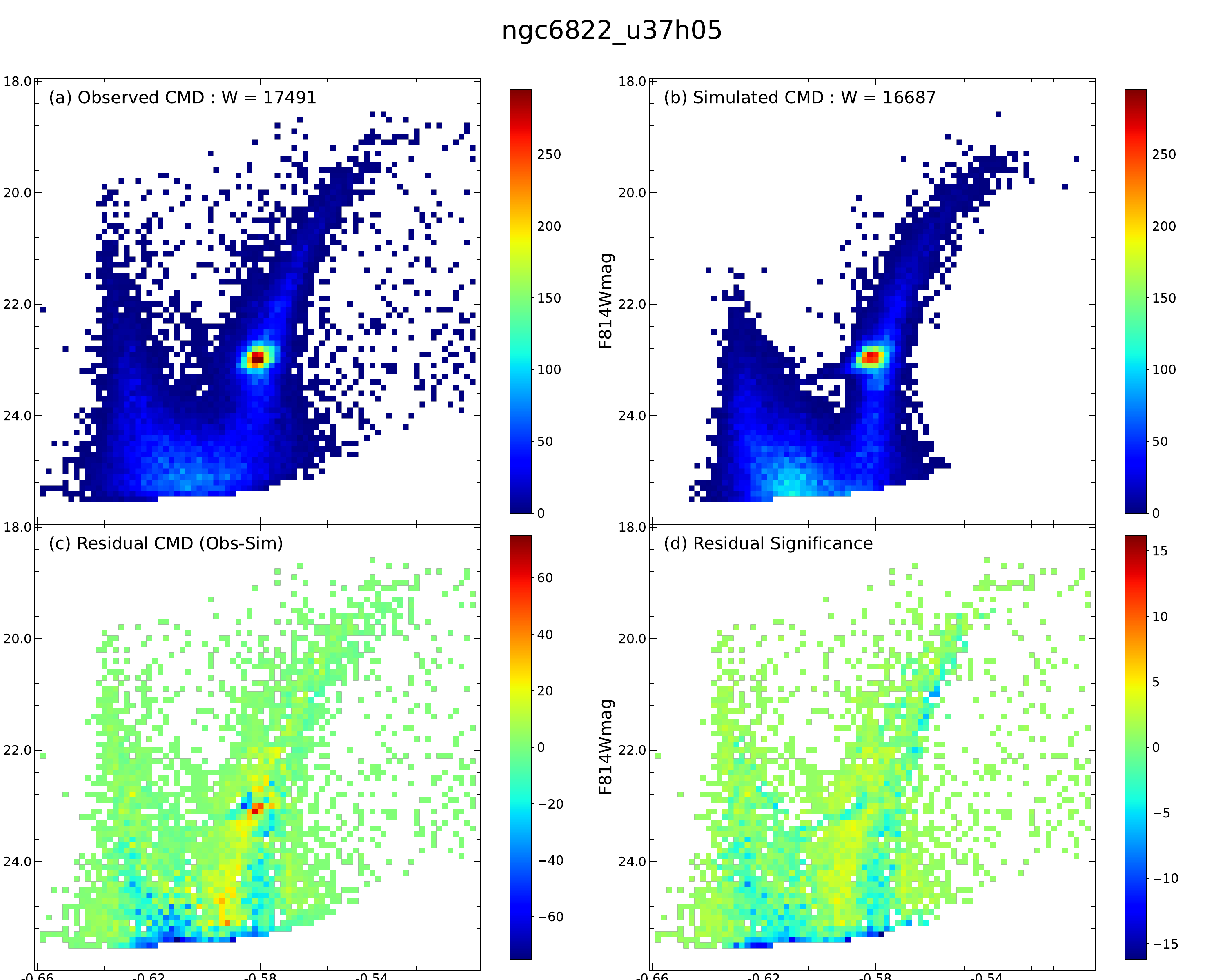}}
\subfigure{\includegraphics[width=0.5\textwidth]{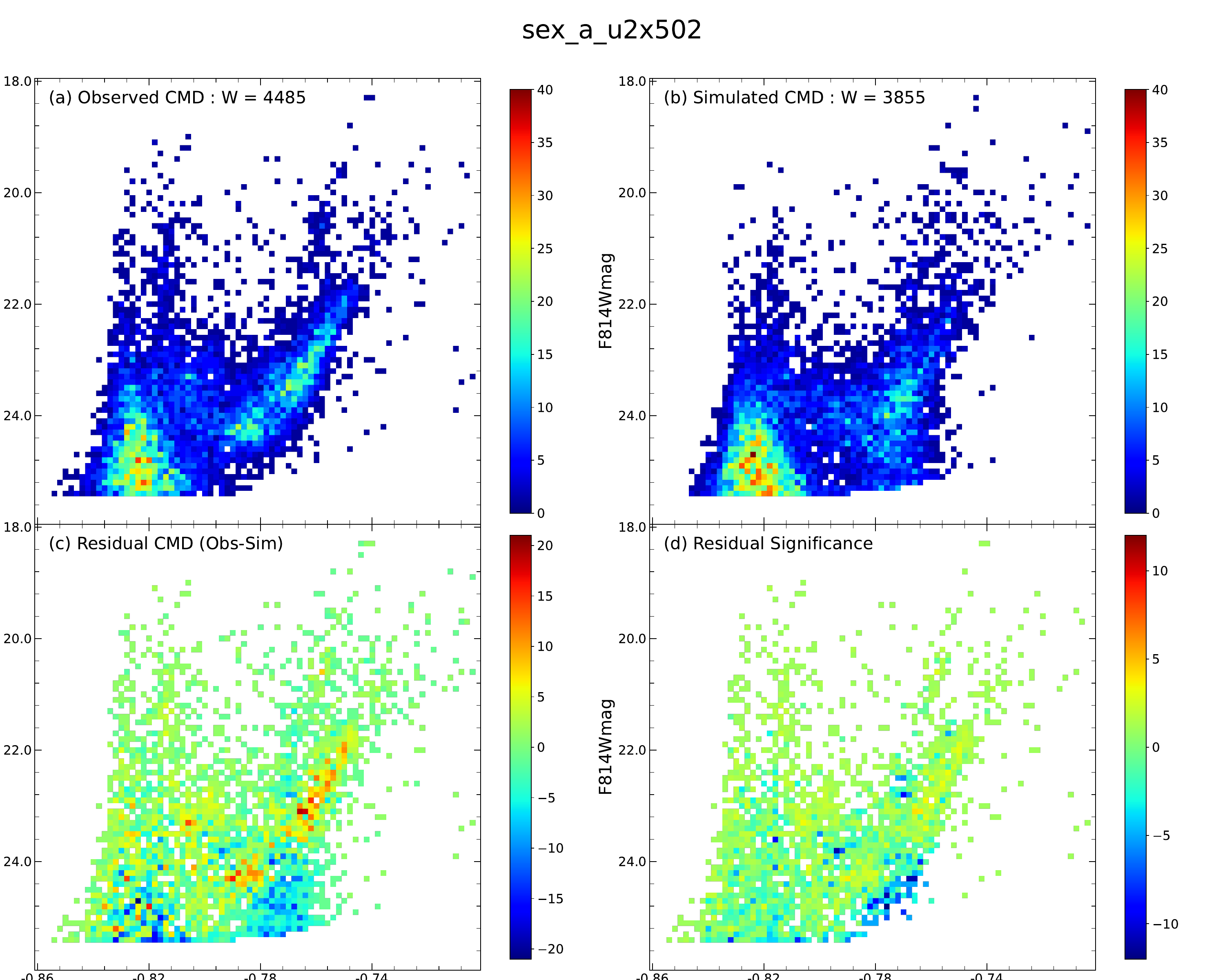}}\\
\subfigure{\includegraphics[width=0.5\textwidth]{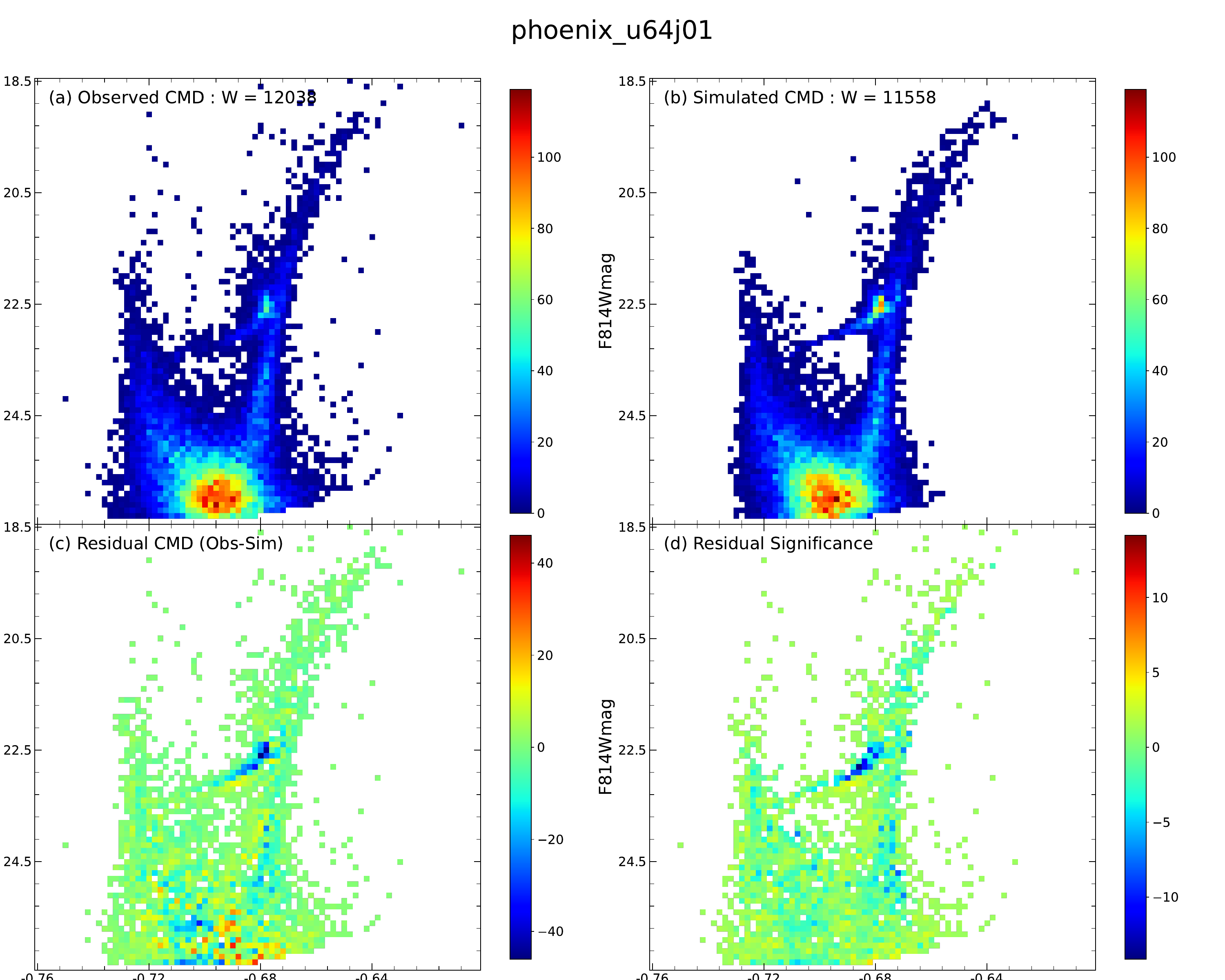}}
\subfigure{\includegraphics[width=0.5\textwidth]{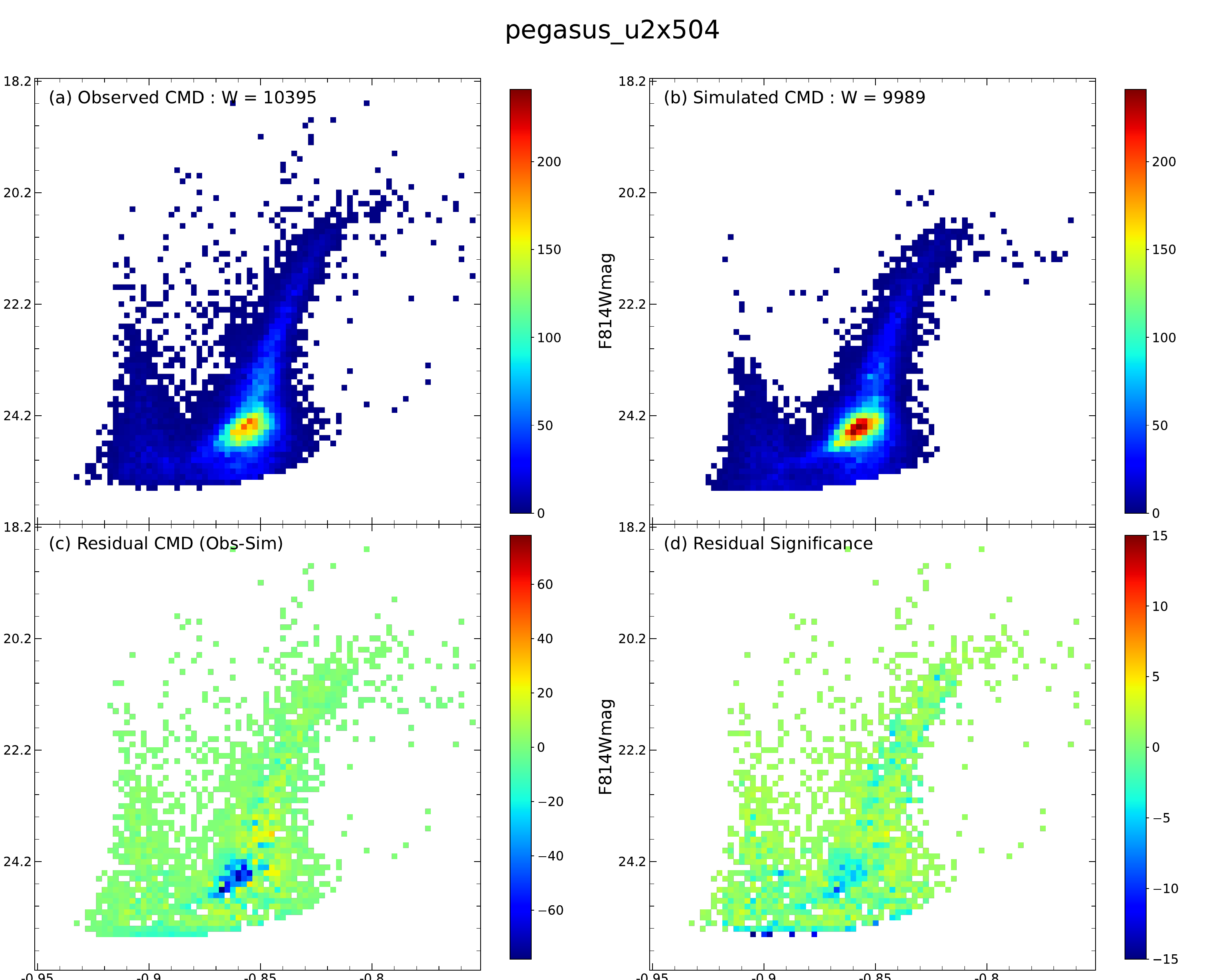}}\\
\caption{Same as Figure \ref{fig:res_num_data1}, but for dIrrs and dTrans.}
\label{fig:res_num_data2}
\end{figure}

We calculate the Saha $W$ values and residual significance to quantitatively evaluate the goodness of fit. Before that, we claim that the simulated stars are randomly selected from synthetic SSPs according to the fitted parameters. The goodness of fit is not only influenced by the fitting process but also influenced by the additional random uncertainties introduced by the simulation procedure. The synthetic CMD method essentially is a probabilistic approach. To show the major populations and minimize the influence of the additional random uncertainties, the simulated CMDs do not contain the stars generated from the age-metallicity cells with extremely small values. The threshold is determined by the 5\% cumulative distribution of the probabilities. After that, we calculate the Saha $W$ value between the observed CMD and itself, benchmarking the goodness of fit. The binned observed CMDs, binned simulated CMDs, their residuals, and residual significance are shown in Figures \ref{fig:res_num_data1} and \ref{fig:res_num_data2}, with Saha $W$ values labeled in the upper left corner. In general, the simulated CMDs align with the observed CMDs well, with the Saha $W$ values close to the benchmark. The majority of pixels have residual/residual significance values close to zero, and a few pixels (around HB/RC and the detection boundary areas) show less good fitting. We note that the simulated CMDs of early-type galaxies (Figure \ref{fig:res_num_data1}) featured blue stars that are not presented in the observed CMDs. Except for the uncertainties from the fitting and simulation processes, the young MS-BSSs ambiguity and the shallow photometric data contribute to the overpredicted young stars in Ursa Minor and Tucana, respectively. For Ursa Minor, as mentioned in Section \ref{subsec:errors}, including detailed synthetic binary populations may mitigate the effect of young MS-BSS ambiguity. For Tucana, the primary reason is the shallow photometric depth, which made the sub-giant stars misinterpreted as young MS stars. While the blue plume stars distribute across a considerable CMD area, their contribution to the total initial stellar mass is very minor. For example, the young populations less than 1 Gyr contribute to the total initial mass are less than 5\% for dSphs and 1\% for dEs, respectively.

\subsection{Analysis for individual galaxies}\label{subsec:csfh}
The synthetic CMD method has been widely used to measure SFHs of nearby galaxies. The test galaxies have been analyzed in many works. For a comprehensive view of the intrinsic properties of these galaxies, we have collected the relevant works and presented them in Table \ref{table:literatures}, with a brief discussion according to their morphological types.

Our primary results include the best-fit SFHs of eight dwarf galaxies (summarized in Figure \ref{fig:result_all}) and the detailed distribution of measured number density ($\rho_{\rm n}$), initial stellar mass ($M_{\rm ini}$), and $SFR$ across the age-metallicity plane and their projection over the age-axis (see the example of NGC 6822 in Figure \ref{fig:ngc6822rms}, that of the other galaxies are presented in the Appendix). We remind the readers to be cautious with the distribution of $SFR$ on the young side. On the one hand, the synthetic CMD method (e.g., the insensitive of MS to metallicity, the young MS-BSSs ambiguity) is not good at measuring the most recent SFR ($<$100 Myr); On the other hand, the short time interval magnifies the uncertainties of $SFR$. As a result, the recent SFRs have significant variation. We calculate the average SFR over the past 100 Myr, i.e., $SFR_{\rm 100}$ to characterize the recent SFR. In general, the $SFR_{\rm 100}$ of most galaxies (except Ursa Minor and Sextans A) is consistent with the $SFR$ of the nearby older stellar populations. 

Galaxies belonging to the same morphology type share common SFH profiles (Figure \ref{fig:result_all}) and display comparable distributions of stellar populations across the age-metallicity plane. Additionally, the differences between early-type and late-type galaxies are obvious. The early-type galaxies are dominated by old populations, with the $\tau_{\rm 50}$ (the time when 50\% of total stellar mass has formed) are older than 9 Gyr. In contrast, the values for late-type galaxies are younger than 8 Gyr.
 
\subsubsection{Dwarf Spheroidal Galaxies}
dSphs are characterized by a spheroidal shape and are predominantly constituted of old, metal-poor stellar populations. The mass maps of Ursa Minor and Tucana show two major groups concentrated at the old and metal-poor corners. The two galaxies have similar cumulative mass profiles (as depicted in Figure \ref{fig:result_all}), with the $\tau_{\rm 50}$ are older than 9 Gyr. We then discuss the properties of individual galaxies.

Figure \ref{fig:3dmass} shows the dominant mass of Ursa Minor formed in two short periods centered at (10.15,-1.9) and (10.0,-1.1) in the age-metallicity plane. The results are consistent with those reported in  \cite{2000MNRAS.317..831H}, \cite{2002AJ....123..813D}, and \cite{2002MNRAS.332...91D}. The derived ages are between 11 and 13 Gyr, and the derived mean metallicities vary from [Fe/H] = -2.2 to [Fe/H] = -1.9, and [M/H] = -1.5. Additionally, the spectroscopic studies \citep{2010ApJ...719..931C,2015MNRAS.449..761U} also yield the same large variation in metallicities, with [Fe/H] range from -1.35 to -3.10 dex. The wide metallicities range argues that Ursa Minor experienced an extended star formation episode that lasted nearly 5 Gyr with low efficiency. 

The stellar populations of Tucana are more complex. Two major groups centering at (10.15,-1.9) and (10.10,-1.3) in the age-metallicity plane, with the second group extending to the younger side (see Figure \ref{fig:tucanarms} in the Appendix). \cite{2005astro.ph..6430D} concluded that around 80 \% of the stars in Tucana were formed 10 Gyr ago, with a subsequent episode of star formation occurring a few Gyr ago. \cite{2010ApJ...722.1864M} employed the IAC, MATCH, and COLE codes to measure the SFH of Tucana. The results from the three numerical codes are consistent, showing the SFR reaches its peak at the age $\geq$ 12.5 Gyr and then gradually decreases until star formation ceased at $\sim$ 9 Gyr ago. At the same time, the mean metallicity steadily increases as a function of the lookback time, from [M/H] = -1.6 to -1.1. \cite{2019AA...630A.116S} further combined information from MSTO and HB to suggest that Tucana has experienced three major phases of star formation. Two of these phases occurred very close together at ancient times, and the last one ended between 6 and 8 Gyr ago. The best-fit SFHs in the age-metallicity plane of both our work (Figure \ref{fig:tucanarms}) and \cite{2019AA...630A.116S} (Figure 8) have similar patterns and maximum SFR values.

\subsubsection{Dwarf Elliptical Galaxies}
dEs have an elliptical shape and are primarily composed of old stars. NGC 185 and NGC 205 are satellites of the large spiral galaxy M31. The dominant populations of the two galaxies are located at (10.15,-0.9) and (10.15,-0.7) in the age-metallicity plane, respectively. Their stellar density, initial stellar mass, and SFR increase monotonically with age. The distribution of these properties over the age-metallicity plane exhibits a similar pattern, with that of NGC 205 displaying larger dispersion (see Figures \ref{fig:ngc185rms} and \ref{fig:ngc205rms}).

The first numerical estimation of NGC 185's SFH is presented in \cite{1999AJ....118.2229M}, showing that the bulk of the stellar populations formed at an early epoch and the star formation has continued at a low rate throughout the galaxy's lifespan. The average SFR is about $8.2 \times 10^{-3} {\rm M}_\odot{\rm yr}^{-1}$. \cite{2014ApJ...789..147W} then arrived at same conclusion. The recent star formation is solidified through analysing stellar content \citep[]{1993AJ....106..964L, 1999AJ....118..366M, 2005AJ....129.2217B,2005AJ....130.2087D,2005A&A...437...61K, 2010ApJ...713..992M}. The latest star formation finished varied from 0.2 Gy \citep{2017MNRAS.466.1764H}, 0.4 Gyr \citep{2005AJ....130.2087D} to 1 Gyr \citep{2005AJ....129.2217B}. Within them, \cite{2017MNRAS.466.1764H} used long-period variable stars to reconstruct the SFH of NGC 185. Their results show an abrupt and brief enhancement around 8.3 Gyr, then followed by a much lower and constant SFR until 200 Myr. The work also concluded that 70-90$\%$ of the total stellar mass was assembled with log($t$) $\sim$ (9.56, 9.90), which is similar to our estimation that log($t$) $\sim$ (9.6, 9.9). However, \cite{2015ApJ...811..114G} showed 70$\%$ of stars were formed 12.5 Gyr ago, with the remaining stars formed between 8 and 10 Gyr. The overestimated old population is likely a consequence of the fitting process excluding the HB and RC stars. In the case of metallicity, \cite{2005AJ....129.2217B} identified ancient stars with [Fe/H] $\lesssim$ -1.5 by the well-defined HB and inferred the median [Fe/H] = -1.11 $\pm$ 0.08. \cite{2014MNRAS.445.3862C} found [Fe/H] = -1.12 $\pm$ 0.35 based on the color information of RGB stars.

Stars in NGC 205 are slightly metal-rich compared to those in NGC 185 \citep[the median iron abundance is $-1.06 \pm 0.04$;][]{2005AJ....129.2217B}, and their metallicity distribution is more concentrated. In the literature, NGC 205's SFH is less mentioned since it is closer to M31. The contamination from M31 stars complicates the composite stellar populations. \cite{2005astro.ph..6430D} showed that half of the stars in NGC 205 were formed in the past 10 Gyr. The $\tau_{\rm 50}$ is consistent with that derived in this work.  Besides that, several works reveal young populations in the center area \citep[]{1996AJ....112.1438L, 1999ApJ...515L..17C,2005AJ....129.2217B,2005AJ....130.2087D,2009A&A...502L...9M}, suggesting the central region has experienced continuous star formation over the last 25 to 650 Myr. The field we analysed here is far from the galaxy center. We do not find any significant young populations in the observed CMDs, and the average SFR over the past 100 Myr is nearly zero. These findings reveal the spatial variations of star formation. 

\begin{table}[!htb]
\footnotesize
\centering
\caption{SFHs of Eight Galaxies in the Literature}
\begin{threeparttable}
\begin{tabular}{llllll}
\hline
\hline
Galaxy Name & Reference & Instrument & Filters & CEH & Method  \\[1ex]
\hline
(1) & (2) & (3) & (4) & (5) & (6) \\[1ex]
\hline
Ursa Minor & \cite{2000MNRAS.317..831H} & WFPC2/HST & F555W,F814W & No  & Maximum Likelihood  \\
           & \cite{2002AJ....123.3199C} & WFC/INT\tnote{1}   & B,V,I       & Yes & $\chi^2$ statistics \\
           & \cite{2002MNRAS.332...91D} & WFPC2/HST & F555W,F814W & Yes & MATCH  \\
           & \cite{2005astro.ph..6430D} & WFPC2/HST & F555W,F814W & Yes & MATCH \\
           & \cite{2014ApJ...789..147W} & WFPC2/HST & F555W,F814W & Yes & MATCH  \\
\hline
Tucana     & \cite{2005astro.ph..6430D} & WFPC2/HST & F555W,F814W & Yes & MATCH\\
           & \cite{2010ApJ...722.1864M} & ACS/HST   & F475W,F814W & Yes & IAC, MATCH, Cole \\
           & \cite{2013ApJ...778..103H} & ACS/HST   & F475W,F814W & Yes & IAC\\
           & \cite{2014ApJ...789..147W} & WFPC2/HST & F555W,F814W & Yes & MATCH \\
           & \cite{2019AA...630A.116S} & ACS/HST  & F555W,F814W & Yes & MORGOTH \\
\hline
\hline

NGC 185    & \cite{1999AJ....118.2229M} & NOT\tnote{2}       & B, V        & Yes & Maximum Likelihood \\
           & \cite{2005astro.ph..6430D} & WFPC2/HST & F555W,F814W & Yes & MATCH\\
           & \cite{2014ApJ...789..147W} & WFPC2/HST & F555W,F814W & Yes & MATCH \\
           & \cite{2015ApJ...811..114G} & ACS/HST   & F606W,F814W & Yes & MATCH (exclude RC,HB) \\
           & \cite{2017MNRAS.466.1764H} &  NOT;CFHT\tnote{3}  & VIK;JHK    & No  & long-period variables\\
           & \cite{2024arXiv240206953R} & UKIRT\tnote{4} & JHK & No & $\chi^2$ statistics \\
        
\hline           
NGC 205     & \cite{2005astro.ph..6430D} & WFPC2/HST & F555W,F814W & Yes & MATCH\\
            & \cite{2014ApJ...789..147W} & WFPC2/HST & F555W,F814W & Yes & MATCH  \\
            
\hline
\hline

NGC6822    & \cite{1996AJ....112.1950G} & INT  & V,I & Yes & Multiple indicators \\
           & \cite{2001AJ....122.2490W} & WFPC2/HST & F555W,F814W & Yes & $\chi^2$ statistics  \\
           & \cite{2003AJ....125.3097W} & WFPC2/HST & F555W,F814W & Yes & MATCH  \\
           & \cite{2005astro.ph..6430D} & WFPC2/HST & F555W,F814W & Yes & MATCH  \\
           & \cite{2012ApJ...747..122C} & ACS/HST   & F475W,F814W & Yes & MATCH  \\
           & \cite{2014ApJ...789..147W} & WFPC2/HST & F555W,F814W & Yes & MATCH  \\
           & \cite{2014AA...572A..26F} & ACS/HST   & F475W,F814W & Yes & IAC \\
           & \cite{2024arXiv240206953R} & UKIRT & JHK & No & $\chi^2$ statistics \\
\hline
Sextans A      & \cite{1997AJ....114.2527D} & WFPC2/HST & V,I         &     & star counts (MS and BHeB) \\
           & \cite{2002AJ....123..813D} & WFPC2/HST & F555W,F814W &     & star counts (MS and BHeB) \\
           & \cite{2003AJ....126..187D} & WFPC2/HST & F555W,F814W & Yes &  MATCH \\
           & \cite{2005astro.ph..6430D} & WFPC2/HST & F555W,F814W & Yes & MATCH \\
           & \cite{2011ApJ...739....5W} & WFPC2/HST & F555W,F814W & Yes & MATCH \\
           & \cite{2014ApJ...789..147W} & WFPC2/HST & F555W,F814W & Yes & MATCH \\
           & \cite{2024arXiv240206953R} & UKIRT & JHK & No & $\chi^2$ statistics \\
              
\hline
\hline

Phoenix    & \cite{2000AJ....120.3060H} & WFPC2/HST & F555W,F814W & Yes & $\chi^2$ statistics \\
           & \cite{2005astro.ph..6430D} & WFPC2/HST & F555W,F814W & Yes & MATCH\\
           & \cite{2009ApJ...705..704H} & WFPC2/HST & F555W,F814W & Yes & IAC \\
           & \cite{2013ApJ...778..103H} & WFPC2/HST & F555W,F814W & Yes & IAC \\
           & \cite{2014ApJ...789..147W} & WFPC2/HST & F555W,F814W & Yes & MATCH \\
\hline

Pegasus    & \cite{1997AJ....114..669A} & NOT       & V,I         & Yes & Multiple indicators\\ 
           & \cite{1998AJ....115.1869G} & WFPC2/HST & F555W,F814W & Yes & Bayesian inference\\
           & \cite{2005astro.ph..6430D} & WFPC2/HST & F555W,F814W & Yes & MATCH\\
           & \cite{2014ApJ...789..147W} & WFPC2/HST & F555W,F814W & Yes & MATCH  \\
           & \cite{2024arXiv240206953R} & UKIRT & JHK & No & $\chi^2$ statistics \\
\hline
\hline

\end{tabular}
\begin{tablenotes}
\item[1] INT: Isaac Newton Telescope.
\item[2] NOT: Nordic Optical Telescope.
\item[3] CHFT: Canada-France-Hawaii Telescope.
\item[4] UKIRT: United Kingdom Infra-Red Telescope.
\end{tablenotes}
\end{threeparttable}
\label{table:literatures}
\end{table}

\subsubsection{Dwarf Irregular Galaxies}
dIrrs have an irregular shape and contain copious star formation. The observed CMDs of NGC 6822 and Sextans A are featured with young and old populations simultaneously. The $\tau_{\rm 50}$ of NGC 6822 is around 8 Gyr, while that of Sextans A is approximately 5 Gyr, indicating that these galaxies contain a large number of intermediate-age stars.

\begin{figure}[!htb]
\includegraphics[width=9cm]{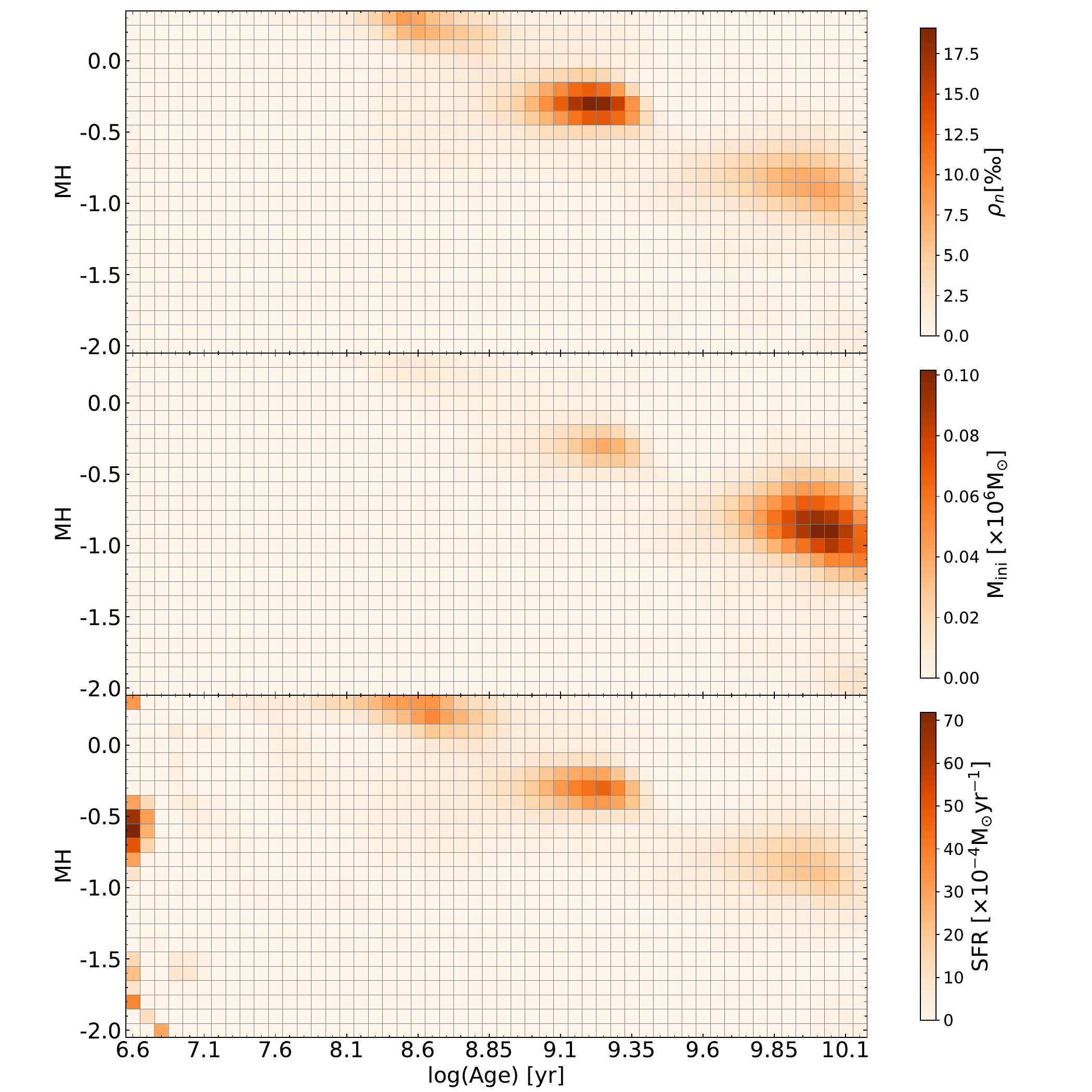}
\includegraphics[width=9cm]{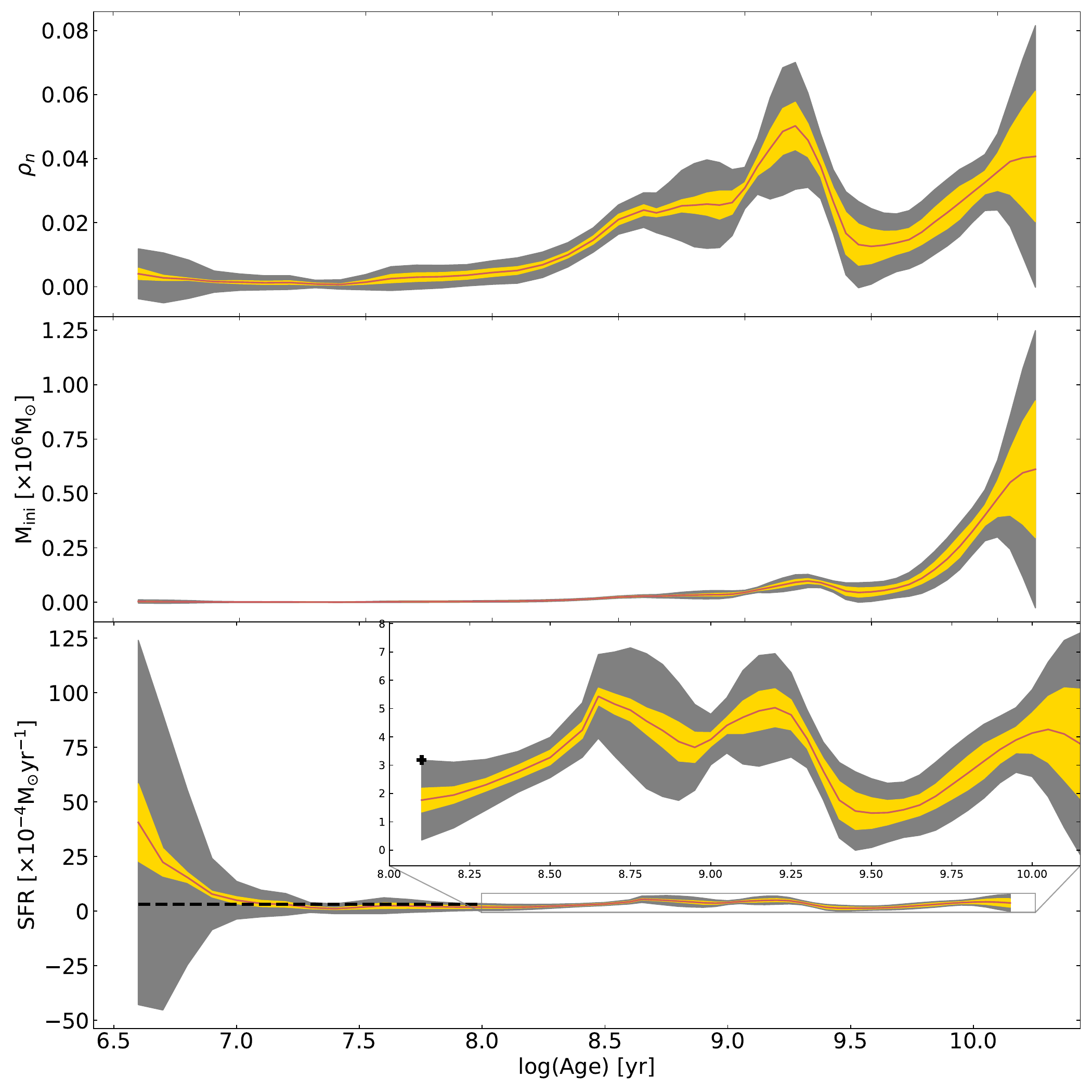}
\caption{The measured number density $\rho_{\rm n}$ (top panel), initial stellar mass $M_{\rm ini}$ (middle panel), and $SFR$ (bottom panel) for NGC 6822 across the age-metallicity plane (left panel) and their projection over the age-axis (right panel). The yellow/gray shadow indicates the random/total (random and systematic) uncertainties, guided by the output of \texttt{SFHNet} (red solid line). A zoom-in view of $SFR$ at the old age side (log($t$)$\geq$ 8.0) is provided in the inset panel. The average SFR over the past 100 Myr, i.e., $SFR_{\rm 100}$ is marked as a black dash line and black plus in the bottom-right panel and its inset panel, respectively.}
\label{fig:ngc6822rms}
\end{figure}

For NGC 6822, the peak in stellar density is observed at coordinates (9.2, -0.3) in the upper left panel of Figure \ref{fig:ngc6822rms}, while the maximum initial stellar mass is at (10.0, -1.0). The intermediate-age stars ( 9.0$ \leq log(t) \leq $9.4) contribute to around 14.5\% of the total stellar mass.  The ratio of the number of stars across the full mass range to the number of stars within the detectable mass range varies considerably between stellar populations. In the old populations, the massive stars have evolved, and the remaining stars have a narrow mass range, resulting in a large coefficient (the last term of Equation \ref{eq:mass}). In the case of NGC 6822, the coefficient for the two populations centered at (9.2, -0.3) and (10.0, -1.0) vary by a factor of up to eight. The gap between the intermediate-age and ancient stellar populations aligns with the SFR dip at 3-5 Gyr observed in \cite{2001AJ....122.2490W}. Another young group of populations centered at (8.5,0.3). In conclusion, we observed three distinct SFR peaks and the recent star formation rate $SFR_{\rm 100} \approx 3 \times 10^{-4} {\rm M}_{\odot} {\rm yr}^{-1}$ for NGC 6822.  

As one of the nearest and most isolated dIrrs, NGC 6822 has been extensively analysed in the literature. The first quantitatively measured SFH of NGC 6822 is provided in \cite{1996AJ....112.1950G, 1996AJ....112.2596G, 1996AJ....112.1928G}. Combined with \cite{2001AJ....122.2490W, 2003AJ....125.3097W}, these works concluded that the SFR of NGC 6822 is relatively constant or increases from 15 Gyr ago to present. After that, \cite{2012ApJ...747..122C} and \cite{2014AA...572A..26F} concluded that NGC 6822 averagely has $\tau_{\rm 50} \sim$ 5 Gyr. Among the six fields analysed in \cite{2012ApJ...747..122C}, the Fields 3 and 4 have $\tau_{\rm 50}$ around 7 Gyr. The two fields have approximately the same distance to the central bar structure as the field analysed here. For the same field, \cite{2014ApJ...789..147W} measured the $\tau_{\rm 50}$ around 3.8 Gyr.

In the context of stellar content analysis, NGC 6822 has young, intermediate-age, and old stellar populations simultaneously. The distinguished young MS, the young stellar objects, red super giants \citep[]{2019MNRAS.490..832J,2020ApJ...892...91H, 2021A&A...647A.167Y,2024arXiv240513499H} solid the recent star formation; the RC, intermediate RGB, the extended AGB \citep[]{1994ApJ...425L...9G,2022ApJ...933..197T}, RR-Lyrae stars \citep[]{2003ApJ...588L..85C, 2003MmSAI..74..860B}, and the weak horizontal branch populations reveal the ancient and intermediate-age stellar populations. In terms of metallicity, the observations of RGB stars yield a mean metallicity [Fe/H] = -1 $\pm$ 0.5 with a large spread \citep{2001MNRAS.327..918T,2003PASP..115..635D} and an average metallicity [Fe/H] = -0.84 $\pm$ 0.04  \citep{2016MNRAS.456.4315S}, respectively. The empirical carbon/M-type stars ratio-metallicity relations indicate that the [Fe/H] varies from -1.0 to -1.4 dex \citep{2022ApJ...933..197T}. Under the assumption of a simple one-to-one correlation between [Fe/H] and [M/H], these values are consistent with the metallicities of old populations detected in this work.

The number density map of Sextans A comprises two distinct regions. The first is a concentration of old and metal-poor populations, situated at (10.1,-1.8). The remaining stars are distributed across a relatively broad area, with log($t$) ranges from 7.6 to 8.85, and [M/H] spans almost the full range (see Figure \ref{fig:sexarms}). The number of young stars is approximately three times higher than old stars, indicating an enhanced recent star formation. The initial star formation occurred at approximately 14 Gyr ago with $SFR \approx 8 \times 10^{-4} {\rm M}_\odot {\rm yr}^{-1}$. The SFR then keeps relatively low until around 1.5 Gyr ago, after which it increases until the present time. 

The low metallicity and high recent SFR discriminate Sextans A from other Local Group galaxies. From the first attempt to measure Sextans A's SFH using the luminosity functions \citep[]{1997AJ....114.2527D, 2002AJ....123..813D} to attempts using CMD-fitting method \citep[]{2003AJ....126..187D, 2005astro.ph..6430D,2011ApJ...739....5W,2014ApJ...789..147W}, the results all agree with that the SFR has increased dramatically in the past few gigayears. The high SFR over the past gigayear has also been detected in \cite{2005astro.ph..6430D},  \cite{2011ApJ...739....5W}, and in this work. Besides that, \cite{2011ApJ...739....5W} yield $\tau_{\rm 50} \sim 4.8$ Gyr, which is consistent with our results 5 Gyr and much younger than $\sim 13$ Gyr from \cite{2014ApJ...789..147W}. One discrepancy we notice is that there is a great metallicity spread in the young populations, which does not agree well with the conclusion that Sextans A reached a metallicity of [M/H]= -1.4 more than 10 Gyr ago and still have that metallicity today \citep{2003AJ....126..187D}. Since the large uncertainties of SFR for young stars, whether a significant chemical enrichment within Sextans A needs to be verified with further observations.

Figure \ref{fig:result_all} shows the SFHs of two dIrrs not only have large uncertainties but also show less consistency with results derived from \cite{2014ApJ...789..147W}. Our measured  $\tau_{\rm 50}$ are much older. The discrepancies drive us to look into the effects of IMF on SFH. Before that, we mention that nearly all CMD-fitting works simulate the CMDs with an invariant IMF for all populations. However, the universality of IMF is under debate \citep{2014prpl.conf...53O, 2020ARA&A..58..577S, 2023Natur.613..460L}. We use the same photometric and ASTs data from \cite{2014ApJ...789..147W} and adopt the same binary fraction, IMF, and age/metallicity grid to straightforwardly compare the deep learning method and the numerical method. The significant difference between the two methods is how to connect the stellar mass with star numbers. In the numerical method, the partial CMDs are extracted from a fixed initial mass budget. Different sampling methods (the ways to discretize the total stellar mass to a list of stars) vary the SFHs \citep{2017A&A...607A.126Y}. The different sampling methods result in varied number of massive stars, the occurrence rate of supernovae, and the chemical enrichment history, which eventually lead to different CMDs. To avoid the effects of different sampling methods, we firstly measure the number of stars in individual populations and then calculate the total mass by integrating the IMF function over the minimum and maximum masses. This method releases the minimum mass limit of individual stellar populations, thereby increasing the variability of fitted populations. Thus, we can reproduce a more diffuse CMD, which is more similar to the observed distribution.

\subsubsection{Transition Dwarf Galaxies}
dTrans are characterized by recent star formation. Their morphology type lies between that of irregular and spheroidal types, which indicate the transition from a phase of active star formation to a more quiescent state. The CMDs of Phoenix and Pegasus are both characterized by young and old stars. The derived density maps have similar patterns. Specifically, Phoenix/Pegasus contains old populations with parameters centered at (10.15,-1.3)/(10.15,-1.2), and intermediate-age populations centered at (9.7, -0.8)/(9.55,-0.7) in the age-metallicity plane. The initial stellar mass increases steadily over the entire lifetime, which agrees with a constant SFR. 

Many studies have arrived at the same conclusion. Phoenix has ongoing and gradually decreasing star formation over time \citep[]{2000AJ....120.3060H,2015arXiv151200567S,2009ApJ...705..704H}. \cite{2009ApJ...705..704H} measured the $\tau_{\rm 50}$ of Phoenix is 10.5 Gyr, which is $\sim$ 7.5 Gyr in this work. Additionally, \cite{2013ApJ...778..103H} statistically measured the SFHs over different radii, which shows the cessation of recent star formation as the galactocentric radius increased. In the aspect of metallicity, the old populations have an overall metallicity around -1.7 dex and that of the young populations centered at around -1.2 dex \citep[]{1999A&A...345..747H, 2000AJ....120.3060H,2009ApJ...705..704H,2013ApJ...778..103H,2015AJ....149..198R,2017MNRAS.466.2006K}. Pegasus has experienced roughly continuous star formation for 15 Gyr \citep {1997AJ....114..669A,1998AJ....115.1869G,2014ApJ...789..147W,2015arXiv151200567S}. \cite{2013ApJ...779..102K} claimed the mean metallicity of Pegasus is [Fe/H]$\approx -1.39 \pm 0.01$.

We also note the difference between the two dTrans. The most notable distinction is the presence of HB features in Phoenix, indicating that this system has a substantial component of old, metal-poor stellar populations. A careful comparison between them may help us to disentangle the effects of individual parameters (such as metallicity, age, and enrichment history) on the strength and morphology of HB, which is beyond the scope of this paper. Besides that, Pegasus has a group of much younger populations at log($t$) = 8.5.

\section{Summary}\label{sec:sum}
Based on the synthetic CMD method, this work uses a deep learning network to measure the star formation history of star-resolved galaxies. We improve the synthetic CMD method with the state-of-the-art stellar evolution model, convolve the theoretical data with observational effects (uncertainties and incompleteness) derived from ASTs, and accelerate the fitting process with deep learning network \texttt{SFHNet}. The fine-tuned model is capable of predicting the distribution of stars across the age-metallicity map. We measure the SFHs of eight nearby dwarf galaxies, which evenly belong to four morphological types. Our measurements align with those reported in the literature. dSphs and dEs are dominated by ancient populations. The projection of stellar density, initial stellar mass, and SFR over the age show a monotonic decline from ancient times to the present. The SFHs of dIrrs and dTrans are more complex. dIrrs are featured with enhanced recent star formation, with the intermediate-age stars dominating the stellar density maps. We identify three discrete star-forming episodes for NGC 6822 and two star-forming episodes for Sextans A (the recent one spreads in a much wider parameter space). dTrans almost keeps a constant SFR over galaxies' entire lifetime.

Results from our deep learning method are generally consistent with those obtained through the numerical method, except for some discrepancies observed in the case of dIrrs. The application of the deep learning network allows for processing large galaxies without an increase in computation time. Our results display a smooth transition between stellar populations. In the big data era, the upcoming space telescopes will provide complete and homogeneous data of large-size galaxies. The well-trained  \texttt{SFHNet} can quickly measure SFHs at varied spatial scales, showing a great advantage at measuring the spatially resolved SFHs of large-size galaxies. The spatial distribution of SFHs not only reveals the history of galaxy mergers and interactions but also advances our understanding of galaxy evolution and feedback mechanisms.

\texttt{SFHNet} is an initial attempt to utilize the deep learning network for SFH measurement. It is expected that future developments will address some of the current limitations of this deep learning approach.  For example, adding binary interaction to the binary population synthesis may mitigate the influence of young MS-BSS ambiguity on the measurement of recent SFR; incorporating the spectroscopic metallicity to reduce the effect of the degeneracy of metallicity, age, and reddening; improving the network to correct the misidentification of subgiant stars in the situation of shallow photometric depth. The high capacity of machine learning networks allows to explore the effects of additional parameters (such as distance, differential reddening, IMF, and binary frequency) on the SFH of galaxies. The high efficiency of machine learning networks may facilitate the study of the SFH of resolved galaxies in the forthcoming era of big data.

\begin{acknowledgments}
We gratefully thank the anonymous referee for their insightful comments and constructive suggestions. Y.Y acknowledges the support from the National Key Research and Development Program of China No.2022YFF0504200, National Natural Science Foundation of China (NSFC) with grant No.12203064. M.Y acknowledges the support from NSFC with grants Nos. 12373048, 12133002. H.T. is supported by NSFC with grants Nos. 12103062, 12173046 and National Key Research and Development Program of China with No. 2019YFA0405500. We acknowledge the science research grants from the China Manned Space Project with No. CMS-CSST-2021-A08. 
\end{acknowledgments}
\facilities{HST}
\software{PyTorch \citep{2019arXiv191201703P}, NumPy \citep{harris2020array}, pandas \citep{reback2020pandas}, Matplotlib \citep{Hunter:2007}, astropy \citep{2022ApJ...935..167A}, dustmaps\citep{2018JOSS....3..695M},PySynphot\citep{2013ascl.soft03023S}, PlotNeuralNet\citep{haris_iqbal_2018_2526396}}

\begin{appendix}
For completeness, we provide the distribution of stellar number density ($\rho_{\rm n}$), initial mass ($M_{\rm ini}$), and $SFR$ over the age-metallicity map and their projection over the age of the other galaxies. They have the same format as Figure \ref{fig:ngc6822rms}.

\begin{figure}[!htb]
\includegraphics[width=9cm]{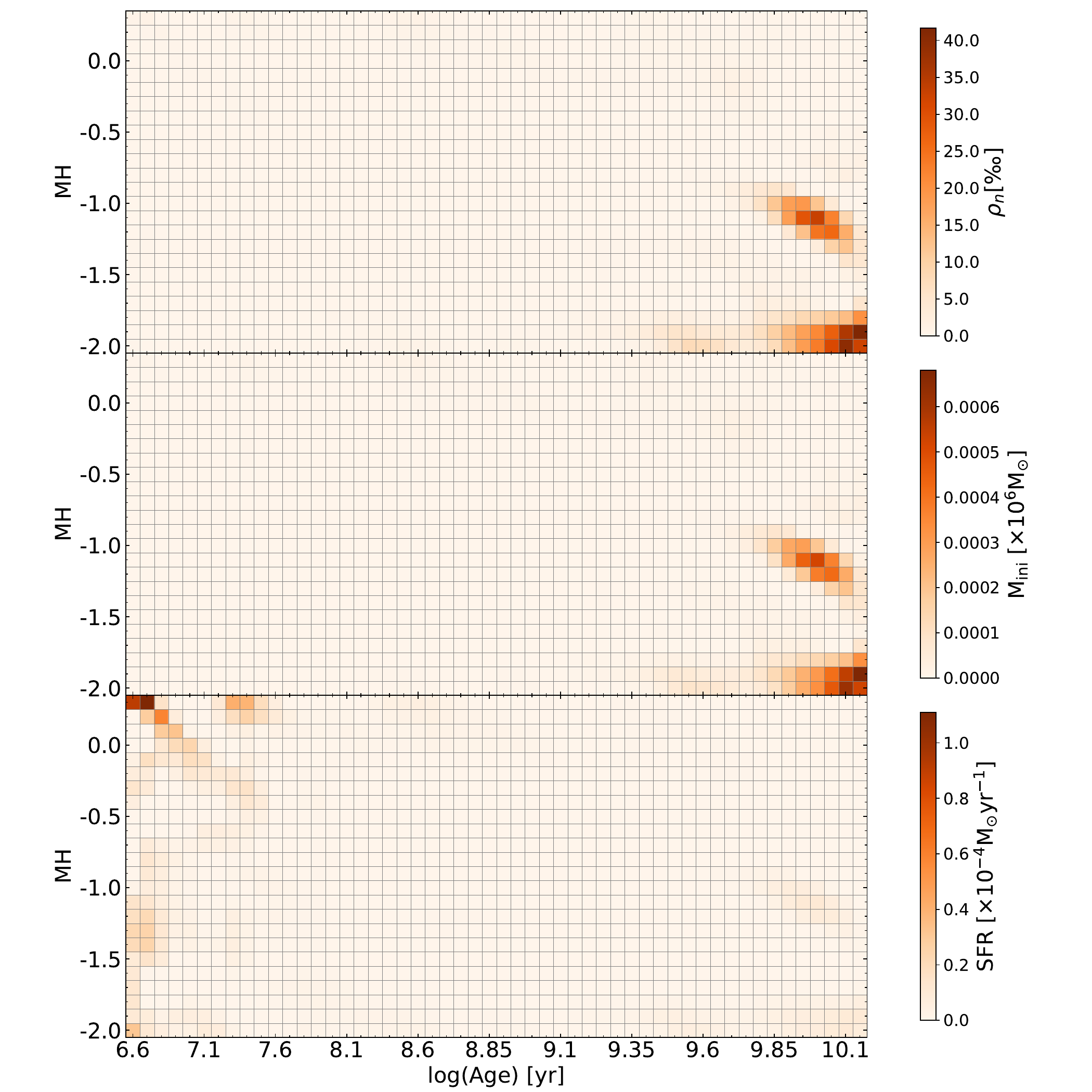}
\includegraphics[width=9cm]{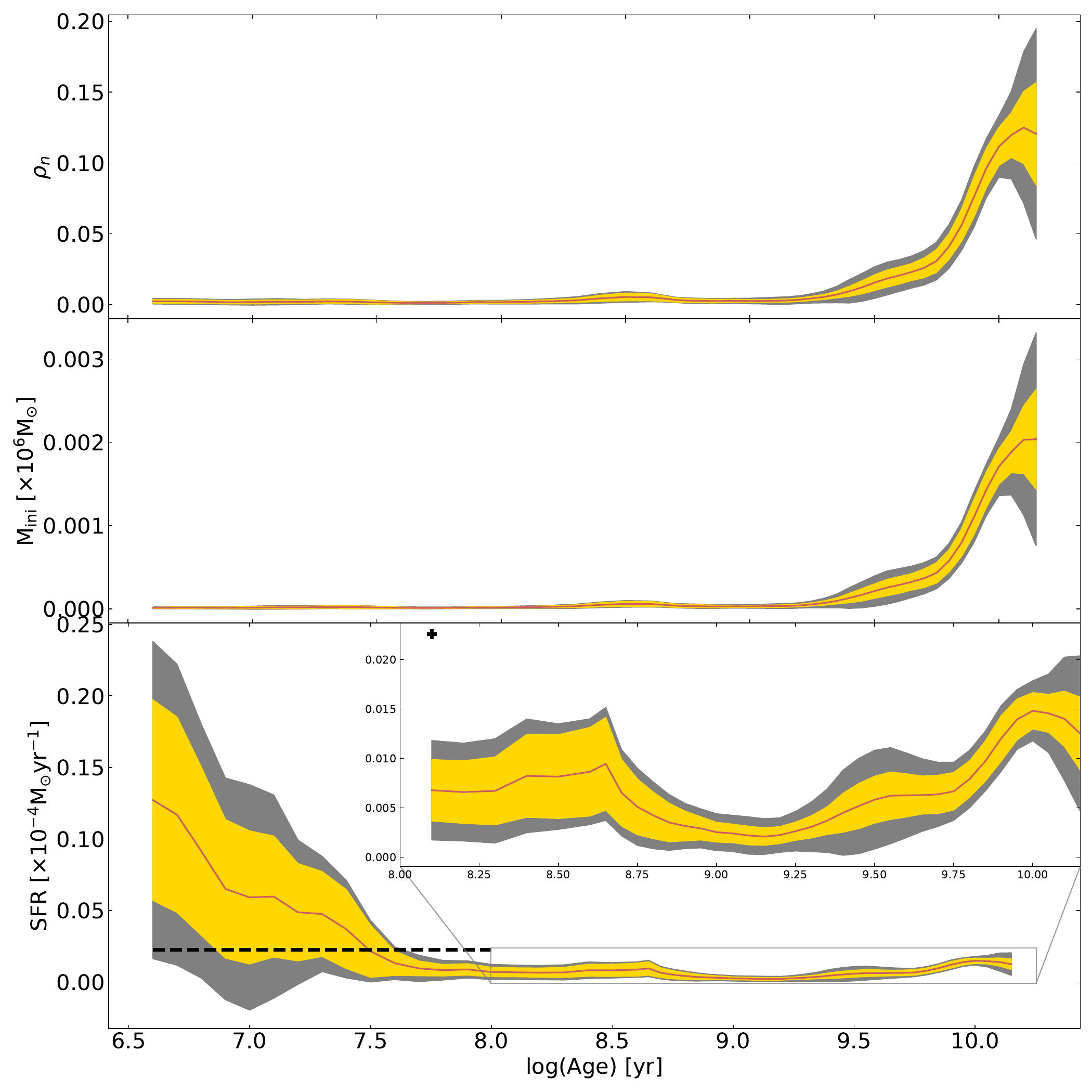}
\caption{Same as Figure \ref{fig:ngc6822rms}, but for Ursa Minor.}
\label{fig:ursaminorrms}
\end{figure}

\begin{figure}[!htb]
\includegraphics[width=9cm]{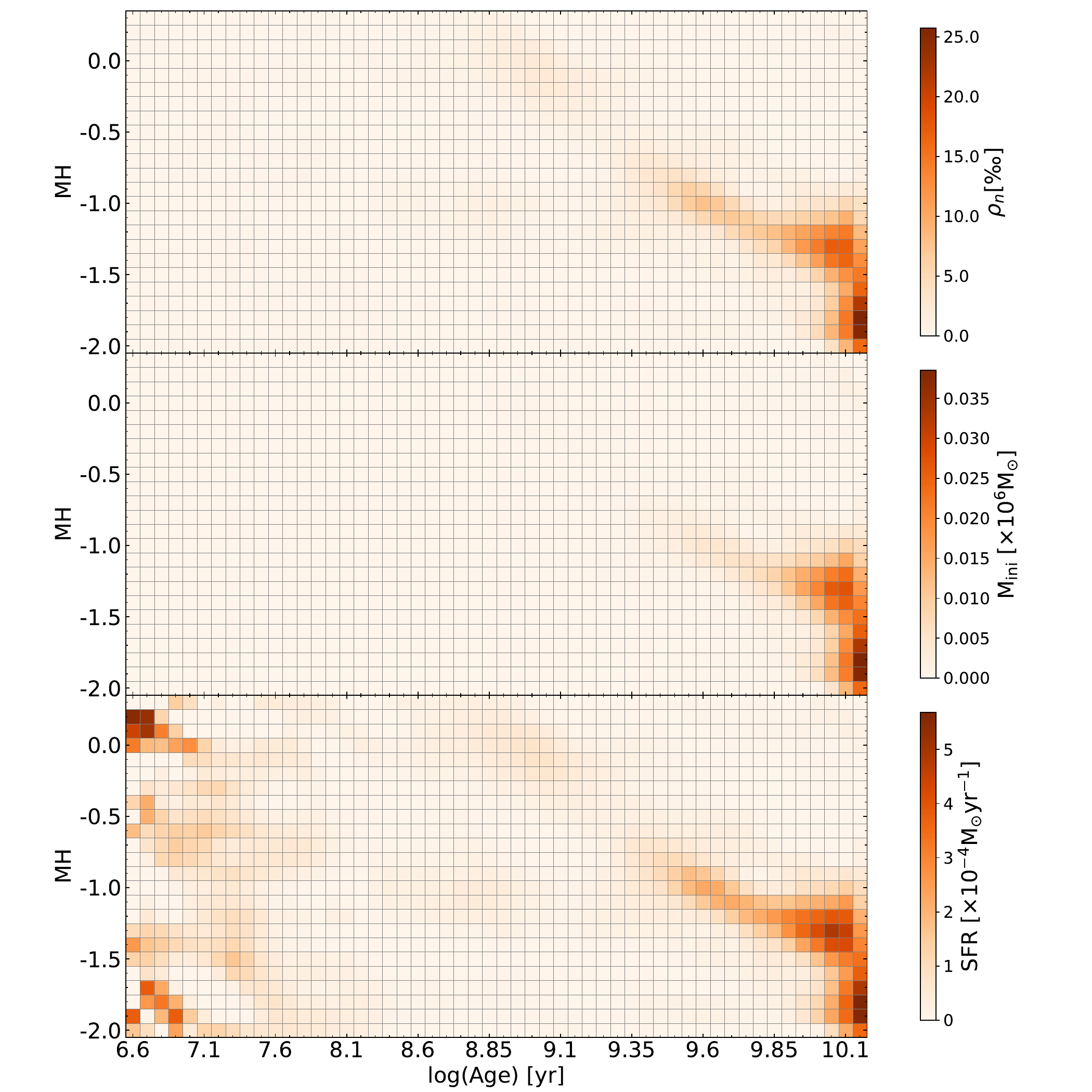}
\includegraphics[width=9cm]{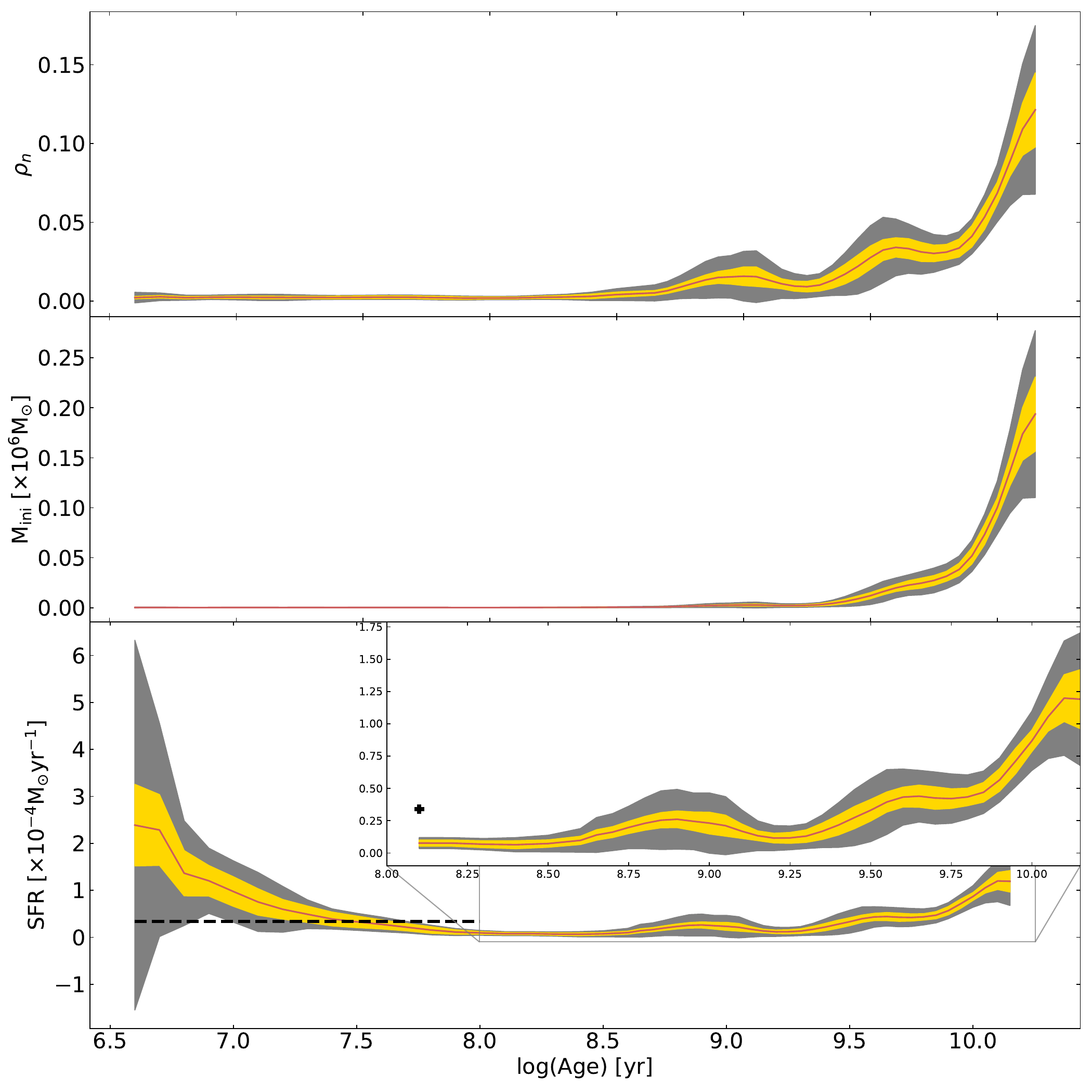}
\caption{Same as Figure \ref{fig:ngc6822rms}, but for Tucana.}
\label{fig:tucanarms}
\end{figure}

\begin{figure}[!htb]
\includegraphics[width=9cm]{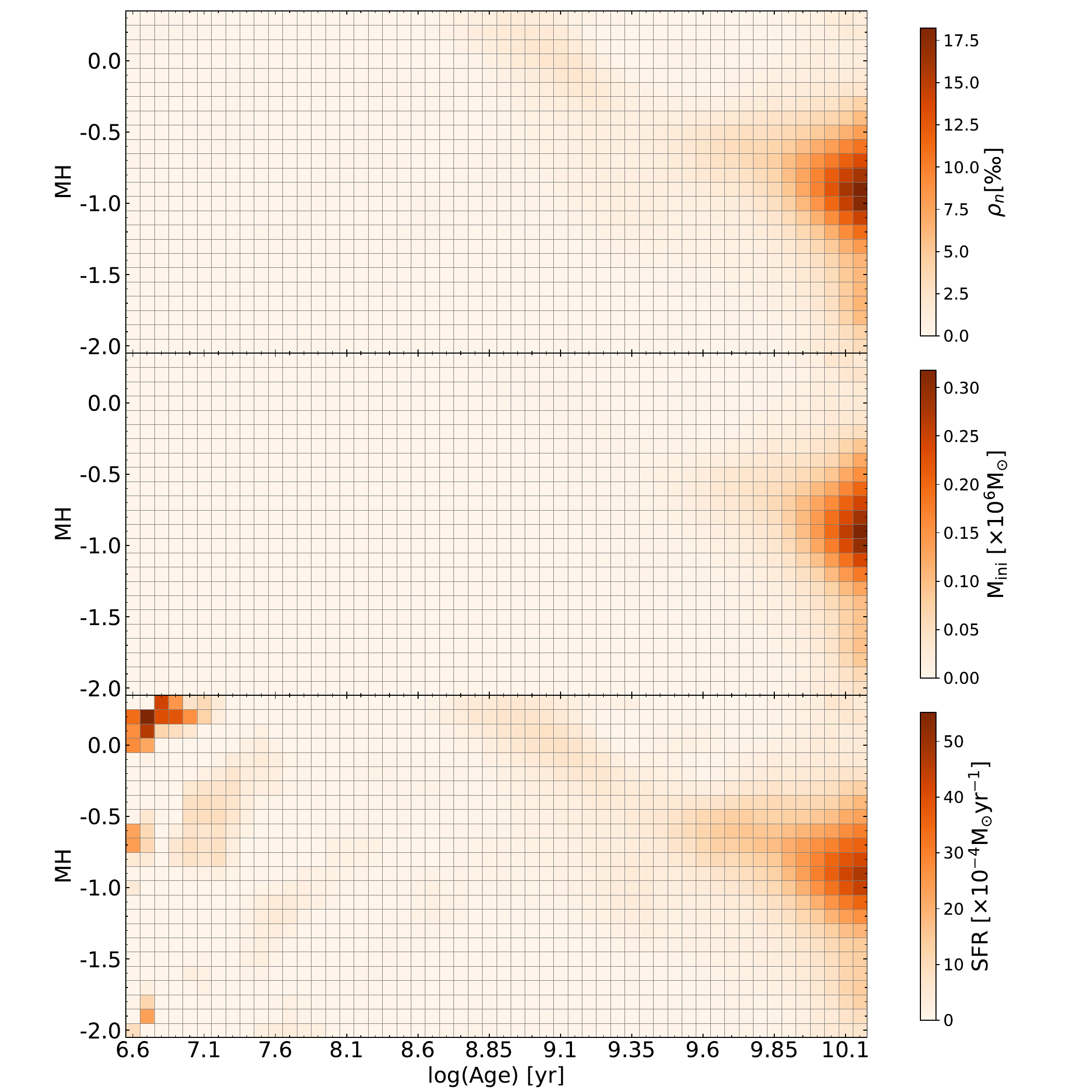}
\includegraphics[width=9cm]{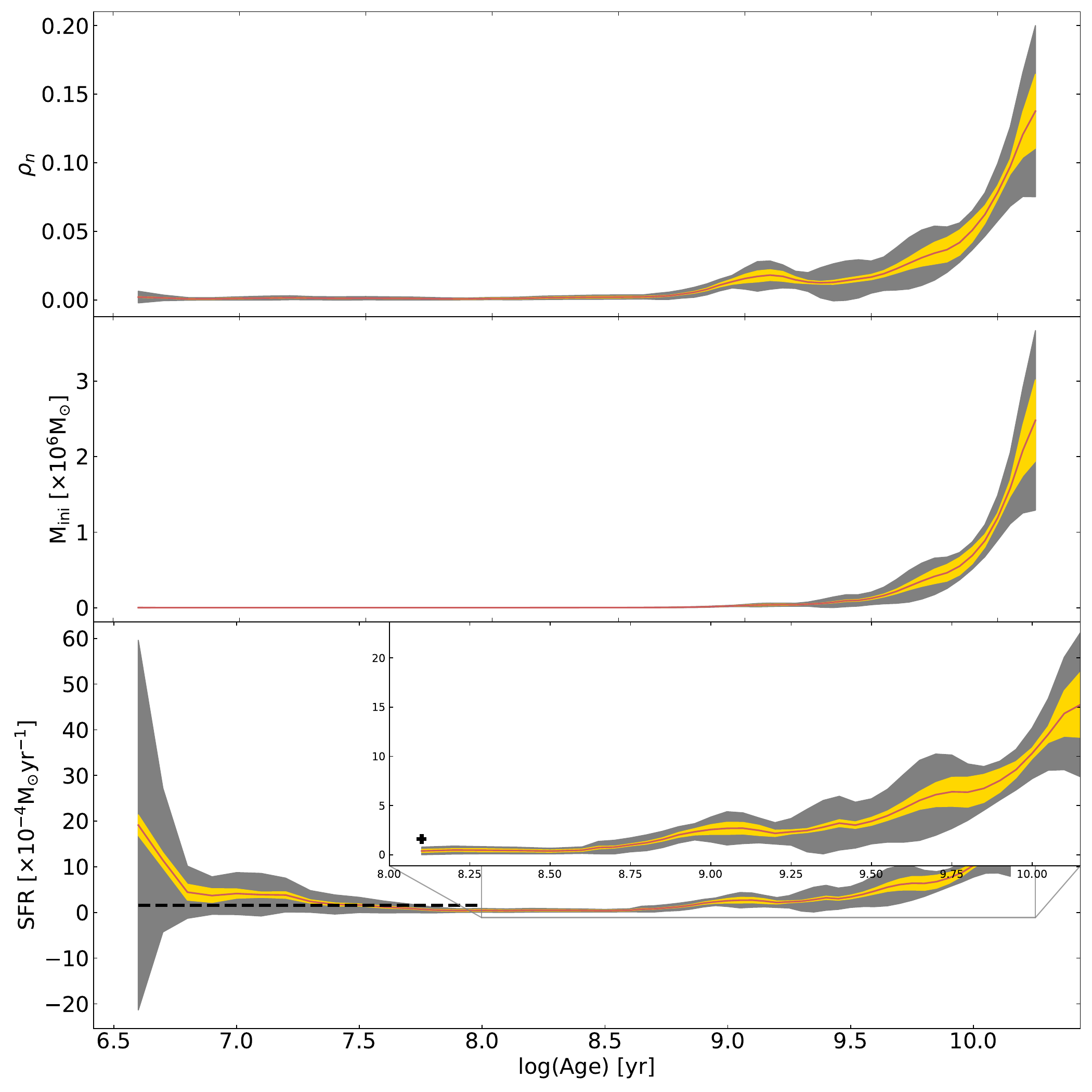}
\caption{Same as Figure \ref{fig:ngc6822rms}, but for NGC 185.}
\label{fig:ngc185rms}
\end{figure}

\begin{figure}[!htb]
\includegraphics[width=9cm]{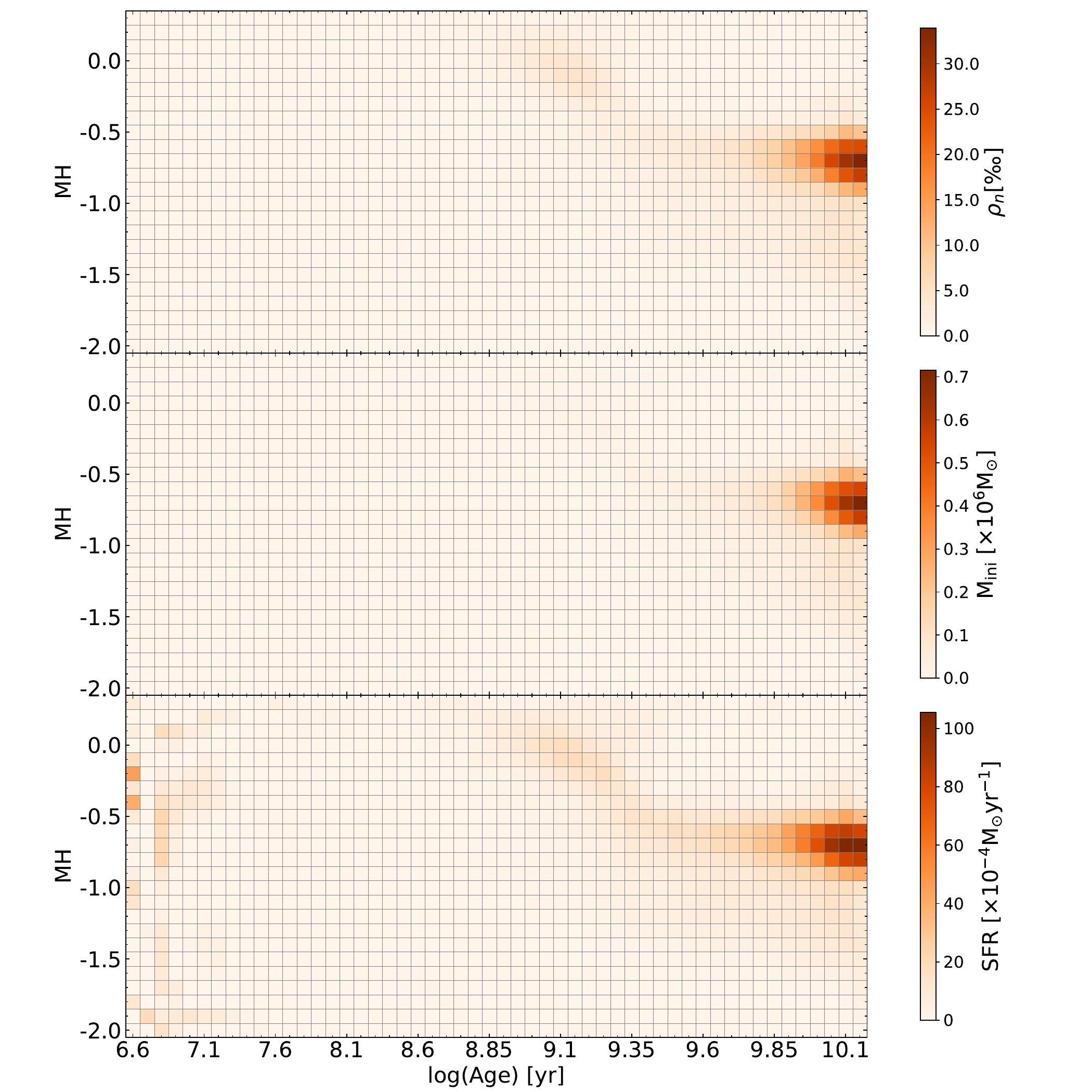}
\includegraphics[width=9cm]{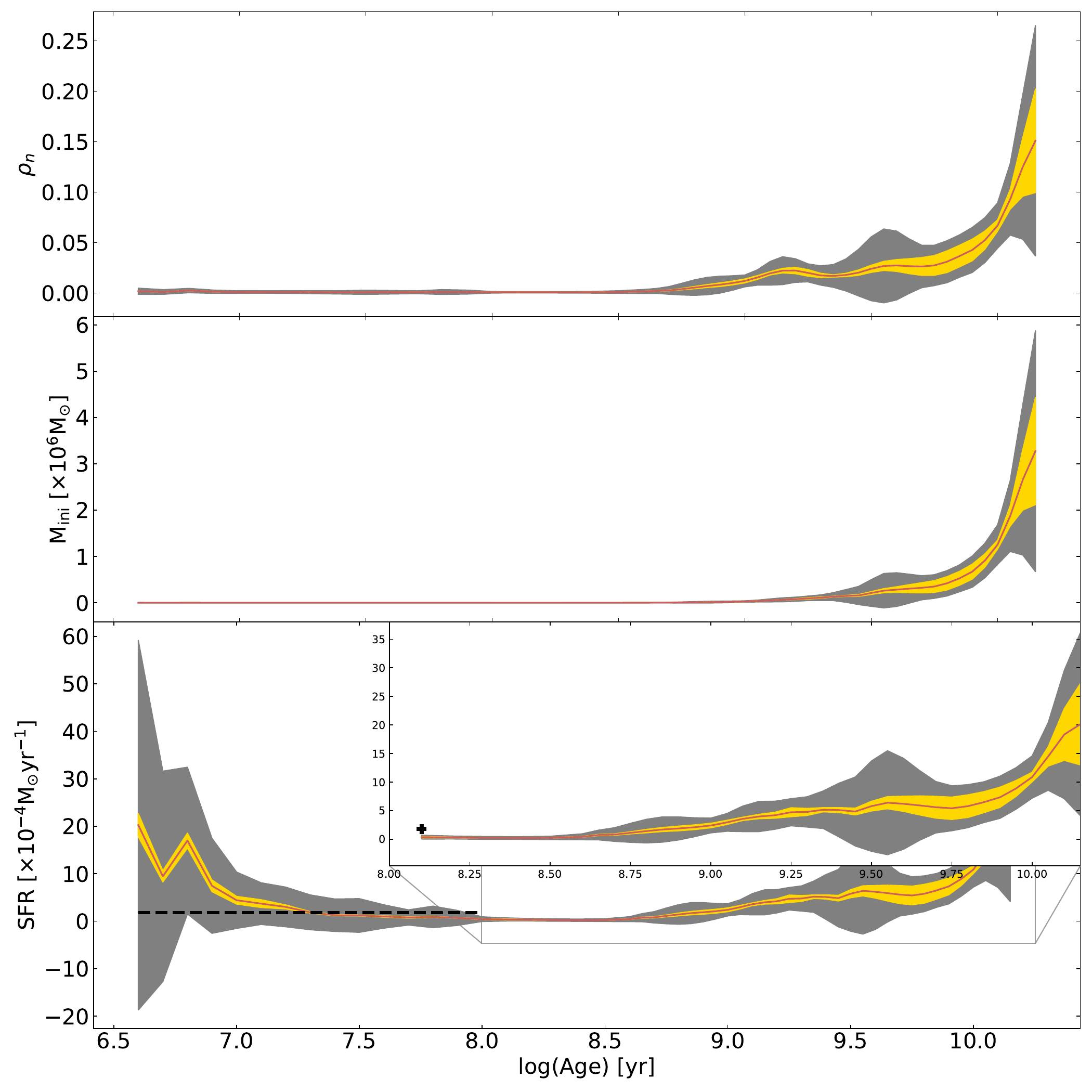}
\caption{Same as Figure \ref{fig:ngc6822rms}, but for NGC 205.}
\label{fig:ngc205rms}
\end{figure}

\begin{figure}[!htb]
\includegraphics[width=9cm]{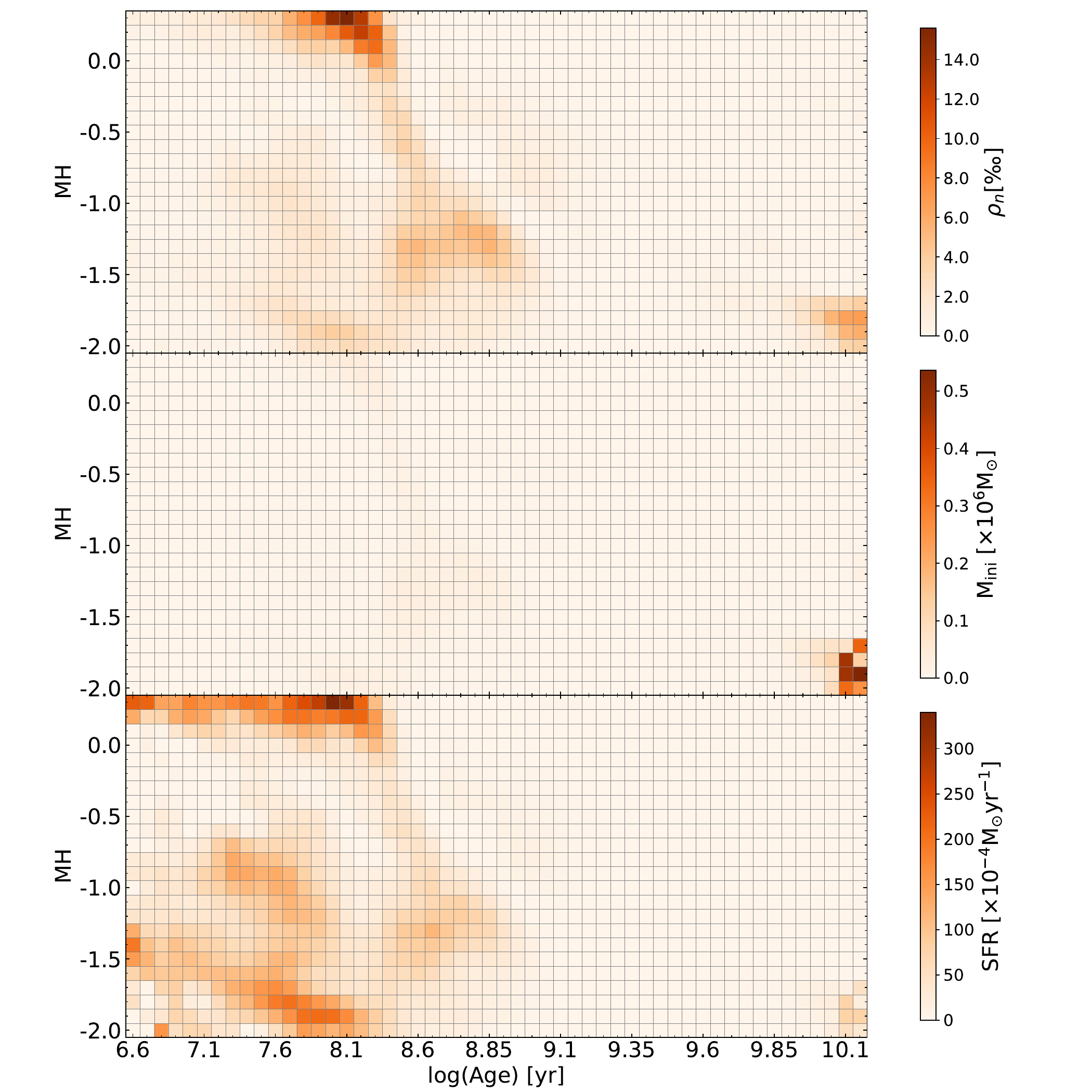}
\includegraphics[width=9cm]{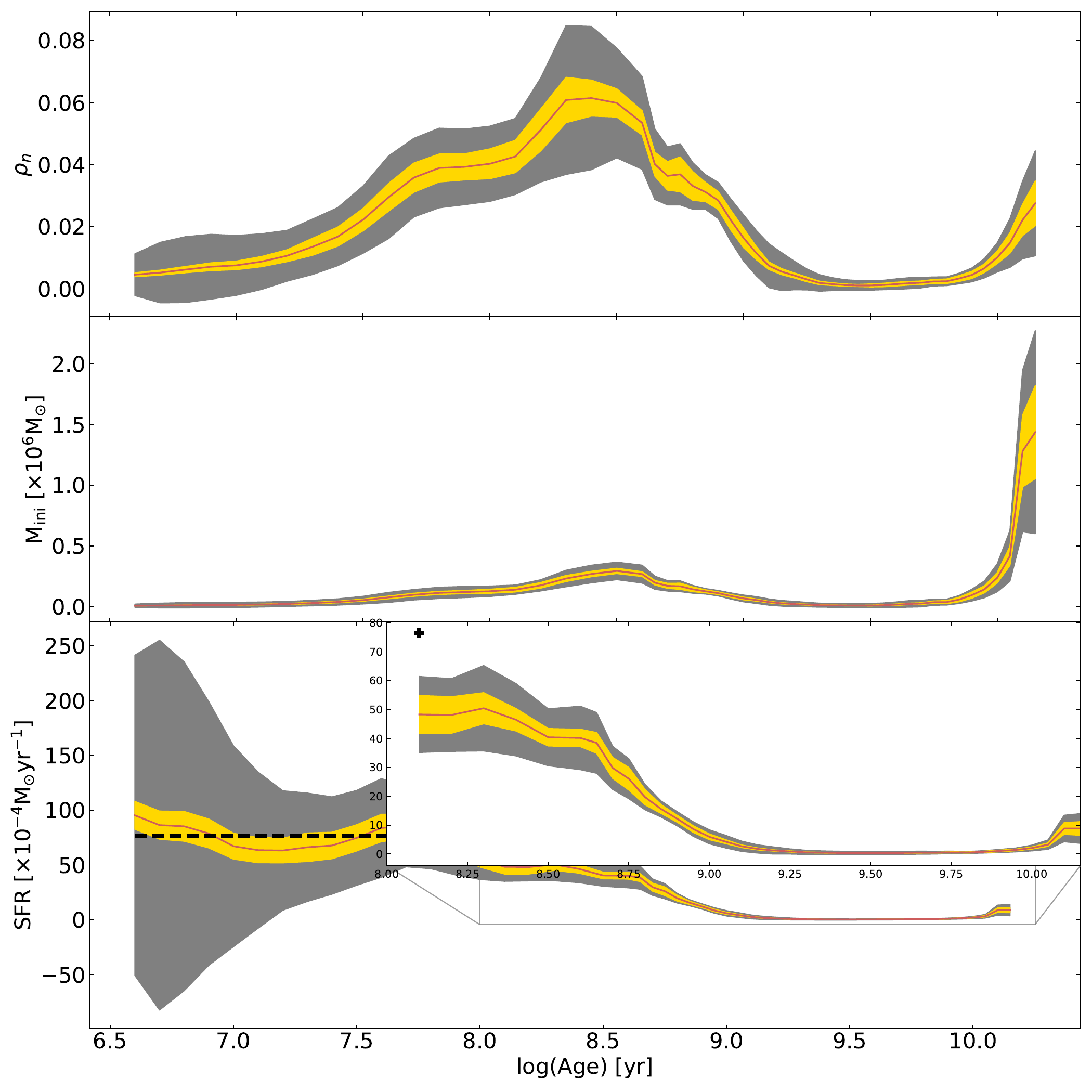}
\caption{Same as Figure \ref{fig:ngc6822rms}, but for Sextans A.}
\label{fig:sexarms}
\end{figure}

\begin{figure}[!htb]
\includegraphics[width=9cm]{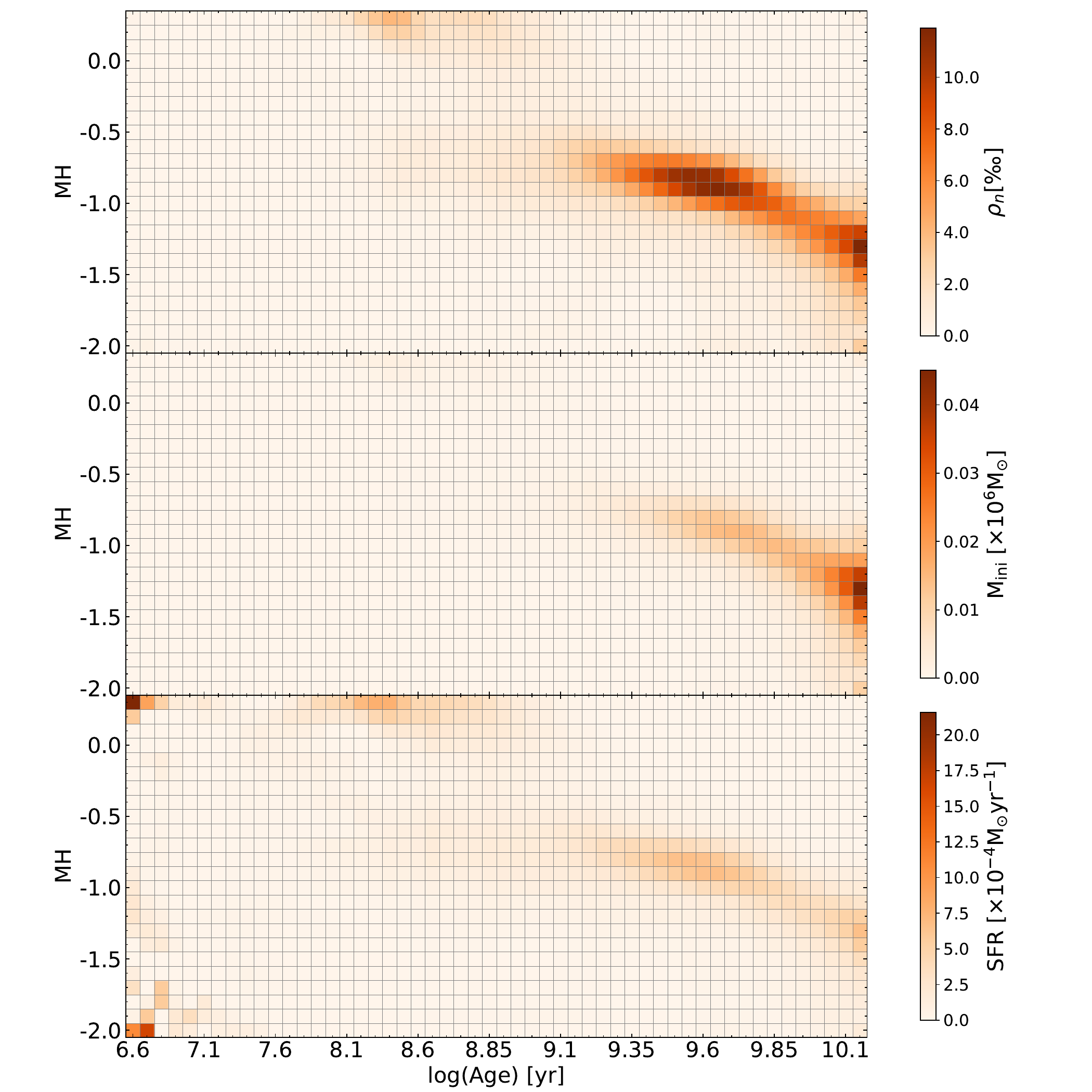}
\includegraphics[width=9cm]{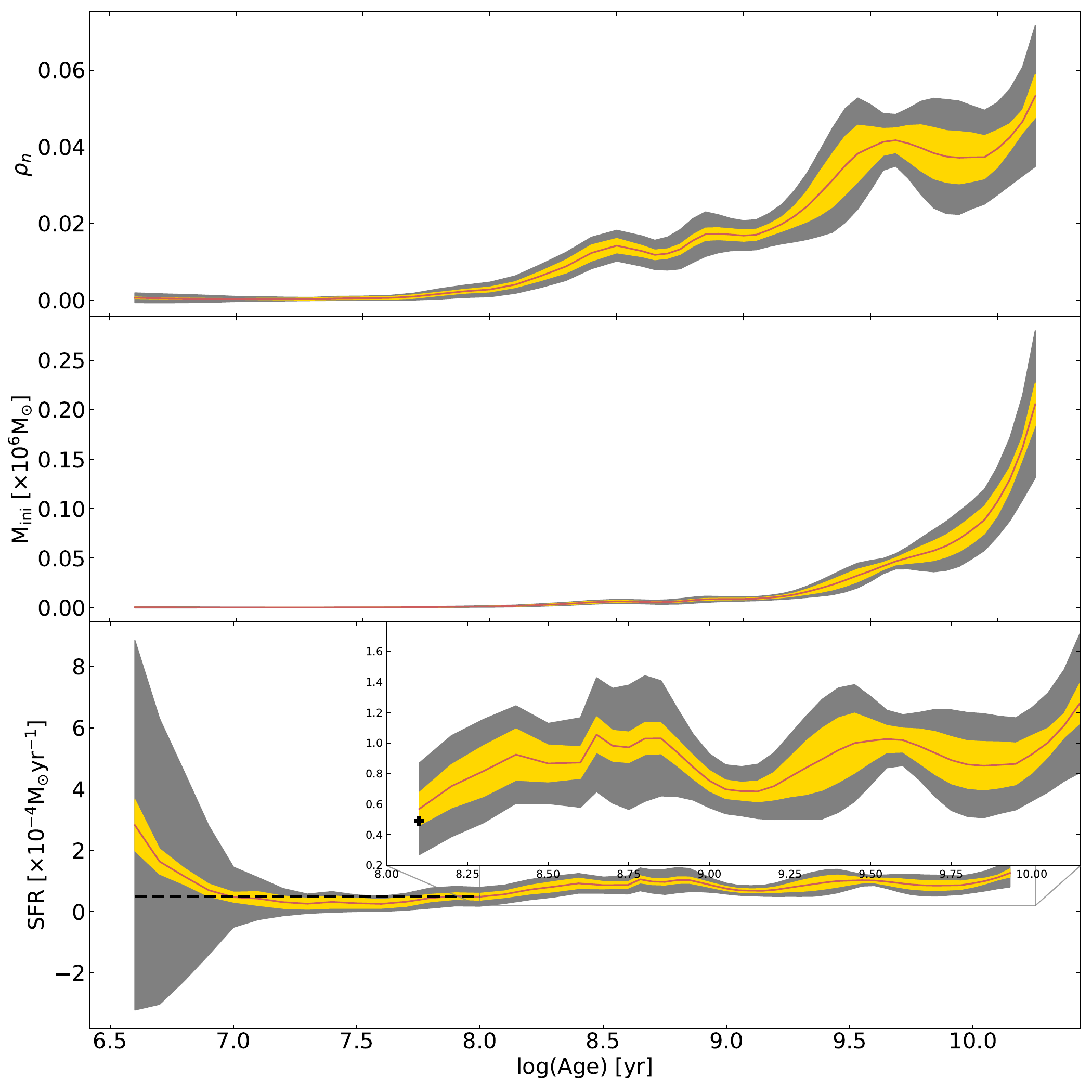}
\caption{Same as Figure \ref{fig:ngc6822rms}, but for Phoenix.}
\label{fig:phoenixrms}
\end{figure}

\begin{figure}[!htb]
\includegraphics[width=9cm]{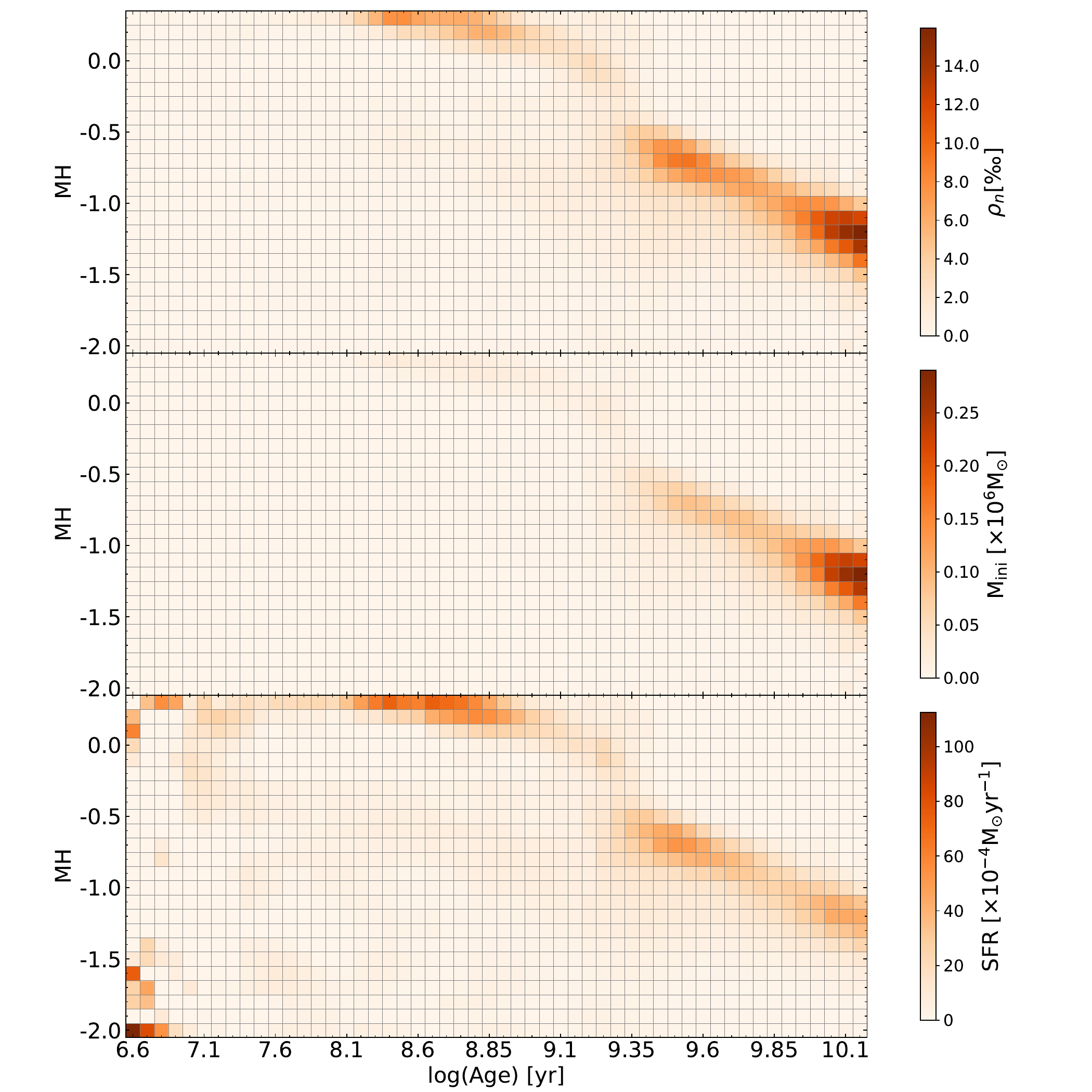}
\includegraphics[width=9cm]{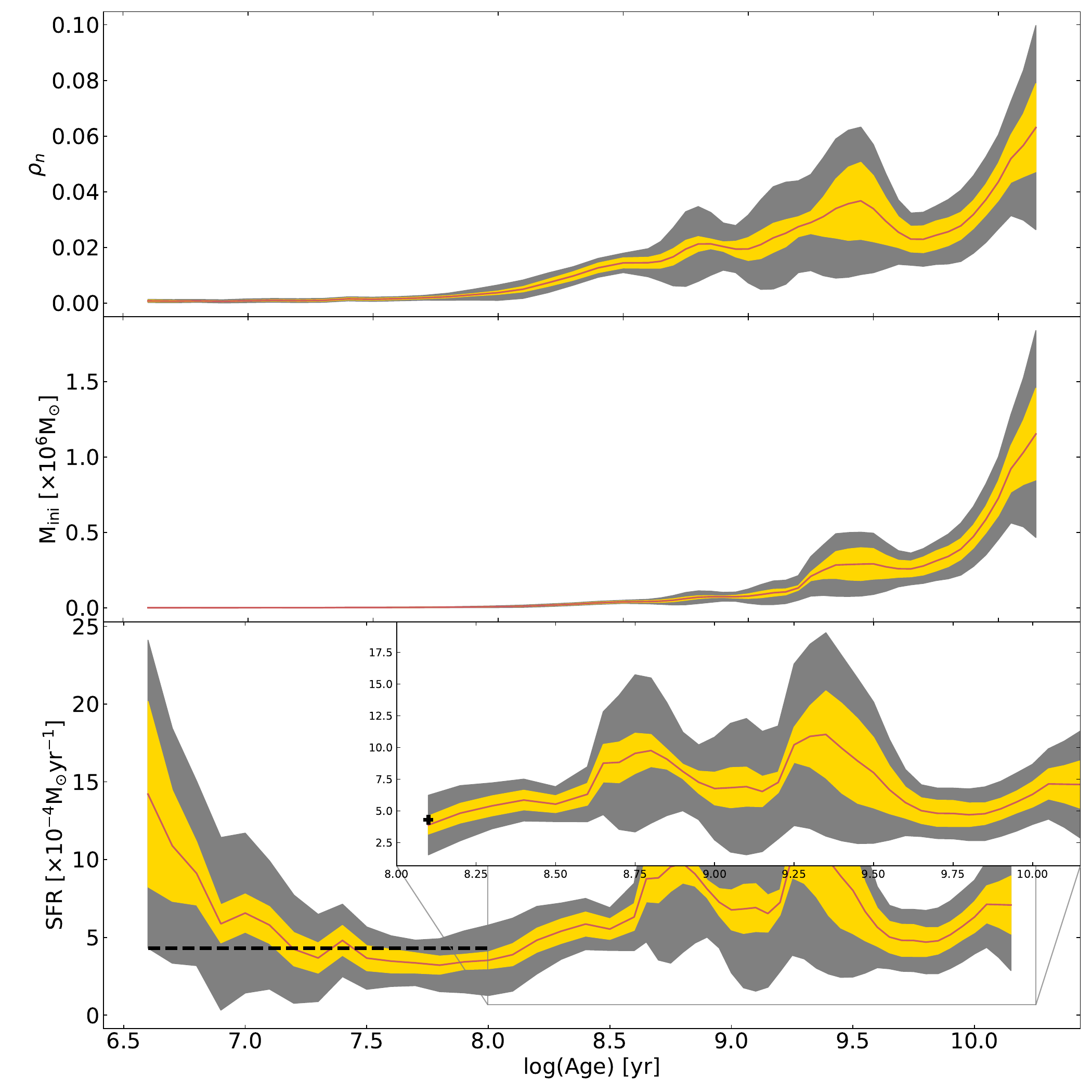}
\caption{Same as Figure \ref{fig:ngc6822rms}, but for Pegasus.}
\label{fig:pegasusrms}
\end{figure}

\end{appendix}
\newpage
\bibliography{sample631}{}

\begin{thebibliography}{}
\expandafter\ifx\csname natexlab\endcsname\relax\def\natexlab#1{#1}\fi
\providecommand{\url}[1]{\href{#1}{#1}}
\providecommand{\dodoi}[1]{doi:~\href{http://doi.org/#1}{\nolinkurl{#1}}}
\providecommand{\doeprint}[1]{\href{http://ascl.net/#1}{\nolinkurl{http://ascl.net/#1}}}
\providecommand{\doarXiv}[1]{\href{https://arxiv.org/abs/#1}{\nolinkurl{https://arxiv.org/abs/#1}}}

\bibitem[{{Akeson} {et~al.}(2019){Akeson}, {Armus}, {Bachelet}, {Bailey},
  {Bartusek}, {Bellini}, {Benford}, {Bennett}, {Bhattacharya}, {Bohlin},
  {Boyer}, {Bozza}, {Bryden}, {Calchi Novati}, {Carpenter}, {Casertano},
  {Choi}, {Content}, {Dayal}, {Dressler}, {Dor{\'e}}, {Fall}, {Fan}, {Fang},
  {Filippenko}, {Finkelstein}, {Foley}, {Furlanetto}, {Kalirai}, {Gaudi},
  {Gilbert}, {Girard}, {Grady}, {Greene}, {Guhathakurta}, {Heinrich},
  {Hemmati}, {Hendel}, {Henderson}, {Henning}, {Hirata}, {Ho}, {Huff},
  {Hutter}, {Jansen}, {Jha}, {Johnson}, {Jones}, {Kasdin}, {Kelly}, {Kirshner},
  {Koekemoer}, {Kruk}, {Lewis}, {Macintosh}, {Madau}, {Malhotra}, {Mandel},
  {Massara}, {Masters}, {McEnery}, {McQuinn}, {Melchior}, {Melton},
  {Mennesson}, {Peeples}, {Penny}, {Perlmutter}, {Pisani}, {Plazas}, {Poleski},
  {Postman}, {Ranc}, {Rauscher}, {Rest}, {Roberge}, {Robertson}, {Rodney},
  {Rhoads}, {Rhodes}, {Ryan}, {Sahu}, {Sand}, {Scolnic}, {Seth}, {Shvartzvald},
  {Siellez}, {Smith}, {Spergel}, {Stassun}, {Street}, {Strolger}, {Szalay},
  {Trauger}, {Troxel}, {Turnbull}, {van der Marel}, {von der Linden}, {Wang},
  {Weinberg}, {Williams}, {Windhorst}, {Wollack}, {Wu}, {Yee}, \&
  {Zimmerman}}]{2019arXiv190205569A}
{Akeson}, R., {Armus}, L., {Bachelet}, E., {et~al.} 2019, arXiv e-prints,
  arXiv:1902.05569, \dodoi{10.48550/arXiv.1902.05569}

\bibitem[{{Annibali} \& {Tosi}(2022)}]{2022NatAs...6...48A}
{Annibali}, F., \& {Tosi}, M. 2022, Nature Astronomy, 6, 48,
  \dodoi{10.1038/s41550-021-01575-x}

\bibitem[{{Aparicio} \& {Gallart}(2004)}]{2004AJ....128.1465A}
{Aparicio}, A., \& {Gallart}, C. 2004, \aj, 128, 1465, \dodoi{10.1086/382836}

\bibitem[{{Aparicio} {et~al.}(1997){Aparicio}, {Gallart}, \&
  {Bertelli}}]{1997AJ....114..669A}
{Aparicio}, A., {Gallart}, C., \& {Bertelli}, G. 1997, \aj, 114, 669,
  \dodoi{10.1086/118501}

\bibitem[{{Aparicio} \& {Hidalgo}(2009)}]{2009AJ....138..558A}
{Aparicio}, A., \& {Hidalgo}, S.~L. 2009, \aj, 138, 558,
  \dodoi{10.1088/0004-6256/138/2/558}

\bibitem[{{Astropy Collaboration} {et~al.}(2022){Astropy Collaboration},
  {Price-Whelan}, {Lim}, {Earl}, {Starkman}, {Bradley}, {Shupe}, {Patil},
  {Corrales}, {Brasseur}, {N{\"o}the}, {Donath}, {Tollerud}, {Morris},
  {Ginsburg}, {Vaher}, {Weaver}, {Tocknell}, {Jamieson}, {van Kerkwijk},
  {Robitaille}, {Merry}, {Bachetti}, {G{\"u}nther}, {Aldcroft},
  {Alvarado-Montes}, {Archibald}, {B{\'o}di}, {Bapat}, {Barentsen},
  {Baz{\'a}n}, {Biswas}, {Boquien}, {Burke}, {Cara}, {Cara}, {Conroy},
  {Conseil}, {Craig}, {Cross}, {Cruz}, {D'Eugenio}, {Dencheva}, {Devillepoix},
  {Dietrich}, {Eigenbrot}, {Erben}, {Ferreira}, {Foreman-Mackey}, {Fox},
  {Freij}, {Garg}, {Geda}, {Glattly}, {Gondhalekar}, {Gordon}, {Grant},
  {Greenfield}, {Groener}, {Guest}, {Gurovich}, {Handberg}, {Hart},
  {Hatfield-Dodds}, {Homeier}, {Hosseinzadeh}, {Jenness}, {Jones}, {Joseph},
  {Kalmbach}, {Karamehmetoglu}, {Ka{\l}uszy{\'n}ski}, {Kelley}, {Kern},
  {Kerzendorf}, {Koch}, {Kulumani}, {Lee}, {Ly}, {Ma}, {MacBride}, {Maljaars},
  {Muna}, {Murphy}, {Norman}, {O'Steen}, {Oman}, {Pacifici}, {Pascual},
  {Pascual-Granado}, {Patil}, {Perren}, {Pickering}, {Rastogi}, {Roulston},
  {Ryan}, {Rykoff}, {Sabater}, {Sakurikar}, {Salgado}, {Sanghi}, {Saunders},
  {Savchenko}, {Schwardt}, {Seifert-Eckert}, {Shih}, {Jain}, {Shukla}, {Sick},
  {Simpson}, {Singanamalla}, {Singer}, {Singhal}, {Sinha}, {Sip{\H{o}}cz},
  {Spitler}, {Stansby}, {Streicher}, {{\v{S}}umak}, {Swinbank}, {Taranu},
  {Tewary}, {Tremblay}, {de Val-Borro}, {Van Kooten}, {Vasovi{\'c}}, {Verma},
  {de Miranda Cardoso}, {Williams}, {Wilson}, {Winkel}, {Wood-Vasey}, {Xue},
  {Yoachim}, {Zhang}, {Zonca}, \& {Astropy Project
  Contributors}}]{2022ApJ...935..167A}
{Astropy Collaboration}, {Price-Whelan}, A.~M., {Lim}, P.~L., {et~al.} 2022,
  \apj, 935, 167, \dodoi{10.3847/1538-4357/ac7c74}

\bibitem[{{Baldacci} {et~al.}(2003){Baldacci}, {Clementini}, {Held}, \&
  {Rizzi}}]{2003MmSAI..74..860B}
{Baldacci}, L., {Clementini}, G., {Held}, E.~V., \& {Rizzi}, L. 2003, \memsai,
  74, 860, \dodoi{10.48550/arXiv.astro-ph/0303094}

\bibitem[{{Bertelli} {et~al.}(1992){Bertelli}, {Mateo}, {Chiosi}, \&
  {Bressan}}]{1992ApJ...388..400B}
{Bertelli}, G., {Mateo}, M., {Chiosi}, C., \& {Bressan}, A. 1992, \apj, 388,
  400, \dodoi{10.1086/171163}

\bibitem[{{Bortolini} {et~al.}(2024){Bortolini}, {Cignoni}, {Sacchi}, {Tosi},
  {Annibali}, {Pascale}, {Bellazzini}, {Calzetti}, {Adamo}, {Dale},
  {Fumagalli}, {Gallagher}, {Grasha}, {Johnson}, {Linden}, {Messa},
  {{\"O}stlin}, {Sabbi}, \& {Wofford}}]{2024MNRAS.527.5339B}
{Bortolini}, G., {Cignoni}, M., {Sacchi}, E., {et~al.} 2024, \mnras, 527, 5339,
  \dodoi{10.1093/mnras/stad3524}

\bibitem[{{Bressan} {et~al.}(2012){Bressan}, {Marigo}, {Girardi}, {Salasnich},
  {Dal Cero}, {Rubele}, \& {Nanni}}]{2012MNRAS.427..127B}
{Bressan}, A., {Marigo}, P., {Girardi}, L., {et~al.} 2012, \mnras, 427, 127,
  \dodoi{10.1111/j.1365-2966.2012.21948.x}

\bibitem[{{Butler} \& {Mart{\'\i}nez-Delgado}(2005)}]{2005AJ....129.2217B}
{Butler}, D.~J., \& {Mart{\'\i}nez-Delgado}, D. 2005, \aj, 129, 2217,
  \dodoi{10.1086/429524}

\bibitem[{{Cannon} {et~al.}(2012){Cannon}, {O'Leary}, {Weisz}, {Skillman},
  {Dolphin}, {Bigiel}, {Cole}, {de Blok}, \& {Walter}}]{2012ApJ...747..122C}
{Cannon}, J.~M., {O'Leary}, E.~M., {Weisz}, D.~R., {et~al.} 2012, \apj, 747,
  122, \dodoi{10.1088/0004-637X/747/2/122}

\bibitem[{{Cappellari} {et~al.}(1999){Cappellari}, {Bertola}, {Burstein},
  {Buson}, {Greggio}, \& {Renzini}}]{1999ApJ...515L..17C}
{Cappellari}, M., {Bertola}, F., {Burstein}, D., {et~al.} 1999, \apjl, 515,
  L17, \dodoi{10.1086/311966}

\bibitem[{{Cardelli} {et~al.}(1989){Cardelli}, {Clayton}, \&
  {Mathis}}]{1989ApJ...345..245C}
{Cardelli}, J.~A., {Clayton}, G.~C., \& {Mathis}, J.~S. 1989, \apj, 345, 245,
  \dodoi{10.1086/167900}

\bibitem[{{Carnall} {et~al.}(2019){Carnall}, {Leja}, {Johnson}, {McLure},
  {Dunlop}, \& {Conroy}}]{2019ApJ...873...44C}
{Carnall}, A.~C., {Leja}, J., {Johnson}, B.~D., {et~al.} 2019, \apj, 873, 44,
  \dodoi{10.3847/1538-4357/ab04a2}

\bibitem[{{Carnall} {et~al.}(2018){Carnall}, {McLure}, {Dunlop}, \&
  {Dav{\'e}}}]{2018MNRAS.480.4379C}
{Carnall}, A.~C., {McLure}, R.~J., {Dunlop}, J.~S., \& {Dav{\'e}}, R. 2018,
  \mnras, 480, 4379, \dodoi{10.1093/mnras/sty2169}

\bibitem[{{Carrera} {et~al.}(2002){Carrera}, {Aparicio},
  {Mart{\'\i}nez-Delgado}, \& {Alonso-Garc{\'\i}a}}]{2002AJ....123.3199C}
{Carrera}, R., {Aparicio}, A., {Mart{\'\i}nez-Delgado}, D., \&
  {Alonso-Garc{\'\i}a}, J. 2002, \aj, 123, 3199, \dodoi{10.1086/340702}

\bibitem[{{Chen} {et~al.}(2024){Chen}, {Liu}, \& {Han}}]{2024PrPNP.13404083C}
{Chen}, X., {Liu}, Z., \& {Han}, Z. 2024, Progress in Particle and Nuclear
  Physics, 134, 104083, \dodoi{10.1016/j.ppnp.2023.104083}

\bibitem[{{Cignoni} {et~al.}(2015){Cignoni}, {Sabbi}, {van der Marel}, {Tosi},
  {Zaritsky}, {Anderson}, {Lennon}, {Aloisi}, {de Marchi}, {Gouliermis},
  {Grebel}, {Smith}, \& {Zeidler}}]{2015ApJ...811...76C}
{Cignoni}, M., {Sabbi}, E., {van der Marel}, R.~P., {et~al.} 2015, \apj, 811,
  76, \dodoi{10.1088/0004-637X/811/2/76}

\bibitem[{{Clementini} {et~al.}(2003){Clementini}, {Held}, {Baldacci}, \&
  {Rizzi}}]{2003ApJ...588L..85C}
{Clementini}, G., {Held}, E.~V., {Baldacci}, L., \& {Rizzi}, L. 2003, \apjl,
  588, L85, \dodoi{10.1086/375633}

\bibitem[{{Cohen} \& {Huang}(2010)}]{2010ApJ...719..931C}
{Cohen}, J.~G., \& {Huang}, W. 2010, \apj, 719, 931,
  \dodoi{10.1088/0004-637X/719/1/931}

\bibitem[{{Cole} {et~al.}(2007){Cole}, {Skillman}, {Tolstoy}, {Gallagher},
  {Aparicio}, {Dolphin}, {Gallart}, {Hidalgo}, {Saha}, {Stetson}, \&
  {Weisz}}]{2007ApJ...659L..17C}
{Cole}, A.~A., {Skillman}, E.~D., {Tolstoy}, E., {et~al.} 2007, \apjl, 659,
  L17, \dodoi{10.1086/516711}

\bibitem[{{Crnojevi{\'c}} {et~al.}(2011){Crnojevi{\'c}}, {Grebel}, \&
  {Cole}}]{2011A&A...530A..59C}
{Crnojevi{\'c}}, D., {Grebel}, E.~K., \& {Cole}, A.~A. 2011, \aap, 530, A59,
  \dodoi{10.1051/0004-6361/201015474}

\bibitem[{{Crnojevi{\'c}} {et~al.}(2014){Crnojevi{\'c}}, {Ferguson}, {Irwin},
  {McConnachie}, {Bernard}, {Fardal}, {Ibata}, {Lewis}, {Martin}, {Navarro},
  {No{\"e}l}, \& {Pasetto}}]{2014MNRAS.445.3862C}
{Crnojevi{\'c}}, D., {Ferguson}, A.~M.~N., {Irwin}, M.~J., {et~al.} 2014,
  \mnras, 445, 3862, \dodoi{10.1093/mnras/stu2003}

\bibitem[{Cybenko(1989)}]{Cybenko1989ApproximationBS}
Cybenko, G.~V. 1989, Mathematics of Control, Signals and Systems, 2, 303.
\newblock \url{https://api.semanticscholar.org/CorpusID:3958369}

\bibitem[{{Dalcanton} {et~al.}(2012){Dalcanton}, {Williams}, {Lang}, {Lauer},
  {Kalirai}, {Seth}, {Dolphin}, {Rosenfield}, {Weisz}, {Bell}, {Bianchi},
  {Boyer}, {Caldwell}, {Dong}, {Dorman}, {Gilbert}, {Girardi}, {Gogarten},
  {Gordon}, {Guhathakurta}, {Hodge}, {Holtzman}, {Johnson}, {Larsen}, {Lewis},
  {Melbourne}, {Olsen}, {Rix}, {Rosema}, {Saha}, {Sarajedini}, {Skillman}, \&
  {Stanek}}]{2012ApJS..200...18D}
{Dalcanton}, J.~J., {Williams}, B.~F., {Lang}, D., {et~al.} 2012, \apjs, 200,
  18, \dodoi{10.1088/0067-0049/200/2/18}

\bibitem[{{Davidge}(2003)}]{2003PASP..115..635D}
{Davidge}, T.~J. 2003, \pasp, 115, 635, \dodoi{10.1086/375389}

\bibitem[{{Davidge}(2005)}]{2005AJ....130.2087D}
---. 2005, \aj, 130, 2087, \dodoi{10.1086/491706}

\bibitem[{{de Boer} {et~al.}(2012{\natexlab{a}}){de Boer}, {Tolstoy}, {Hill},
  {Saha}, {Olsen}, {Starkenburg}, {Lemasle}, {Irwin}, \&
  {Battaglia}}]{2012A&A...539A.103D}
{de Boer}, T.~J.~L., {Tolstoy}, E., {Hill}, V., {et~al.} 2012{\natexlab{a}},
  \aap, 539, A103, \dodoi{10.1051/0004-6361/201118378}

\bibitem[{{de Boer} {et~al.}(2012{\natexlab{b}}){de Boer}, {Tolstoy}, {Hill},
  {Saha}, {Olszewski}, {Mateo}, {Starkenburg}, {Battaglia}, \&
  {Walker}}]{2012A&A...544A..73D}
---. 2012{\natexlab{b}}, \aap, 544, A73, \dodoi{10.1051/0004-6361/201219547}

\bibitem[{{Dohm-Palmer} {et~al.}(2002){Dohm-Palmer}, {Skillman}, {Mateo},
  {Saha}, {Dolphin}, {Tolstoy}, {Gallagher}, \& {Cole}}]{2002AJ....123..813D}
{Dohm-Palmer}, R.~C., {Skillman}, E.~D., {Mateo}, M., {et~al.} 2002, \aj, 123,
  813, \dodoi{10.1086/324635}

\bibitem[{{Dohm-Palmer} {et~al.}(1997){Dohm-Palmer}, {Skillman}, {Saha},
  {Tolstoy}, {Mateo}, {Gallagher}, {Hoessel}, {Chiosi}, \&
  {Dufour}}]{1997AJ....114.2527D}
{Dohm-Palmer}, R.~C., {Skillman}, E.~D., {Saha}, A., {et~al.} 1997, \aj, 114,
  2527, \dodoi{10.1086/118665}

\bibitem[{{Dolphin}(1997)}]{1997NewA....2..397D}
{Dolphin}, A. 1997, \na, 2, 397, \dodoi{10.1016/S1384-1076(97)00029-8}

\bibitem[{{Dolphin}(2000)}]{2000PASP..112.1383D}
{Dolphin}, A.~E. 2000, \pasp, 112, 1383, \dodoi{10.1086/316630}

\bibitem[{{Dolphin}(2002)}]{2002MNRAS.332...91D}
---. 2002, \mnras, 332, 91, \dodoi{10.1046/j.1365-8711.2002.05271.x}

\bibitem[{{Dolphin}(2012)}]{2012ApJ...751...60D}
---. 2012, \apj, 751, 60, \dodoi{10.1088/0004-637X/751/1/60}

\bibitem[{{Dolphin}(2013)}]{2013ApJ...775...76D}
---. 2013, \apj, 775, 76, \dodoi{10.1088/0004-637X/775/1/76}

\bibitem[{{Dolphin} {et~al.}(2005){Dolphin}, {Weisz}, {Skillman}, \&
  {Holtzman}}]{2005astro.ph..6430D}
{Dolphin}, A.~E., {Weisz}, D.~R., {Skillman}, E.~D., \& {Holtzman}, J.~A. 2005,
  arXiv e-prints, astro, \dodoi{10.48550/arXiv.astro-ph/0506430}

\bibitem[{{Dolphin} {et~al.}(2003){Dolphin}, {Saha}, {Skillman}, {Dohm-Palmer},
  {Tolstoy}, {Cole}, {Gallagher}, {Hoessel}, \& {Mateo}}]{2003AJ....126..187D}
{Dolphin}, A.~E., {Saha}, A., {Skillman}, E.~D., {et~al.} 2003, \aj, 126, 187,
  \dodoi{10.1086/375761}

\bibitem[{{Eldridge} {et~al.}(2008){Eldridge}, {Izzard}, \&
  {Tout}}]{2008MNRAS.384.1109E}
{Eldridge}, J.~J., {Izzard}, R.~G., \& {Tout}, C.~A. 2008, \mnras, 384, 1109,
  \dodoi{10.1111/j.1365-2966.2007.12738.x}

\bibitem[{Escoufier(1973)}]{Escoufier1973LETD}
Escoufier, Y. 1973, Biometrics, 29, 751.
\newblock \url{https://api.semanticscholar.org/CorpusID:124051421}

\bibitem[{{Euclid Collaboration} {et~al.}(2024){Euclid Collaboration},
  {Mellier}, {Abdurro'uf}, {Acevedo Barroso}, {Ach{\'u}carro}, {Adamek},
  {Adam}, {Addison}, {Aghanim}, {Aguena}, {Ajani}, {Akrami}, {Al-Bahlawan},
  {Alavi}, {Albuquerque}, {Alestas}, {Alguero}, {Allaoui}, {Allen}, {Allevato},
  {Alonso-Tetilla}, {Altieri}, {Alvarez-Candal}, {Amara}, {Amendola}, {Amiaux},
  {Andika}, {Andreon}, {Andrews}, {Angora}, {Angulo}, {Annibali}, {Anselmi},
  {Anselmi}, {Arcari}, {Archidiacono}, {Aric{\`o}}, {Arnaud}, {Arnouts},
  {Asgari}, {Asorey}, {Atayde}, {Atek}, {Atrio-Barandela}, {Aubert}, {Aubourg},
  {Auphan}, {Auricchio}, {Aussel}, {Aussel}, {Avelino}, {Avgoustidis}, {Avila},
  {Awan}, {Azzollini}, {Baccigalupi}, {Bachelet}, {Bacon}, {Baes}, {Bagley},
  {Bahr-Kalus}, {Balaguera-Antolinez}, {Balbinot}, {Balcells}, {Baldi},
  {Baldry}, {Balestra}, {Ballardini}, {Ballester}, {Balogh}, {Ba{\~n}ados},
  {Barbier}, {Bardelli}, {Barreiro}, {Barriere}, {Barros}, {Barthelemy},
  {Bartolo}, {Basset}, {Battaglia}, {Battisti}, {Baugh}, {Baumont},
  {Bazzanini}, {Beaulieu}, {Beckmann}, {Belikov}, {Bel}, {Bellagamba}, {Bella},
  {Bellini}, {Benabed}, {Bender}, {Benevento}, {Bennett}, {Benson},
  {Bergamini}, {Bermejo-Climent}, {Bernardeau}, {Bertacca}, {Berthe},
  {Berthier}, {Bethermin}, {Beutler}, {Bevillon}, {Bhargava}, {Bhatawdekar},
  {Bisigello}, {Biviano}, {Blake}, {Blanchard}, {Blazek}, {Blot}, {Bosco},
  {Bodendorf}, {Boenke}, {B{\"o}hringer}, {Bolzonella}, {Bonchi}, {Bonici},
  {Bonino}, {Bonino}, {Bonvin}, {Bon}, {Booth}, {Borgani}, {Borlaff},
  {Borsato}, {Bosco}, {Bose}, {Botticella}, {Boucaud}, {Bouche}, {Boucher},
  {Boutigny}, {Bouvard}, {Bouy}, {Bowler}, {Bozza}, {Bozzo}, {Branchini},
  {Brau-Nogue}, {Brekke}, {Bremer}, {Brescia}, {Breton}, {Brinchmann},
  {Brinckmann}, {Brockley-Blatt}, {Brodwin}, {Brouard}, {Brown}, {Bruton},
  {Bucko}, {Buddelmeijer}, {Buenadicha}, {Buitrago}, {Burger}, {Burigana},
  {Busillo}, {Busonero}, {Cabanac}, {Cabayol-Garcia}, {Cagliari}, {Caillat},
  {Caillat}, {Calabrese}, {Calabro}, {Calderone}, {Calura}, {Camacho Quevedo},
  {Camera}, {Campos}, {Canas-Herrera}, {Candini}, {Cantiello}, {Capobianco},
  {Cappellaro}, {Cappelluti}, {Cappi}, {Caputi}, {Cara}, {Carbone}, {Cardone},
  {Carella}, {Carlberg}, {Carle}, {Carminati}, {Caro}, {Carrasco}, {Carretero},
  {Carrilho}, {Carron Duque}, {Carry}, {Carvalho}, {Carvalho}, {Casas},
  {Casas}, {Casenove}, {Casey}, {Cassata}, {Castander}, {Castelao},
  {Castellano}, {Castiblanco}, {Castignani}, {Castro}, {Cavet}, {Cavuoti},
  {Chabaud}, {Chambers}, {Charles}, {Charlot}, {Chartab}, {Chary}, {Chaumeil},
  {Cho}, {Chon}, {Ciancetta}, {Ciliegi}, {Cimatti}, {Cimino}, {Cioni},
  {Claydon}, {Cleland}, {Cl{\'e}ment}, {Clements}, {Clerc}, {Clesse}, {Codis},
  {Cogato}, {Colbert}, {Cole}, {Coles}, {Collett}, {Collins}, {Colodro-Conde},
  {Colombo}, {Combes}, {Conforti}, {Congedo}, {Conseil}, {Conselice},
  {Contarini}, {Contini}, {Conversi}, {Cooray}, {Copin}, {Corasaniti},
  {Corcho-Caballero}, {Corcione}, {Cordes}, {Corpace}, {Correnti}, {Costanzi},
  {Costille}, {Courbin}, {Courcoult Mifsud}, {Courtois}, {Cousinou}, {Covone},
  {Cowell}, {Cragg}, {Cresci}, {Cristiani}, {Crocce}, {Cropper}, {E Crouzet},
  {Csizi}, {Cuby}, {Cucchetti}, {Cucciati}, {Cuillandre}, {Cunha}, {Cuozzo},
  {Daddi}, {D'Addona}, {Dafonte}, {Dagoneau}, {Dalessandro}, {Dalton},
  {D'Amico}, {Dannerbauer}, {Danto}, {Das}, {Da Silva}, {da Silva}, {Daste},
  {Davies}, {Davini}, {de Boer}, {Decarli}, {De Caro}, {Degaudenzi}, {Degni},
  {de Jong}, {de la Bella}, {de la Torre}, {Delhaise}, {Delley}, {Delucchi},
  {De Lucia}, {Denniston}, {De Paolis}, {De Petris}, {Derosa}, {Desai},
  {Desjacques}, {Despali}, {Desprez}, {De Vicente-Albendea}, {Deville}, {Dias},
  {D{\'\i}az-S{\'a}nchez}, {Diaz}, {Di Domizio}, {Diego}, {Di Ferdinando}, {Di
  Giorgio}, {Dimauro}, {Dinis}, {Dolag}, {Dolding}, {Dole}, {Dom{\'\i}nguez
  S{\'a}nchez}, {Dor{\'e}}, {Dournac}, {Douspis}, {Dreihahn}, {Droge}, {Dryer},
  {Dubath}, {Duc}, {Ducret}, {Duffy}, {Dufresne}, {Duncan}, {Dupac}, {Duret},
  {Durrer}, {Durret}, {Dusini}, {Ealet}, {Eggemeier}, {Eisenhardt}, {Elbaz},
  {Elkhashab}, {Ellien}, {Endicott}, {Enia}, {Erben}, {Escartin Vigo},
  {Escoffier}, {Escudero Sanz}, {Essert}, {Ettori}, {Ezziati}, {Fabbian},
  {Fabricius}, {Fang}, {Farina}, {Farina}, {Farinelli}, {Farrens}, {Faustini},
  {Feltre}, {Ferguson}, {Ferrando}, {Ferrari}, {Ferr{\'e}-Mateu}, {Ferreira},
  {Ferreras}, {Ferrero}, {Ferriol}, {Ferruit}, {Filleul}, {Finelli},
  {Finkelstein}, {Finoguenov}, {Fiorini}, {Flentge}, {Focardi}, {Fonseca},
  {Fontana}, {Fontanot}, {Fornari}, {Fosalba}, {Fossati}, {Fotopoulou},
  {Fouchez}, {Fourmanoit}, {Frailis}, {Fraix-Burnet}, {Franceschi}, {Franco},
  {Franzetti}, {Freihoefer}, {Frittoli}, {Frugier}, {Frusciante}, {Fumagalli},
  {Fumagalli}, {Fumana}, {Fu}, {Gabarra}, {Galeotta}, {Galluccio}, {Ganga},
  {Gao}, {Garc{\'\i}a-Bellido}, {Garcia}, {Gardner}, {Garilli},
  {Gaspar-Venancio}, {Gasparetto}, {Gautard}, {Gavazzi}, {Gaztanaga},
  {Genolet}, {Genova Santos}, {Gentile}, {George}, {Ghaffari}, {Giacomini},
  {Gianotti}, {Gibb}, {Gillard}, {Gillis}, {Ginolfi}, {Giocoli}, {Girardi},
  {Giri}, {Goh}, {G{\'o}mez-Alvarez}, {Gonzalez}, {Gonzalez}, {Gonzalez},
  {Gouyou Beauchamps}, {Gozaliasl}, {Gracia-Carpio}, {Grandis}, {Granett},
  {Granvik}, {Grazian}, {Gregorio}, {Grenet}, {Grillo}, {Grupp}, {Gruppioni},
  {Gruppuso}, {Guerbuez}, {Guerrini}, {Guidi}, {Guillard}, {Gutierrez},
  {Guttridge}, {Guzzo}, {Gwyn}, {Haapala}, {Haase}, {Haddow}, {Hailey}, {Hall},
  {Hall}, {Hamaus}, {Haridasu}, {Harnois-D{\'e}raps}, {Harper}, {Hartley},
  {Hasinger}, {Hassani}, {Hatch}, {Haugan}, {H{\"a}u{\ss}ler}, {Heavens},
  {Heisenberg}, {Helmi}, {Helou}, {Hemmati}, {Henares}, {Herent},
  {Hern{\'a}ndez-Monteagudo}, {Heuberger}, {Hewett}, {Heydenreich},
  {Hildebrandt}, {Hirschmann}, {Hjorth}, {Hoar}, {Hoekstra}, {Holland},
  {Holliman}, {Holmes}, {Hook}, {Horeau}, {Hormuth}, {Hornstrup}, {Hosseini},
  {Hu}, {Hudelot}, {Hudson}, {Huertas-Company}, {Huff}, {Hughes}, {Humphrey},
  {Hunt}, {Huynh}, {Ibata}, {Ichikawa}, {Iglesias-Groth}, {Ilbert}, {Ili{\'c}},
  {Ingoglia}, {Iodice}, {Israel}, {Israelsson}, {Izzo}, {Jablonka}, {Jackson},
  {Jacobson}, {Jafariyazani}, {Jahnke}, {Jansen}, {Jarvis}, {Jasche}, {Jauzac},
  {Jeffrey}, {Jhabvala}, {Jimenez-Teja}, {Jimenez Mu{\~n}oz}, {Joachimi},
  {Johansson}, {Joudaki}, {Jullo}, {Kajava}, {Kang}, {Kannawadi}, {Kansal},
  {Karagiannis}, {K{\"a}rcher}, {Kashlinsky}, {Kazandjian}, {Keck},
  {Keih{\"a}nen}, {Kerins}, {Kermiche}, {Khalil}, {Kiessling}, {Kiiveri},
  {Kilbinger}, {Kim}, {King}, {Kirkpatrick}, {Kitching}, {Kluge}, {Knabenhans},
  {Knapen}, {Knebe}, {Kneib}, {Kohley}, {Koopmans}, {Koskinen}, {Koulouridis},
  {Kou}, {Kov{\'a}cs}, {Kova\{{\v{c}}\}i{\'c}}, {Kowalczyk}, {Koyama},
  {Kraljic}, {Krause}, {Kruk}, {Kubik}, {Kuchner}, {Kuijken}, {K{\"u}mmel},
  {Kunz}, {Kurki-Suonio}, {Lacasa}, {Lacey}, {La Franca}, {Lagarde}, {Lahav},
  {Laigle}, {La Marca}, {La Marle}, {Lamine}, {Lam}, {Lan{\c{c}}on}, {Landt},
  {Langer}, {Lapi}, {Larcheveque}, {Larsen}, {Lattanzi}, {Laudisio}, {Laugier},
  {Laureijs}, {Lavaux}, {Lawrenson}, {Lazanu}, {Lazeyras}, {Le Boulc'h}, {Le
  Brun}, {Le Brun}, {Leclercq}, {Lee}, {Le Graet}, {Legrand}, {Leirvik}, {Le
  Jeune}, {Lembo}, {Le Mignant}, {Lepinzan}, {Lepori}, {Lesci}, {Lesgourgues},
  {Leuzzi}, {Levi}, {Liaudat}, {Libet}, {Liebing}, {Ligori}, {Lilje}, {Lin},
  {Linde}, {Linder}, {Lindholm}, {Linke}, {Li}, {Liu}, {Lloro}, {Lobo},
  {Lodieu}, {Lombardi}, {Lombriser}, {Lonare}, {Longo}, {L{\'o}pez-Caniego},
  {Lopez Lopez}, {Alvarez}, {Loureiro}, {Loveday}, {Lusso}, {Macias-Perez},
  {Maciaszek}, {Magliocchetti}, {Magnard}, {Magnier}, {Magro}, {Mahler},
  {Mainetti}, {Maino}, {Maiorano}, {Maiorano}, {Malavasi}, {Mamon}, {Mancini},
  {Mandelbaum}, {Manera}, {Manj{\'o}n-Garc{\'\i}a}, {Mannucci}, {Mansutti},
  {Manteiga Outeiro}, {Maoli}, {Maraston}, {Marcin}, {Marcos-Arenal},
  {Margalef-Bentabol}, {Marggraf}, {Marinucci}, {Marinucci}, {Markovic},
  {Marleau}, {Marpaud}, {Martignac}, {Mart{\'\i}n-Fleitas}, {Martin-Moruno},
  {Martin}, {Martinelli}, {Martinet}, {Martin}, {Martins}, {Marulli},
  {Massari}, {Massey}, {Masters}, {Matarrese}, {Matsuoka}, {Matthew},
  {Maughan}, {Mauri}, {Maurin}, {Maurogordato}, {McCarthy}, {McConnachie},
  {McCracken}, {McDonald}, {McEwen}, {McPartland}, {Medinaceli}, {Mehta},
  {Mei}, {Melchior}, {Melin}, {M{\'e}nard}, {Mendes}, {Mendez-Abreu},
  {Meneghetti}, {Mercurio}, {Merlin}, {Metcalf}, {Meylan}, {Migliaccio},
  {Mignoli}, {Miller}, {Miluzio}, {Milvang-Jensen}, {Mimoso}, {Miquel},
  {Miyatake}, {Mobasher}, {Mohr}, {Monaco}, {Mongui{\'o}}, {Montoro}, {Mora},
  {Moradinezhad Dizgah}, {Moresco}, {Moretti}, {Morgante}, {Morisset},
  {Moriya}, {Morris}, {Mortlock}, {Moscardini}, {Mota}, {Moustakas}, {Moutard},
  {M{\"u}ller}, {Munari}, {Murphree}, {Murray}, {Murray}, {Musi}, {Nadathur},
  {Nagam}, {Nagao}, {Naidoo}, {Nakajima}, {Nally}, {Natoli}, {Navarro-Alsina},
  {Navarro Girones}, {Neissner}, {Nersesian}, {Nesseris}, {Nguyen-Kim},
  {Nicastro}, {Nichol}, {Nielbock}, {Niemi}, {Nieto}, {Nilsson}, {Noller},
  {Norberg}, {Nourizonoz}, {Ntelis}, {Nucita}, {Nugent}, {Nunes}, {Nutma},
  {Ocampo}, {Odier}, {Oesch}, {Oguri}, {Magalhaes Oliveira}, {Onoue},
  {Oosterbroek}, {Oppizzi}, {Ordenovic}, {Osato}, {Pacaud}, {Pace}, {Padilla},
  {Paech}, {Pagano}, {Page}, {Palazzi}, {Paltani}, {Pamuk}, {Pandolfi},
  {Paoletti}, {Paolillo}, {Papaderos}, {Pardede}, {Parimbelli}, {Parmar},
  {Partmann}, {Pasian}, {Passalacqua}, {Paterson}, {Patrizii}, {Pattison},
  {Paulino-Afonso}, {Paviot}, {Peacock}, {Pearce}, {Pedersen}, {Peel},
  {Peletier}, {Pellejero Ibanez}, {Pello}, {Penny}, {Percival},
  {Perez-Garrido}, {Perotto}, {Pettorino}, {Pezzotta}, {Pezzuto}, {Philippon},
  {Piersanti}, {Pietroni}, {Piga}, {Pilo}, {Pires}, {Pisani}, {Pizzella},
  {Pizzuti}, {Plana}, {Polenta}, {Pollack}, {Poncet}, {P{\"o}ntinen}, {Pool},
  {Popa}, {Popa}, {Popp}, {Porciani}, {Porth}, {Potter}, {Poulain},
  {Pourtsidou}, {Pozzetti}, {Prandoni}, {Pratt}, {Prezelus}, {Prieto}, {Pugno},
  {Quai}, {Quilley}, {Racca}, {Raccanelli}, {R{\'a}cz}, {Radinovi{\'c}},
  {Radovich}, {Ragagnin}, {Ragnit}, {Raison}, {Ramos-Chernenko}, {Ranc},
  {Raylet}, {Rebolo}, {Refregier}, {Reimberg}, {Reiprich}, {Renk}, {Renzi},
  {Retre}, {Revaz}, {Reyl{\'e}}, {Reynolds}, {Rhodes}, {Ricci}, {Ricci},
  {Riccio}, {Ricken}, {Rissanen}, {Risso}, {Rix}, {Robin}, {Rocca-Volmerange},
  {Rocci}, {Rodenhuis}, {Rodighiero}, {Rodriguez Monroy}, {Rollins},
  {Romanello}, {Roman}, {Romelli}, {Romero-Gomez}, {Roncarelli}, {Rosati},
  {Rosset}, {Rossetti}, {Roster}, {Rottgering}, {Rozas-Fern{\'a}ndez}, {Ruane},
  {Rubino-Martin}, {Rudolph}, {Ruppin}, {Rusholme}, {Sacquegna},
  {S{\'a}ez-Casares}, {Saga}, {Saglia}, {Sahl{\'e}n}, {Saifollahi}, {Sakr},
  {Salvalaggio}, {Salvaterra}, {Salvati}, {Salvato}, {Salvignol},
  {S{\'a}nchez}, {Sanchez}, {Sanders}, {Sapone}, {Saponara}, {Sarpa}, {Sarron},
  {Sartori}, {Sassolas}, {Sauniere}, {Sauvage}, {Sawicki}, {Scaramella},
  {Scarlata}, {Scharr{\'e}}, {Schaye}, {Schewtschenko}, {Schindler},
  {Schinnerer}, {Schirmer}, {Schmidt}, {Schmidt}, {Schmidt}, {Schneider},
  {Schneider}, {Schneider}, {Sch{\"o}neberg}, {Schrabback}, {Schultheis},
  {Schulz}, {Schwartz}, {Sciotti}, {Scodeggio}, {Scognamiglio}, {Scott},
  {Scottez}, {Secroun}, {Sefusatti}, {Seidel}, {Seiffert}, {Sellentin},
  {Selwood}, {Semboloni}, {Sereno}, {Serjeant}, {Serrano}, {Shankar},
  {Sharples}, {Short}, {Shulevski}, {Shuntov}, {Sias}, {Sikkema}, {Silvestri},
  {Simon}, {Sirignano}, {Sirri}, {Skottfelt}, {Slezak}, {Sluse}, {Smith},
  {Smith}, {Smith}, {Smit}, {Soldano}, {Solheim}, {Sorce}, {Sorrenti},
  {Soubrie}, {Spinoglio}, {Spurio Mancini}, {Stadel}, {Stagnaro}, {Stanco},
  {Stanford}, {Starck}, {Stassi}, {Steinwagner}, {Stern}, {Stone}, {Strada},
  {Strafella}, {Stramaccioni}, {Surace}, {Sureau}, {Suyu}, {Swindells},
  {Szafraniec}, {Szapudi}, {Taamoli}, {Talia}, {Tallada-Cresp{\'\i}},
  {Tanidis}, {Tao}, {Tarr{\'\i}o}, {Tavagnacco}, {Taylor}, {Taylor}, {Taylor},
  {Teixeira}, {Tenti}, {Teodoro Idiago}, {Teplitz}, {Tereno}, {Tessore},
  {Testa}, {Testera}, {Tewes}, {Teyssier}, {Theret}, {Thizy}, {Thomas}, {Toba},
  {Toft}, {Toledo-Moreo}, {Tolstoy}, {Tommasi}, {Torbaniuk}, {Torradeflot},
  {Tortora}, {Tosi}, {Tosti}, {Trifoglio}, {Troja}, {Trombetti}, {Tronconi},
  {Tsedrik}, {Tsyganov}, {Tucci}, {Tutusaus}, {Uhlemann}, {Ulivi}, {Urbano},
  {Vacher}, {Vaillon}, {Valdes}, {Valentijn}, {Valenziano}, {Valieri},
  {Valiviita}, {Van den Broeck}, {Vassallo}, {Vavrek}, {Venemans}, {Venhola},
  {Ventura}, {Verdoes Kleijn}, {Vergani}, {Verma}, {Vernizzi}, {Veropalumbo},
  {Verza}, {Vescovi}, {Vibert}, {Viel}, {Vielzeuf}, {Viglione}, {Viitanen},
  {Villaescusa-Navarro}, {Vinciguerra}, {Visticot}, {Voggel}, {von
  Wietersheim-Kramsta}, {Vriend}, {Wachter}, {Walmsley}, {Walth}, {Walton},
  {Walton}, {Wander}, {Wang}, {Wang}, {Weaver}, {Weller}, {Whalen}, {Wiesmann},
  {Wilde}, {Williams}, {Winther}, {Wittje}, {Wong}, {Wright}, {Yankelevich},
  {Yeung}, {Youles}, {Yung}, {Zacchei}, {Zalesky}, {Zamorani}, {Zamorano
  Vitorelli}, {Zanoni Marc}, {Zennaro}, {Zerbi}, {Zinchenko}, {Zoubian},
  {Zucca}, \& {Zumalacarregui}}]{2024arXiv240513491E}
{Euclid Collaboration}, {Mellier}, Y., {Abdurro'uf}, {et~al.} 2024, arXiv
  e-prints, arXiv:2405.13491, \dodoi{10.48550/arXiv.2405.13491}

\bibitem[{Fukushima \& Miyake(1982)}]{FUKUSHIMA1982455}
Fukushima, K., \& Miyake, S. 1982, Pattern Recognition, 15, 455,
  \dodoi{https://doi.org/10.1016/0031-3203(82)90024-3}

\bibitem[{Funahashi(1989)}]{FUNAHASHI1989183}
Funahashi, K.-I. 1989, Neural Networks, 2, 183,
  \dodoi{https://doi.org/10.1016/0893-6080(89)90003-8}

\bibitem[{{Fusco} {et~al.}(2014){Fusco}, {Buonanno}, {Hidalgo}, {Aparicio},
  {Pietrinferni}, {Bono}, {Monelli}, \& {Cassisi}}]{2014AA...572A..26F}
{Fusco}, F., {Buonanno}, R., {Hidalgo}, S.~L., {et~al.} 2014, \aap, 572, A26,
  \dodoi{10.1051/0004-6361/201323075}

\bibitem[{{Gallagher} {et~al.}(1998){Gallagher}, {Tolstoy}, {Dohm-Palmer},
  {Skillman}, {Cole}, {Hoessel}, {Saha}, \& {Mateo}}]{1998AJ....115.1869G}
{Gallagher}, J.~S., {Tolstoy}, E., {Dohm-Palmer}, R.~C., {et~al.} 1998, \aj,
  115, 1869, \dodoi{10.1086/300328}

\bibitem[{{Gallart} {et~al.}(1996{\natexlab{a}}){Gallart}, {Aparicio},
  {Bertelli}, \& {Chiosi}}]{1996AJ....112.1950G}
{Gallart}, C., {Aparicio}, A., {Bertelli}, G., \& {Chiosi}, C.
  1996{\natexlab{a}}, \aj, 112, 1950, \dodoi{10.1086/118154}

\bibitem[{{Gallart} {et~al.}(1996{\natexlab{b}}){Gallart}, {Aparicio},
  {Bertelli}, \& {Chiosi}}]{1996AJ....112.2596G}
---. 1996{\natexlab{b}}, \aj, 112, 2596, \dodoi{10.1086/118205}

\bibitem[{{Gallart} {et~al.}(1994){Gallart}, {Aparicio}, {Chiosi}, {Bertelli},
  \& {Vilchez}}]{1994ApJ...425L...9G}
{Gallart}, C., {Aparicio}, A., {Chiosi}, C., {Bertelli}, G., \& {Vilchez},
  J.~M. 1994, \apjl, 425, L9, \dodoi{10.1086/187298}

\bibitem[{{Gallart} {et~al.}(1996{\natexlab{c}}){Gallart}, {Aparicio}, \&
  {Vilchez}}]{1996AJ....112.1928G}
{Gallart}, C., {Aparicio}, A., \& {Vilchez}, J.~M. 1996{\natexlab{c}}, \aj,
  112, 1928, \dodoi{10.1086/118153}

\bibitem[{{Gallart} {et~al.}(2005){Gallart}, {Zoccali}, \&
  {Aparicio}}]{2005ARA&A..43..387G}
{Gallart}, C., {Zoccali}, M., \& {Aparicio}, A. 2005, \araa, 43, 387,
  \dodoi{10.1146/annurev.astro.43.072103.150608}

\bibitem[{{Geha} {et~al.}(2015){Geha}, {Weisz}, {Grocholski}, {Dolphin}, {van
  der Marel}, \& {Guhathakurta}}]{2015ApJ...811..114G}
{Geha}, M., {Weisz}, D., {Grocholski}, A., {et~al.} 2015, \apj, 811, 114,
  \dodoi{10.1088/0004-637X/811/2/114}

\bibitem[{{Girardi} {et~al.}(2005){Girardi}, {Groenewegen}, {Hatziminaoglou},
  \& {da Costa}}]{2005A&A...436..895G}
{Girardi}, L., {Groenewegen}, M.~A.~T., {Hatziminaoglou}, E., \& {da Costa}, L.
  2005, \aap, 436, 895, \dodoi{10.1051/0004-6361:20042352}

\bibitem[{{Green}(2018)}]{2018JOSS....3..695M}
{Green}, G. 2018, The Journal of Open Source Software, 3, 695,
  \dodoi{10.21105/joss.00695}

\bibitem[{{Gupta} \& {Akin}(2020)}]{2020arXiv200302838G}
{Gupta}, S., \& {Akin}, B. 2020, arXiv e-prints, arXiv:2003.02838,
  \dodoi{10.48550/arXiv.2003.02838}

\bibitem[{{Hamedani Golshan} {et~al.}(2017){Hamedani Golshan}, {Javadi}, {van
  Loon}, {Khosroshahi}, \& {Saremi}}]{2017MNRAS.466.1764H}
{Hamedani Golshan}, R., {Javadi}, A., {van Loon}, J.~T., {Khosroshahi}, H., \&
  {Saremi}, E. 2017, \mnras, 466, 1764, \dodoi{10.1093/mnras/stw3174}

\bibitem[{{Han} {et~al.}(2003){Han}, {Podsiadlowski}, {Maxted}, \&
  {Marsh}}]{2003MNRAS.341..669H}
{Han}, Z., {Podsiadlowski}, P., {Maxted}, P.~F.~L., \& {Marsh}, T.~R. 2003,
  \mnras, 341, 669, \dodoi{10.1046/j.1365-8711.2003.06451.x}

\bibitem[{{Han} {et~al.}(2002){Han}, {Podsiadlowski}, {Maxted}, {Marsh}, \&
  {Ivanova}}]{2002MNRAS.336..449H}
{Han}, Z., {Podsiadlowski}, P., {Maxted}, P.~F.~L., {Marsh}, T.~R., \&
  {Ivanova}, N. 2002, \mnras, 336, 449,
  \dodoi{10.1046/j.1365-8711.2002.05752.x}

\bibitem[{Harris {et~al.}(2020)Harris, Millman, van~der Walt, Gommers,
  Virtanen, Cournapeau, Wieser, Taylor, Berg, Smith, Kern, Picus, Hoyer, van
  Kerkwijk, Brett, Haldane, del R{\'{i}}o, Wiebe, Peterson,
  G{\'{e}}rard-Marchant, Sheppard, Reddy, Weckesser, Abbasi, Gohlke, \&
  Oliphant}]{harris2020array}
Harris, C.~R., Millman, K.~J., van~der Walt, S.~J., {et~al.} 2020, Nature, 585,
  357, \dodoi{10.1038/s41586-020-2649-2}

\bibitem[{{He} {et~al.}(2015){He}, {Zhang}, {Ren}, \&
  {Sun}}]{2015arXiv151203385H}
{He}, K., {Zhang}, X., {Ren}, S., \& {Sun}, J. 2015, arXiv e-prints,
  arXiv:1512.03385, \dodoi{10.48550/arXiv.1512.03385}

\bibitem[{{Held} {et~al.}(1999){Held}, {Saviane}, \&
  {Momany}}]{1999A&A...345..747H}
{Held}, E.~V., {Saviane}, I., \& {Momany}, Y. 1999, \aap, 345, 747,
  \dodoi{10.48550/arXiv.astro-ph/9903017}

\bibitem[{{Hern\'andez} {et~al.}(2000){Hern\'andez}, {Gilmore}, \&
  {Valls-Gabaud}}]{2000MNRAS.317..831H}
{Hern\'andez}, X., {Gilmore}, G., \& {Valls-Gabaud}, D. 2000, \mnras, 317, 831,
  \dodoi{10.1046/j.1365-8711.2000.03809.x}

\bibitem[{{Hern\'andez} {et~al.}(1999){Hern\'andez}, {Valls-Gabaud}, \&
  {Gilmore}}]{1999MNRAS.304..705H}
{Hern\'andez}, X., {Valls-Gabaud}, D., \& {Gilmore}, G. 1999, \mnras, 304, 705,
  \dodoi{10.1046/j.1365-8711.1999.02102.x}

\bibitem[{{Hidalgo} {et~al.}(2009){Hidalgo}, {Aparicio},
  {Mart{\'\i}nez-Delgado}, \& {Gallart}}]{2009ApJ...705..704H}
{Hidalgo}, S.~L., {Aparicio}, A., {Mart{\'\i}nez-Delgado}, D., \& {Gallart}, C.
  2009, \apj, 705, 704, \dodoi{10.1088/0004-637X/705/1/704}

\bibitem[{{Hidalgo} {et~al.}(2013){Hidalgo}, {Monelli}, {Aparicio}, {Gallart},
  {Skillman}, {Cassisi}, {Bernard}, {Mayer}, {Stetson}, {Cole}, \&
  {Dolphin}}]{2013ApJ...778..103H}
{Hidalgo}, S.~L., {Monelli}, M., {Aparicio}, A., {et~al.} 2013, \apj, 778, 103,
  \dodoi{10.1088/0004-637X/778/2/103}

\bibitem[{{Hirschauer} {et~al.}(2020){Hirschauer}, {Gray}, {Meixner}, {Jones},
  {Srinivasan}, {Boyer}, \& {Sargent}}]{2020ApJ...892...91H}
{Hirschauer}, A.~S., {Gray}, L., {Meixner}, M., {et~al.} 2020, \apj, 892, 91,
  \dodoi{10.3847/1538-4357/ab7b60}

\bibitem[{{Holtzman} {et~al.}(2006){Holtzman}, {Afonso}, \&
  {Dolphin}}]{2006ApJS..166..534H}
{Holtzman}, J.~A., {Afonso}, C., \& {Dolphin}, A. 2006, \apjs, 166, 534,
  \dodoi{10.1086/507074}

\bibitem[{{Holtzman} {et~al.}(2000){Holtzman}, {Smith}, \&
  {Grillmair}}]{2000AJ....120.3060H}
{Holtzman}, J.~A., {Smith}, G.~H., \& {Grillmair}, C. 2000, \aj, 120, 3060,
  \dodoi{10.1086/316844}

\bibitem[{{Hunt} {et~al.}(2024){Hunt}, {Annibali}, {Cuillandre}, {Ferguson},
  {Jablonka}, {Larsen}, {Marleau}, {Schinnerer}, {Schirmer}, {Stone},
  {Tortora}, {Saifollahi}, {Lan{\c{c}}on}, {Bolzonella}, {Gwyn}, {Kluge},
  {Laureijs}, {Carollo}, {Collins}, {Dimauro}, {Duc}, {Erkal}, {Howell},
  {Nally}, {Saremi}, {Scaramella}, {Belokurov}, {Conselice}, {Knapen},
  {McConnachie}, {McDonald}, {Miro Carretero}, {Roman}, {Sauvage}, {Sola},
  {Aghanim}, {Altieri}, {Andreon}, {Auricchio}, {Awan}, {Azzollini}, {Baldi},
  {Balestra}, {Bardelli}, {Basset}, {Bender}, {Bonino}, {Branchini}, {Brescia},
  {Brinchmann}, {Camera}, {Candini}, {Capobianco}, {Carbone}, {Carretero},
  {Casas}, {Castellano}, {Cavuoti}, {Cimatti}, {Congedo}, {Conversi}, {Copin},
  {Corcione}, {Courbin}, {Courtois}, {Cropper}, {Da Silva}, {Degaudenzi}, {Di
  Giorgio}, {Dinis}, {Dubath}, {Dupac}, {Dusini}, {Farina}, {Farrens},
  {Ferriol}, {Fosalba}, {Frailis}, {Franceschi}, {Fumana}, {Galeotta},
  {Garilli}, {Gillard}, {Gillis}, {Giocoli}, {G{\'o}mez-Alvarez}, {Granett},
  {Grazian}, {Grupp}, {Guzzo}, {Haugan}, {Hoar}, {Hoekstra}, {Holliman},
  {Holmes}, {Hook}, {Hormuth}, {Hornstrup}, {Hudelot}, {Jahnke},
  {Keih{\"a}nen}, {Kermiche}, {Kiessling}, {Kilbinger}, {Kitching}, {Kohley},
  {Kubik}, {Kuijken}, {K{\"u}mmel}, {Kunz}, {Kurki-Suonio}, {Lahav}, {Le
  Mignant}, {Lilje}, {Lindholm}, {Lloro}, {Maiorano}, {Mansutti}, {Marggraf},
  {Markovic}, {Martinet}, {Marulli}, {Massey}, {Maurogordato}, {McCracken},
  {Medinaceli}, {Mei}, {Mellier}, {Meneghetti}, {Merlin}, {Meylan}, {Moresco},
  {Moscardini}, {Munari}, {Nakajima}, {Nichol}, {Niemi}, {Nightingale},
  {Padilla}, {Paltani}, {Pasian}, {Pedersen}, {Percival}, {Pettorino}, {Pires},
  {Polenta}, {Poncet}, {Popa}, {Pozzetti}, {Racca}, {Raison}, {Rebolo},
  {Refregier}, {Renzi}, {Rhodes}, {Riccio}, {Romelli}, {Roncarelli},
  {Rossetti}, {Saglia}, {Sapone}, {Sartoris}, {Schneider}, {Schrabback},
  {Scodeggio}, {Secroun}, {Seidel}, {Serrano}, {Sirignano}, {Sirri},
  {Skottfelt}, {Stanco}, {Tallada-Cresp{\'\i}}, {Tavagnacco}, {Taylor},
  {Teplitz}, {Tereno}, {Toledo-Moreo}, {Torradeflot}, {Tutusaus}, {Valentijn},
  {Valenziano}, {Vassallo}, {Verdoes Kleijn}, {Veropalumbo}, {Wang}, {Weller},
  {Williams}, {Zamorani}, {Zucca}, {Burigana}, {De Lucia}, {George}, {Scottez},
  {Miluzio}, {Simon}, {Mora}, {Mart{\'\i}n-Fleitas}, \&
  {Scott}}]{2024arXiv240513499H}
{Hunt}, L.~K., {Annibali}, F., {Cuillandre}, J.~C., {et~al.} 2024, arXiv
  e-prints, arXiv:2405.13499, \dodoi{10.48550/arXiv.2405.13499}

\bibitem[{Hunter(2007)}]{Hunter:2007}
Hunter, J.~D. 2007, Computing in Science \& Engineering, 9, 90,
  \dodoi{10.1109/MCSE.2007.55}

\bibitem[{{Hurley} {et~al.}(2002){Hurley}, {Tout}, \&
  {Pols}}]{2002MNRAS.329..897H}
{Hurley}, J.~R., {Tout}, C.~A., \& {Pols}, O.~R. 2002, \mnras, 329, 897,
  \dodoi{10.1046/j.1365-8711.2002.05038.x}

\bibitem[{{Ioffe} \& {Szegedy}(2015)}]{2015arXiv150203167I}
{Ioffe}, S., \& {Szegedy}, C. 2015, arXiv e-prints, arXiv:1502.03167,
  \dodoi{10.48550/arXiv.1502.03167}

\bibitem[{Iqbal(2018)}]{haris_iqbal_2018_2526396}
Iqbal, H. 2018, HarisIqbal88/PlotNeuralNet v1.0.0, v1.0.0,  Zenodo,
  \dodoi{10.5281/zenodo.2526396}

\bibitem[{{Iyer} \& {Gawiser}(2017)}]{2017ApJ...838..127I}
{Iyer}, K., \& {Gawiser}, E. 2017, \apj, 838, 127,
  \dodoi{10.3847/1538-4357/aa63f0}

\bibitem[{{Iyer} {et~al.}(2019){Iyer}, {Gawiser}, {Faber}, {Ferguson},
  {Kartaltepe}, {Koekemoer}, {Pacifici}, \& {Somerville}}]{2019ApJ...879..116I}
{Iyer}, K.~G., {Gawiser}, E., {Faber}, S.~M., {et~al.} 2019, \apj, 879, 116,
  \dodoi{10.3847/1538-4357/ab2052}

\bibitem[{{Jones} {et~al.}(2019){Jones}, {Sharp}, {Reiter}, {Hirschauer},
  {Meixner}, \& {Srinivasan}}]{2019MNRAS.490..832J}
{Jones}, O.~C., {Sharp}, M.~J., {Reiter}, M., {et~al.} 2019, \mnras, 490, 832,
  \dodoi{10.1093/mnras/stz2560}

\bibitem[{{Kacharov} {et~al.}(2017){Kacharov}, {Battaglia}, {Rejkuba}, {Cole},
  {Carrera}, {Fraternali}, {Wilkinson}, {Gallart}, {Irwin}, \&
  {Tolstoy}}]{2017MNRAS.466.2006K}
{Kacharov}, N., {Battaglia}, G., {Rejkuba}, M., {et~al.} 2017, \mnras, 466,
  2006, \dodoi{10.1093/mnras/stw3188}

\bibitem[{{Kang} {et~al.}(2005){Kang}, {Sohn}, {Rhee}, {Shin}, {Chun}, \&
  {Kim}}]{2005A&A...437...61K}
{Kang}, A., {Sohn}, Y.~J., {Rhee}, J., {et~al.} 2005, \aap, 437, 61,
  \dodoi{10.1051/0004-6361:20052692}

\bibitem[{{Kirby} {et~al.}(2013){Kirby}, {Cohen}, {Guhathakurta}, {Cheng},
  {Bullock}, \& {Gallazzi}}]{2013ApJ...779..102K}
{Kirby}, E.~N., {Cohen}, J.~G., {Guhathakurta}, P., {et~al.} 2013, \apj, 779,
  102, \dodoi{10.1088/0004-637X/779/2/102}

\bibitem[{{Kouwenhoven} {et~al.}(2009){Kouwenhoven}, {Brown}, {Goodwin},
  {Portegies Zwart}, \& {Kaper}}]{2009A&A...493..979K}
{Kouwenhoven}, M.~B.~N., {Brown}, A.~G.~A., {Goodwin}, S.~P., {Portegies
  Zwart}, S.~F., \& {Kaper}, L. 2009, \aap, 493, 979,
  \dodoi{10.1051/0004-6361:200810234}

\bibitem[{Krizhevsky {et~al.}(2017)Krizhevsky, Sutskever, \&
  Hinton}]{10.1145/3065386}
Krizhevsky, A., Sutskever, I., \& Hinton, G.~E. 2017, Commun. ACM, 60, 84–90,
  \dodoi{10.1145/3065386}

\bibitem[{{Kroupa}(2001)}]{2001MNRAS.322..231K}
{Kroupa}, P. 2001, \mnras, 322, 231, \dodoi{10.1046/j.1365-8711.2001.04022.x}

\bibitem[{Lecun {et~al.}(1998)Lecun, Bottou, Bengio, \& Haffner}]{726791}
Lecun, Y., Bottou, L., Bengio, Y., \& Haffner, P. 1998, Proceedings of the
  IEEE, 86, 2278, \dodoi{10.1109/5.726791}

\bibitem[{{Lee}(1996)}]{1996AJ....112.1438L}
{Lee}, M.~G. 1996, \aj, 112, 1438, \dodoi{10.1086/118112}

\bibitem[{{Lee} {et~al.}(1993){Lee}, {Freedman}, \&
  {Madore}}]{1993AJ....106..964L}
{Lee}, M.~G., {Freedman}, W.~L., \& {Madore}, B.~F. 1993, \aj, 106, 964,
  \dodoi{10.1086/116697}

\bibitem[{{Leja} {et~al.}(2019){Leja}, {Carnall}, {Johnson}, {Conroy}, \&
  {Speagle}}]{2019ApJ...876....3L}
{Leja}, J., {Carnall}, A.~C., {Johnson}, B.~D., {Conroy}, C., \& {Speagle},
  J.~S. 2019, \apj, 876, 3, \dodoi{10.3847/1538-4357/ab133c}

\bibitem[{{Lewis} {et~al.}(2015){Lewis}, {Dolphin}, {Dalcanton}, {Weisz},
  {Williams}, {Bell}, {Seth}, {Simones}, {Skillman}, {Choi}, {Fouesneau},
  {Guhathakurta}, {Johnson}, {Kalirai}, {Leroy}, {Monachesi}, {Rix}, \&
  {Schruba}}]{2015ApJ...805..183L}
{Lewis}, A.~R., {Dolphin}, A.~E., {Dalcanton}, J.~J., {et~al.} 2015, \apj, 805,
  183, \dodoi{10.1088/0004-637X/805/2/183}

\bibitem[{{Li} {et~al.}(2023){Li}, {Liu}, {Zhang}, {Tian}, {Fu}, {Li}, \&
  {Yan}}]{2023Natur.613..460L}
{Li}, J., {Liu}, C., {Zhang}, Z.-Y., {et~al.} 2023, \nat, 613, 460,
  \dodoi{10.1038/s41586-022-05488-1}

\bibitem[{{Marleau} {et~al.}(2010){Marleau}, {Noriega-Crespo}, \&
  {Misselt}}]{2010ApJ...713..992M}
{Marleau}, F.~R., {Noriega-Crespo}, A., \& {Misselt}, K.~A. 2010, \apj, 713,
  992, \dodoi{10.1088/0004-637X/713/2/992}

\bibitem[{{Mart{\'\i}nez-Delgado} {et~al.}(1999){Mart{\'\i}nez-Delgado},
  {Aparicio}, \& {Gallart}}]{1999AJ....118.2229M}
{Mart{\'\i}nez-Delgado}, D., {Aparicio}, A., \& {Gallart}, C. 1999, \aj, 118,
  2229, \dodoi{10.1086/301077}

\bibitem[{{McQuinn} {et~al.}(2010){McQuinn}, {Skillman}, {Cannon}, {Dalcanton},
  {Dolphin}, {Hidalgo-Rodr{\'\i}guez}, {Holtzman}, {Stark}, {Weisz}, \&
  {Williams}}]{2010ApJ...721..297M}
{McQuinn}, K. B.~W., {Skillman}, E.~D., {Cannon}, J.~M., {et~al.} 2010, \apj,
  721, 297, \dodoi{10.1088/0004-637X/721/1/297}

\bibitem[{{McQuinn} {et~al.}(2024){McQuinn}, {B. Newman}, {Savino}, {Dolphin},
  {Weisz}, {Williams}, {Boyer}, {Cohen}, {Correnti}, {Cole}, {Geha}, {Gennaro},
  {Kallivayalil}, {Sandstrom}, {Skillman}, {Anderson}, {Bolatto},
  {Boylan-Kolchin}, {Garling}, {Gilbert}, {Girardi}, {Kalirai}, {Mazzi},
  {Pastorelli}, {Richstein}, \& {Warfield}}]{2024ApJ...961...16M}
{McQuinn}, K. B.~W., {B. Newman}, M.~J., {Savino}, A., {et~al.} 2024, \apj,
  961, 16, \dodoi{10.3847/1538-4357/ad1105}

\bibitem[{{Mighell} \& {Burke}(1999)}]{1999AJ....118..366M}
{Mighell}, K.~J., \& {Burke}, C.~J. 1999, \aj, 118, 366, \dodoi{10.1086/300923}

\bibitem[{{Momany}(2015)}]{2015ASSL..413..129M}
{Momany}, Y. 2015, in Astrophysics and Space Science Library, Vol. 413,
  Astrophysics and Space Science Library, ed. H.~M.~J. {Boffin}, G.~{Carraro},
  \& G.~{Beccari}, 129, \dodoi{10.1007/978-3-662-44434-4_6}

\bibitem[{{Momany} {et~al.}(2007){Momany}, {Held}, {Saviane}, {Zaggia},
  {Rizzi}, \& {Gullieuszik}}]{2007A&A...468..973M}
{Momany}, Y., {Held}, E.~V., {Saviane}, I., {et~al.} 2007, \aap, 468, 973,
  \dodoi{10.1051/0004-6361:20067024}

\bibitem[{{Monaco} {et~al.}(2009){Monaco}, {Saviane}, {Perina}, {Bellazzini},
  {Buzzoni}, {Federici}, {Fusi Pecci}, \& {Galleti}}]{2009A&A...502L...9M}
{Monaco}, L., {Saviane}, I., {Perina}, S., {et~al.} 2009, \aap, 502, L9,
  \dodoi{10.1051/0004-6361/200912412}

\bibitem[{{Monelli} {et~al.}(2010{\natexlab{a}}){Monelli}, {Gallart},
  {Hidalgo}, {Aparicio}, {Skillman}, {Cole}, {Weisz}, {Mayer}, {Bernard},
  {Cassisi}, {Dolphin}, {Drozdovsky}, \& {Stetson}}]{2010ApJ...722.1864M}
{Monelli}, M., {Gallart}, C., {Hidalgo}, S.~L., {et~al.} 2010{\natexlab{a}},
  \apj, 722, 1864, \dodoi{10.1088/0004-637X/722/2/1864}

\bibitem[{{Monelli} {et~al.}(2010{\natexlab{b}}){Monelli}, {Hidalgo},
  {Stetson}, {Aparicio}, {Gallart}, {Dolphin}, {Cole}, {Weisz}, {Skillman},
  {Bernard}, {Mayer}, {Navarro}, {Cassisi}, {Drozdovsky}, \&
  {Tolstoy}}]{2010ApJ...720.1225M}
{Monelli}, M., {Hidalgo}, S.~L., {Stetson}, P.~B., {et~al.} 2010{\natexlab{b}},
  \apj, 720, 1225, \dodoi{10.1088/0004-637X/720/2/1225}

\bibitem[{{O'Donnell}(1994)}]{1994ApJ...422..158O}
{O'Donnell}, J.~E. 1994, \apj, 422, 158, \dodoi{10.1086/173713}

\bibitem[{{Offner} {et~al.}(2014){Offner}, {Clark}, {Hennebelle}, {Bastian},
  {Bate}, {Hopkins}, {Moraux}, \& {Whitworth}}]{2014prpl.conf...53O}
{Offner}, S.~S.~R., {Clark}, P.~C., {Hennebelle}, P., {et~al.} 2014, in
  Protostars and Planets VI, ed. H.~{Beuther}, R.~S. {Klessen}, C.~P.
  {Dullemond}, \& T.~{Henning}, 53--75,
  \dodoi{10.2458/azu_uapress_9780816531240-ch003}

\bibitem[{pandas~development team(2020)}]{reback2020pandas}
pandas~development team, T. 2020, pandas-dev/pandas: Pandas, latest,  Zenodo,
  \dodoi{10.5281/zenodo.3509134}

\bibitem[{{Paszke} {et~al.}(2019){Paszke}, {Gross}, {Massa}, {Lerer},
  {Bradbury}, {Chanan}, {Killeen}, {Lin}, {Gimelshein}, {Antiga}, {Desmaison},
  {K{\"o}pf}, {Yang}, {DeVito}, {Raison}, {Tejani}, {Chilamkurthy}, {Steiner},
  {Fang}, {Bai}, \& {Chintala}}]{2019arXiv191201703P}
{Paszke}, A., {Gross}, S., {Massa}, F., {et~al.} 2019, arXiv e-prints,
  arXiv:1912.01703, \dodoi{10.48550/arXiv.1912.01703}

\bibitem[{{Ren} {et~al.}(2024){Ren}, {Jiang}, {Wang}, {Yang}, \&
  {Yan}}]{2024arXiv240206953R}
{Ren}, Y., {Jiang}, B., {Wang}, Y., {Yang}, M., \& {Yan}, Z. 2024, arXiv
  e-prints, arXiv:2402.06953, \dodoi{10.48550/arXiv.2402.06953}

\bibitem[{Robert \& Escoufier(1976)}]{f46022f3-ac12-367a-9986-5b7830d42a7b}
Robert, P., \& Escoufier, Y. 1976, Journal of the Royal Statistical Society.
  Series C (Applied Statistics), 25, 257.
\newblock \url{http://www.jstor.org/stable/2347233}

\bibitem[{{Ross} {et~al.}(2015){Ross}, {Holtzman}, {Saha}, \&
  {Anthony-Twarog}}]{2015AJ....149..198R}
{Ross}, T.~L., {Holtzman}, J., {Saha}, A., \& {Anthony-Twarog}, B.~J. 2015,
  \aj, 149, 198, \dodoi{10.1088/0004-6256/149/6/198}

\bibitem[{{Sacchi} {et~al.}(2018){Sacchi}, {Cignoni}, {Aloisi}, {Tosi},
  {Calzetti}, {Lee}, {Adamo}, {Annibali}, {Dale}, {Elmegreen}, {Gouliermis},
  {Grasha}, {Grebel}, {Hunter}, {Sabbi}, {Smith}, {Thilker}, {Ubeda}, \&
  {Whitmore}}]{2018ApJ...857...63S}
{Sacchi}, E., {Cignoni}, M., {Aloisi}, A., {et~al.} 2018, \apj, 857, 63,
  \dodoi{10.3847/1538-4357/aab844}

\bibitem[{{Saha}(1998)}]{1998AJ....115.1206S}
{Saha}, P. 1998, \aj, 115, 1206, \dodoi{10.1086/300247}

\bibitem[{{Sandler} {et~al.}(2018){Sandler}, {Howard}, {Zhu}, {Zhmoginov}, \&
  {Chen}}]{2018arXiv180104381S}
{Sandler}, M., {Howard}, A., {Zhu}, M., {Zhmoginov}, A., \& {Chen}, L.-C. 2018,
  arXiv e-prints, arXiv:1801.04381, \dodoi{10.48550/arXiv.1801.04381}

\bibitem[{{Savino} {et~al.}(2018){Savino}, {de Boer}, {Salaris}, \&
  {Tolstoy}}]{2018MNRAS.480.1587S}
{Savino}, A., {de Boer}, T.~J.~L., {Salaris}, M., \& {Tolstoy}, E. 2018,
  \mnras, 480, 1587, \dodoi{10.1093/mnras/sty1954}

\bibitem[{{Savino} {et~al.}(2019){Savino}, {Tolstoy}, {Salaris}, {Monelli}, \&
  {de Boer}}]{2019AA...630A.116S}
{Savino}, A., {Tolstoy}, E., {Salaris}, M., {Monelli}, M., \& {de Boer},
  T.~J.~L. 2019, \aap, 630, A116, \dodoi{10.1051/0004-6361/201936077}

\bibitem[{{Savino} {et~al.}(2023){Savino}, {Weisz}, {Skillman}, {Dolphin},
  {Cole}, {Kallivayalil}, {Wetzel}, {Anderson}, {Besla}, {Boylan-Kolchin},
  {Brown}, {Bullock}, {Collins}, {Cooper}, {Deason}, {Dotter}, {Fardal},
  {Ferguson}, {Fritz}, {Geha}, {Gilbert}, {Guhathakurta}, {Ibata}, {Irwin},
  {Jeon}, {Kirby}, {Lewis}, {Mackey}, {Majewski}, {Martin}, {McConnachie},
  {Patel}, {Rich}, {Simon}, {Sohn}, {Tollerud}, \& {van der
  Marel}}]{2023ApJ...956...86S}
{Savino}, A., {Weisz}, D.~R., {Skillman}, E.~D., {et~al.} 2023, \apj, 956, 86,
  \dodoi{10.3847/1538-4357/acf46f}

\bibitem[{{Schlegel, Finkbeiner \& Davis}(1998)}]{1998ApJ...500..525S}
{Schlegel, Finkbeiner \& Davis}. 1998, \apj, 500, 525, \dodoi{10.1086/305772}

\bibitem[{{Skillman} {et~al.}(2003){Skillman}, {Tolstoy}, {Cole}, {Dolphin},
  {Saha}, {Gallagher}, {Dohm-Palmer}, \& {Mateo}}]{2003ApJ...596..253S}
{Skillman}, E.~D., {Tolstoy}, E., {Cole}, A.~A., {et~al.} 2003, \apj, 596, 253,
  \dodoi{10.1086/377635}

\bibitem[{{Skillman} {et~al.}(2017){Skillman}, {Monelli}, {Weisz}, {Hidalgo},
  {Aparicio}, {Bernard}, {Boylan-Kolchin}, {Cassisi}, {Cole}, {Dolphin},
  {Ferguson}, {Gallart}, {Irwin}, {Martin}, {Mart{\'\i}nez-V{\'a}zquez},
  {Mayer}, {McConnachie}, {McQuinn}, {Navarro}, \&
  {Stetson}}]{2017ApJ...837..102S}
{Skillman}, E.~D., {Monelli}, M., {Weisz}, D.~R., {et~al.} 2017, \apj, 837,
  102, \dodoi{10.3847/1538-4357/aa60c5}

\bibitem[{{Smith}(2020)}]{2020ARA&A..58..577S}
{Smith}, R.~J. 2020, \araa, 58, 577,
  \dodoi{10.1146/annurev-astro-032620-020217}

\bibitem[{{STScI Development Team}(2013)}]{2013ascl.soft03023S}
{STScI Development Team}. 2013, {pysynphot: Synthetic photometry software
  package}, Astrophysics Source Code Library, record ascl:1303.023

\bibitem[{{Swan} {et~al.}(2016){Swan}, {Cole}, {Tolstoy}, \&
  {Irwin}}]{2016MNRAS.456.4315S}
{Swan}, J., {Cole}, A.~A., {Tolstoy}, E., \& {Irwin}, M.~J. 2016, \mnras, 456,
  4315, \dodoi{10.1093/mnras/stv2774}

\bibitem[{{Szegedy} {et~al.}(2016){Szegedy}, {Ioffe}, {Vanhoucke}, \&
  {Alemi}}]{2016arXiv160207261S}
{Szegedy}, C., {Ioffe}, S., {Vanhoucke}, V., \& {Alemi}, A. 2016, arXiv
  e-prints, arXiv:1602.07261, \dodoi{10.48550/arXiv.1602.07261}

\bibitem[{{Szegedy} {et~al.}(2015){Szegedy}, {Vanhoucke}, {Ioffe}, {Shlens}, \&
  {Wojna}}]{2015arXiv151200567S}
{Szegedy}, C., {Vanhoucke}, V., {Ioffe}, S., {Shlens}, J., \& {Wojna}, Z. 2015,
  arXiv e-prints, arXiv:1512.00567, \dodoi{10.48550/arXiv.1512.00567}

\bibitem[{{Szegedy} {et~al.}(2014){Szegedy}, {Liu}, {Jia}, {Sermanet}, {Reed},
  {Anguelov}, {Erhan}, {Vanhoucke}, \& {Rabinovich}}]{2014arXiv1409.4842S}
{Szegedy}, C., {Liu}, W., {Jia}, Y., {et~al.} 2014, arXiv e-prints,
  arXiv:1409.4842, \dodoi{10.48550/arXiv.1409.4842}

\bibitem[{{Tan} \& {Le}(2019)}]{2019arXiv190511946T}
{Tan}, M., \& {Le}, Q.~V. 2019, arXiv e-prints, arXiv:1905.11946,
  \dodoi{10.48550/arXiv.1905.11946}

\bibitem[{{Tan} \& {Le}(2021)}]{2021arXiv210400298T}
---. 2021, arXiv e-prints, arXiv:2104.00298, \dodoi{10.48550/arXiv.2104.00298}

\bibitem[{{Tantalo} {et~al.}(2022){Tantalo}, {Dall'Ora}, {Bono}, {Stetson},
  {Fabrizio}, {Ferraro}, {Nonino}, {Braga}, {da Silva}, {Fiorentino},
  {Iannicola}, {Marengo}, {Monelli}, {Mullen}, {Pietrinferni}, \&
  {Salaris}}]{2022ApJ...933..197T}
{Tantalo}, M., {Dall'Ora}, M., {Bono}, G., {et~al.} 2022, \apj, 933, 197,
  \dodoi{10.3847/1538-4357/ac7468}

\bibitem[{{Tolstoy} {et~al.}(2009){Tolstoy}, {Hill}, \&
  {Tosi}}]{2009ARA&A..47..371T}
{Tolstoy}, E., {Hill}, V., \& {Tosi}, M. 2009, \araa, 47, 371,
  \dodoi{10.1146/annurev-astro-082708-101650}

\bibitem[{{Tolstoy} {et~al.}(2001){Tolstoy}, {Irwin}, {Cole}, {Pasquini},
  {Gilmozzi}, \& {Gallagher}}]{2001MNRAS.327..918T}
{Tolstoy}, E., {Irwin}, M.~J., {Cole}, A.~A., {et~al.} 2001, \mnras, 327, 918,
  \dodoi{10.1046/j.1365-8711.2001.04785.x}

\bibitem[{{Tolstoy} \& {Saha}(1996)}]{1996ApJ...462..672T}
{Tolstoy}, E., \& {Saha}, A. 1996, \apj, 462, 672, \dodoi{10.1086/177181}

\bibitem[{{Tosi} {et~al.}(1991){Tosi}, {Greggio}, {Marconi}, \&
  {Focardi}}]{1991AJ....102..951T}
{Tosi}, M., {Greggio}, L., {Marconi}, G., \& {Focardi}, P. 1991, \aj, 102, 951,
  \dodoi{10.1086/115925}

\bibitem[{{Ural} {et~al.}(2015){Ural}, {Cescutti}, {Koch}, {Kleyna},
  {Feltzing}, \& {Wilkinson}}]{2015MNRAS.449..761U}
{Ural}, U., {Cescutti}, G., {Koch}, A., {et~al.} 2015, \mnras, 449, 761,
  \dodoi{10.1093/mnras/stv294}

\bibitem[{{Walmswell} {et~al.}(2013){Walmswell}, {Eldridge}, {Brewer}, \&
  {Tout}}]{2013MNRAS.435.2171W}
{Walmswell}, J.~J., {Eldridge}, J.~J., {Brewer}, B.~J., \& {Tout}, C.~A. 2013,
  \mnras, 435, 2171, \dodoi{10.1093/mnras/stt1444}

\bibitem[{{Weisz} {et~al.}(2014){Weisz}, {Dolphin}, {Skillman}, {Holtzman},
  {Gilbert}, {Dalcanton}, \& {Williams}}]{2014ApJ...789..147W}
{Weisz}, D.~R., {Dolphin}, A.~E., {Skillman}, E.~D., {et~al.} 2014, \apj, 789,
  147, \dodoi{10.1088/0004-637X/789/2/147}

\bibitem[{{Weisz} {et~al.}(2008){Weisz}, {Skillman}, {Cannon}, {Dolphin},
  {Kennicutt}, {Lee}, \& {Walter}}]{2008ApJ...689..160W}
{Weisz}, D.~R., {Skillman}, E.~D., {Cannon}, J.~M., {et~al.} 2008, \apj, 689,
  160, \dodoi{10.1086/592323}

\bibitem[{{Weisz} {et~al.}(2011){Weisz}, {Dalcanton}, {Williams}, {Gilbert},
  {Skillman}, {Seth}, {Dolphin}, {McQuinn}, {Gogarten}, {Holtzman}, {Rosema},
  {Cole}, {Karachentsev}, \& {Zaritsky}}]{2011ApJ...739....5W}
{Weisz}, D.~R., {Dalcanton}, J.~J., {Williams}, B.~F., {et~al.} 2011, \apj,
  739, 5, \dodoi{10.1088/0004-637X/739/1/5}

\bibitem[{{Williams} {et~al.}(2014){Williams}, {Lang}, {Dalcanton}, {Dolphin},
  {Weisz}, {Bell}, {Bianchi}, {Byler}, {Gilbert}, {Girardi}, {Gordon},
  {Gregersen}, {Johnson}, {Kalirai}, {Lauer}, {Monachesi}, {Rosenfield},
  {Seth}, \& {Skillman}}]{2014ApJS..215....9W}
{Williams}, B.~F., {Lang}, D., {Dalcanton}, J.~J., {et~al.} 2014, \apjs, 215,
  9, \dodoi{10.1088/0067-0049/215/1/9}

\bibitem[{{Williams} {et~al.}(2017){Williams}, {Dolphin}, {Dalcanton}, {Weisz},
  {Bell}, {Lewis}, {Rosenfield}, {Choi}, {Skillman}, \&
  {Monachesi}}]{2017ApJ...846..145W}
{Williams}, B.~F., {Dolphin}, A.~E., {Dalcanton}, J.~J., {et~al.} 2017, \apj,
  846, 145, \dodoi{10.3847/1538-4357/aa862a}

\bibitem[{{Williams} {et~al.}(2021){Williams}, {Durbin}, {Dalcanton}, {Lang},
  {Girardi}, {Smercina}, {Dolphin}, {Weisz}, {Choi}, {Bell}, {Rosolowsky},
  {Skillman}, {Koch}, {Lindberg}, {Hagen}, {Gordon}, {Seth}, {Gilbert},
  {Guhathakurta}, {Lauer}, \& {Bianchi}}]{2021ApJS..253...53W}
{Williams}, B.~F., {Durbin}, M.~J., {Dalcanton}, J.~J., {et~al.} 2021, \apjs,
  253, 53, \dodoi{10.3847/1538-4365/abdf4e}

\bibitem[{{Williams} {et~al.}(2023){Williams}, {Durbin}, {Lang}, {Dalcanton},
  {Dolphin}, {Smercina}, {Yanchulova Merica-Jones}, {Weisz}, {Bell}, {Gilbert},
  {Girardi}, {Gordon}, {Guhathakurta}, {Johnson}, {Lauer}, {Seth}, \&
  {Skillman}}]{2023ApJS..268...48W}
{Williams}, B.~F., {Durbin}, M., {Lang}, D., {et~al.} 2023, \apjs, 268, 48,
  \dodoi{10.3847/1538-4365/acea61}

\bibitem[{{Wyder}(2001)}]{2001AJ....122.2490W}
{Wyder}, T.~K. 2001, \aj, 122, 2490, \dodoi{10.1086/323706}

\bibitem[{{Wyder}(2003)}]{2003AJ....125.3097W}
---. 2003, \aj, 125, 3097, \dodoi{10.1086/375208}

\bibitem[{{Yan} {et~al.}(2017){Yan}, {Jerabkova}, \&
  {Kroupa}}]{2017A&A...607A.126Y}
{Yan}, Z., {Jerabkova}, T., \& {Kroupa}, P. 2017, \aap, 607, A126,
  \dodoi{10.1051/0004-6361/201730987}

\bibitem[{{Yang} {et~al.}(2021){Yang}, {Bonanos}, {Jiang}, {Lam}, {Gao},
  {Gavras}, {Maravelias}, {Wang}, {Chen}, {Tramper}, {Ren}, \&
  {Spetsieri}}]{2021A&A...647A.167Y}
{Yang}, M., {Bonanos}, A.~Z., {Jiang}, B., {et~al.} 2021, \aap, 647, A167,
  \dodoi{10.1051/0004-6361/202039596}

\bibitem[{{Zhan}(2011)}]{2011SSPMA..41.1441Z}
{Zhan}, H. 2011, Scientia Sinica Physica, Mechanica \& Astronomica, 41, 1441,
  \dodoi{10.1360/132011-961}

\bibitem[{{Zhan}(2018)}]{2018cosp...42E3821Z}
{Zhan}, H. 2018, in 42nd COSPAR Scientific Assembly, Vol.~42, E1.16--4--18

\bibitem[{Zhan(2021)}]{Zhan2021}
Zhan, H. 2021, Chinese Science Bulletin, 66, 1290,
  \dodoi{https://doi.org/10.1360/TB-2021-0016}

\end{thebibliography}
\bibliographystyle{aasjournal}

\end{document}